\documentstyle[preprint,eqsecnum,aps]{revtex}
\begin{document}


\def\Rnum{{\bf R}}
\def\Cnum{{\bf C}}

\def\eqref#1{(\ref{#1})}
\def\eqrefs#1#2{(\ref{#1}) and~(\ref{#2})}
\def\eqsref#1#2{(\ref{#1}) to~(\ref{#2})}

\def\Eqref#1{Eq.~(\ref{#1})}
\def\Eqrefs#1#2{Eqs.~(\ref{#1}) and~(\ref{#2})}
\def\Eqsref#1#2{Eqs.~(\ref{#1}) to~(\ref{#2})}

\def\secref#1{Sec.~\ref{#1}}
\def\secrefs#1#2{Secs.~\ref{#1} and~\ref{#2}}
\def\secsref#1#2{Secs.~\ref{#1} to~\ref{#2}}

\def\appref#1{App.~\ref{#1}}

\def\Ref#1{Ref.\cite{#1}}
\def\Refs#1{Refs.\cite{#1}}

\def\Cite#1{${\mathstrut}^{\cite{#1}}$}

\def\tableref#1{Table~\ref{#1}}

\def\figref#1{Fig.~\ref{#1}}

\hyphenation{Eq Eqs Sec App Ref Fig}


\def\EQ{\begin{equation}}
\def\EQs{\begin{eqnarray}}
\def\endEQ{\end{equation}}
\def\endEQs{\end{eqnarray}}

\def\eqtext#1{\hbox{\rm{#1}}}

\def\proclaim#1{\medbreak
\noindent{\it {#1}}\par\medbreak}
\def\Proclaim#1#2{\medbreak
\noindent{\bf {#1}}{\it {#2}}\par\medbreak}


\def\fewquad{\qquad\qquad}
\def\severalquad{\qquad\fewquad}
\def\manyquad{\qquad\severalquad}
\def\manymanyquad{\manyquad\manyquad}

\def\sub#1{
\setbox1=\hbox{{$\scriptscriptstyle #1$}} 
\dimen1=0.6\ht1
\mkern-2mu \lower\dimen1\box1 \hbox to\dimen1{\box1\hfill} }

\def\eqtext#1{\hbox{\rm{#1}}}

\def\endproof{
\setbox2=\hbox{{$\sqcup$}} \setbox1=\hbox{{$\sqcap$}} 
\dimen1=\wd1
\box2\kern-\dimen1 \hbox to\dimen1{\box1} }

\def\mstrut{\mathstrut}
\def\hp#1{\hphantom{#1}}

\def\mixedindices#1#2{{\mstrut}^{\mstrut #1}_{\mstrut #2}}
\def\downindex#1{{\mstrut}^{\mstrut}_{\mstrut #1}}
\def\upindex#1{{\mstrut}_{\mstrut}^{\mstrut #1}}
\def\downupindices#1#2{{\mstrut}_{\mstrut #1}^{\hp{#1}\mstrut #2}}
\def\updownindices#1#2{{\mstrut}^{\mstrut #1}_{\hp{#1}\mstrut #2}}

\def\index#1{{\scriptstyle #1}}

\def\Parder#1#2{
\mathchoice{\partial{#1} \over\partial{#2}}{\partial{#1}/\partial{#2}}{}{} }
\def\parder#1{\partial/\partial{#1}}

\def\F#1#2{F\updownindices{#1}{#2}}
\def\duF#1#2{{*F}\updownindices{#1}{#2}}
\def\nF#1#2#3{F\mixedindices{(#1)}{#2}\downindex{#3}}
\def\dunF#1#2#3{{*F}\mixedindices{(#1)}{#2}\downindex{#3}}
\def\der#1{\partial\downindex{#1}}
\def\coder#1{\partial\upindex{#1}}
\def\D#1{D\downindex{#1}}
\def\coD#1{D\upindex{#1}}
\def\covder#1{\nabla\downindex{#1}}
\def\covcoder#1{\nabla\upindex{#1}}
\def\g#1{g\downindex{#1}}
\def\flat#1{\eta\downindex{#1}}
\def\invflat#1{\eta\upindex{#1}}
\def\vol#1#2{\epsilon\downupindices{#1}{#2}}
\def\e#1#2{e\downupindices{#1}{#2}}
\def\inve#1#2{e\updownindices{#1}{#2}}
\def\x#1#2{x\mixedindices{#1}{#2}}
\def\id#1#2{\delta\mixedindices{#2}{#1}}

\def\Fder#1#2#3{F\updownindices{#1}{#2,}\upindex{#3}}
\def\Fjetder#1#2#3#4{\der{F}\mixedindices{#1,\hp{#2}#3}{\hp{#1,}#2#4}}
\def\Fzerojetder#1#2{\der{F}\mixedindices{#1}{#2}}

\def\Q#1{Q\downindex{#1}}
\def\tQ#1{{\widetilde Q}\downindex{#1}}
\def\Qtens#1#2{q\downupindices{#1}{#2}}
\def\tQtens#1#2{{\widetilde q}\downupindices{#1}{#2}}
\def\bigQ{{\bf Q}}
\def\X{\chi}
\def\tX{{\widetilde\chi}}

\def\curr#1#2{\Psi\mixedindices{#1}{#2}}
\def\dens#1{\Psi\upindex{#1}}
\def\denshat#1{{\hat\Psi}\upindex{#1}}
\def\H#1#2{\Phi\mixedindices{#1}{#2}}
\def\TH#1{\Phi\mixedindices{#1}{T}}
\def\ZH#1{\Phi\mixedindices{#1}{Z}}
\def\VH#1{\Phi\mixedindices{#1}{V}}
\def\WH#1{\Phi\mixedindices{#1}{W}}
\def\curl#1{\Theta\upindex{#1}}
\def\triv#1#2{\Upsilon\mixedindices{#1}{#2}}
\def\cons#1{\der{#1}\Psi\upindex{#1}}

\def\E#1{E\mixedindices{#1}{F}}

\def\P#1{P\downindex{#1}}
\def\tP#1{{\widetilde P}\downindex{#1}}
\def\Ptens#1#2{P\mixedindices{#2}{#1}}
\def\tPtens#1#2{{\widetilde P}\mixedindices{#2}{#1}}
\def\bigP{{\bf P}}

\def\sP#1#2{{\rm P}\mixedindices{#1}{#2}}
\def\csP#1#2{\bar {\rm P}\mixedindices{#1}{#2}}
\def\zeroP#1#2{{U}\mixedindices{#2}{#1}}
\def\firstP#1#2{{V}\mixedindices{#2}{#1}}
\def\zerocP#1#2{{\bar U}\mixedindices{#2}{#1}}
\def\firstcP#1#2{{\bar V}\mixedindices{#2}{#1}}
\def\sPhat#1{\hat{\rm P}\downupindices{#1}{}}

\def\sQ#1{{\rm Q}\downupindices{#1}{}}
\def\csQ#1{\bar {\rm Q}\downupindices{#1}{}}
\def\S#1{S\downindex{#1}}
\def\tS#1{\tilde S\downindex{#1}}

\def\sTQ#1{{\rm Q}\mixedindices{\rm T}{#1}}
\def\sZQ#1{{\rm Q}\mixedindices{\rm Z}{#1}}
\def\sVQ#1{{\rm Q}\mixedindices{\rm V}{#1}}
\def\sWQ#1{{\rm Q}\mixedindices{\rm W}{#1}}

\def\p#1#2#3#4{p\mixedindices{\hp{#1}#2 #3}{#1\hp{#2} #4}}
\def\tp#1#2#3#4{{\tilde p}\mixedindices{\hp{#1}#2 #3}{#1\hp{#2} #4}}
\def\phat#1#2#3#4{{\hat p}\mixedindices{\hp{#1}#2 #3}{#1\hp{#2} #4}}
\def\tphat#1#2#3#4{\hat{\tilde p}\mixedindices{\hp{#1}#2 #3}{#1\hp{#2} #4}}

\def\sF#1#2{\phi\mixedindices{#1}{#2}}
\def\csF#1#2{{\bar\phi}\mixedindices{#1}{#2}}
\def\sFder#1#2#3#4{\phi\mixedindices{#1\hp{#2,}#3}{\hp{#1}#2#4}}
\def\csFder#1#2#3#4{{\bar\phi}\mixedindices{#1\hp{#2}#3}{\hp{#1}#2#4}}
\def\snF#1#2#3#4{\phi\mixedindices{(#1)}{#2}\mixedindices{#3}{#4}}
\def\csnF#1#2#3#4{\bar\phi\mixedindices{(#1)}{#2}\mixedindices{#3}{#4}}

\def\sder#1#2{\partial\mixedindices{#1}{#2}}
\def\covsder#1#2{\nabla\mixedindices{#1}{#2}}

\def\sjetder#1#2#3#4{\der{\phi}\mixedindices{#1\hp{#2,}#3}{\hp{#1}#2,#4}}
\def\csjetder#1#2#3#4{\der{\bar\phi}\mixedindices{#1\hp{#2,}#3}{\hp{#1}#2,#4}}
\def\szerojetder#1#2{\der{\phi}\mixedindices{#1}{#2}}
\def\cszerojetder#1#2{\der{\bar\phi}\mixedindices{#1}{#2}}
\def\sD#1#2{D\mixedindices{#2}{#1}}

\def\sFsol#1#2{\varphi\mixedindices{#1}{#2}}
\def\csFsol#1#2{{\bar\varphi}\mixedindices{#1}{#2}}
\def\sFdersol#1#2#3#4{\varphi\mixedindices{#1\hp{#2}#3}{\hp{#1}#2#4}}
\def\csFdersol#1#2#3#4{{\bar\varphi}\mixedindices{#1\hp{#2}#3}{\hp{#1}#2#4}}
\def\spinor#1#2#3{{#1}\mixedindices{#2}{#3}}

\def\ME#1#2{\Delta\mixedindices{#2}{#1}}
\def\cME#1#2{\bar\Delta\mixedindices{#2}{#1}}

\def\KV#1#2{\xi\mixedindices{#1}{#2}}
\def\othKV#1#2{\zeta\mixedindices{#1}{#2}}
\def\cKV#1#2{{\bar\xi}\mixedindices{#1}{#2}}
\def\cothKV#1#2{\bar\zeta\mixedindices{#1}{#2}}
\def\KS#1#2{\kappa\mixedindices{#1}{#2}}
\def\cKS#1#2{{\bar\kappa}\mixedindices{#1}{#2}}
\def\K#1#2{K\updownindices{#1}{#2}}
\def\KY#1#2{Y\mixedindices{#2}{#1}}
\def\cKY#1#2{\bar Y\mixedindices{#2}{#1}}
\def\W#1{W\downindex{#1}}
\def\tW#1{{\tilde W}\downindex{#1}}
\def\bigW{{\bf W}}

\def\so#1#2{o\mixedindices{#1}{#2}}
\def\si#1#2{\iota\mixedindices{#1}{#2}}
\def\cso#1#2{\bar o\mixedindices{#1}{#2}}
\def\csi#1#2{\bar\iota\mixedindices{#1}{#2}}
\def\l#1{\ell\upindex{#1}}
\def\n#1{n\upindex{#1}}
\def\m#1{m\upindex{#1}}
\def\cm#1{\bar m\upindex{#1}}

\def\a#1#2{\alpha\mixedindices{#1}{#2}}
\def\b#1#2{\beta\mixedindices{#2}{#1}}
\def\ca#1#2{\bar\alpha\mixedindices{#1}{#2}}
\def\cb#1#2{\bar\beta\mixedindices{#2}{#1}}
\def\ta#1{\tilde\alpha\upindex{#1}}
\def\tb#1{\tilde\beta\downindex{#1}}

\def\w#1{\omega\upindex{#1}}
\def\cw#1{\bar\omega\upindex{#1}}

\def\T#1#2#3{{#1}\updownindices{#2}{#3}}

\def\linop#1{L_{#1}}
\def\Lie#1{{\cal L}_{#1}}
\def\jetLie#1{{\cal L}^{F}_{#1}}
\def\sLie#1#2#3{\pi_{#1}\mixedindices{#2}{#3}}
\def\csLie#1#2#3{{\bar\pi}_{#1}\mixedindices{#2}{#3}}
\def\gen#1{{\bf X}_{#1}}

\def\vs#1#2{{\cal V}\mixedindices{#2}{#1}}
\def\ks#1#2{{\cal K}\mixedindices{#2}{#1}}
\def\qvs#1#2{{\cal N}\mixedindices{#2}{#1}}

\def\i{{\rm i}}
\def\tot{{\rm tot\,}}
\def\pr{{\rm pr\,}}
\def\div{{\rm div\,}}

\def\C#1#2{C\mixedindices{#1}{#2}}

\def\s{s}

\def\Meq/{Maxwell's equations}
\def\conslaw/{conservation law}
\def\Kvec/{Killing vector}
\def\Kspin/{Killing spinor}
\def\Kten/{Killing tensor}
\def\KYten/{Killing-Yano tensor}

\def\ie/{i.e.}
\def\eg/{e.g.}

\hyphenation{
}

\title{ Classification of local conservation laws of Maxwell's equations }

\author{ Stephen C. Anco${}^1$\cite{email1} 
and Juha Pohjanpelto${}^2$\cite{email2} }

\address{
${}^1$Department of Mathematics\\
Brock University, 
St Catharines, ON L2S 3A1 Canada\\
${}^2$Department of Mathematics\\
Oregon State University, 
Corvallis, OR 97331-4605 U.S.A }

\maketitle

\begin{abstract}
A complete and explicit classification of
all independent local \conslaw/s of
\Meq/ in four dimensional Minkowski space is given. 
Besides the elementary linear \conslaw/s,
and the well-known quadratic \conslaw/s associated to 
the conserved stress-energy and zilch tensors,
there are also chiral quadratic \conslaw/s which are associated to
a new conserved tensor. 
The chiral \conslaw/s possess odd parity 
under the electric-magnetic duality transformation of \Meq/,
in contrast to the even parity 
of the stress-energy and zilch \conslaw/s.
The main result of the classification establishes that
every local \conslaw/ of \Meq/ 
is equivalent to a linear combination of 
the elementary \conslaw/s, 
the stress-energy and zilch \conslaw/s, 
the chiral \conslaw/s,
and their higher order extensions obtained by 
replacing the electromagnetic field tensor by 
its repeated Lie derivatives with respect to the conformal Killing vectors 
on Minkowski space. 
The classification is based on spinorial methods
and provides a direct, unified characterization of the \conslaw/s
in terms of \Kspin/s. 
\end{abstract}

\newpage

\section{Introduction}

Conservation laws play an important role in physical field theories 
by determining conserved quantities 
for the time evolution of the fields. 
For free electromagnetic fields, 
\Meq/ exhibit a rich structure of \conslaw/s. 
The well-known Maxwell stress-energy tensor yields 
local  \conslaw/s 
for energy, momentum, angular and boost momentum 
\cite{Bessel-Hagen}
which arise from \Kvec/s associated to the Poincar\'e symmetries 
of flat spacetime.
In addition, because of the conformal invariance of \Meq/,
the stress-energy tensor also yields local  \conslaw/s 
arising from conformal \Kvec/s
associated to conformal symmetries of the spacetime. 
Interestingly, \Meq/ possess another physically significant
conserved tensor, given in its original form 
by Lipkin's ``zilch'' tensor \cite{Lipkin},
and subsequently generalized in \Refs{Kibble,Fairlie,Morgan,survey}. 
This tensor yields additional local  \conslaw/s 
and corresponding conserved quantities arising from conformal \Kvec/s. 
The physical meaning of these conserved quantities is discussed
in \Ref{zilch}. 

More recently, new conserved quantities 
have been found by Fushchich and Nikitin \cite{FushchichNikitin}
through an analysis of quadratic expressions 
in the electromagnetic field variables 
whose integrals are constant in time on the solutions of \Meq/. 
These quantities correspond to local  \conslaw/s
associated with a new conserved tensor 
which is independent of the stress-energy and zilch tensors.
The new conserved tensor is physically interesting since,
as we point out here, 
it possesses odd parity, \ie/ chirality, 
under the duality transformation
interchanging the electric and magnetic fields.
Hence, 
in a remarkable contrast to the stress-energy and zilch \conslaw/s,
which are invariant under the duality transformation,
the new chiral \conslaw/s distinguish between pure 
electric and pure magnetic fields.

All these \conslaw/s and underlying conserved tensors 
have extensions obtained by 
replacing the electromagnetic fields by their repeated Lie derivatives
with respect to conformal \Kvec/s.
(See, e.g. \Ref{survey}). 
The resulting set of all such higher order local \conslaw/s 
yields an infinite number of conserved quantities.
This proliferation of \conslaw/s 
raises the immediate questions:
Do \Meq/ admit any other independent local  \conslaw/s? 
Can a unified account be given of all the different local \conslaw/s
and associated conserved tensors?

In this paper we answer these questions by 
presenting a direct, unified classification of all 
local \conslaw/s of \Meq/ in flat spacetime.
As a result of our classification, we are able to show that 
every local \conslaw/ which is quadratic in the electromagnetic fields
can be expressed as a 
linear combination of the stress-energy and zilch \conslaw/s, 
the chiral \conslaw/s, and their extensions. 
Moreover, we show that \Meq/ have no other local \conslaw/s
apart from elementary ones which are linear in the electromagnetic fields. 

Our method is based on the general approach described
in \Refs{Olver,AncoBluman1,AncoBluman2} for constructing 
local \conslaw/s for any field equations.
In this approach, all local \conslaw/s can
be obtained from adjoint symmetries which
are solutions of the formal adjoint equations of the 
determining equations for symmetries.
Ordinarily, there are additional constraint conditions which 
an adjoint symmetry must satisfy in order to yield a \conslaw/; 
however, we show that as a consequence of linearity of \Meq/
these conditions can be by-passed in the present case.

The determining equations for adjoint symmetries of 
\Meq/ can be elegantly solved by spinorial methods.
The solutions are characterized in terms of 
symmetric spinorial tensors, called \Kspin/s,  
which were first used by Penrose \cite{WalkerPenrose} to
construct first integrals for null geodesics 
in black-hole spacetimes.  
Killing spinors also play a central role in
twistor theory as the principal parts of
trace-free symmetric twistors \cite{Penrose}. 
In flat spacetime, 
\Kspin/s have an important factorization property in terms of twistors. 
This allows for a simple classification of all adjoint symmetries,
which is pivotal in our analysis of local \conslaw/s of \Meq/.
In particular, \Kspin/s lead to a unified derivation of 
the stress-energy and zilch \conslaw/s 
together with the chiral \conslaw/s.

In \secref{results} we establish some notation 
and summarize our main classification results. 
In \secref{method} we describe our method. 
We present the details of the classification analysis 
in \secrefs{adjointsymms}{currents}. 
In \secref{conversion} we translate between the
tensor form and spinor form of our classification results. 
We make some concluding remarks in \secref{conclude}.
Throughout we use the index notation and conventions of \Ref{Penrose}.

\section{Main Results}
\label{results}

The free \Meq/ for the electromagnetic field tensor
$\F{}{\mu\nu}(x) = -\F{}{\nu\mu}(x)$ 
in four dimensional Minkowski space 
$M^4=(R^4,\eta)$
are given by 
\EQ
\coder{\mu}\F{}{\mu\nu}(x) =0, \qquad
\coder{\mu}\duF{}{\mu\nu}(x) =0,
\label{Meq}
\endEQ
where, in the standard Minkowski coordinates $\x{\mu}{}$, 
$\der{\mu} = \parder{}{\x{\mu}{}}$ 
is the coordinate derivative,
$\duF{}{\mu\nu} = 
\frac{1}{2} \vol{\mu\nu\sigma\tau}{} \F{\sigma\tau}{}$ 
is the dual of $\F{}{\mu\nu}$, 
$\vol{\alpha\beta\sigma\tau}{}$ 
is the spacetime volume form,
and indices are raised and lowered using 
the spacetime metric $\flat{\mu\nu}$
and its inverse $\invflat{\mu\nu}$. 
The structure of \eqref{Meq} 
explicitly displays the symmetry of the field equations 
under the duality transformation
\EQ\label{duality}
\F{}{\mu\nu} \rightarrow \duF{}{\mu\nu}, \quad
\duF{}{\mu\nu} \rightarrow -\F{}{\mu\nu} . 
\endEQ

Let $J^q(F)$, $0\leq q\leq\infty$, 
denote the coordinate space 
\EQ
J^q(F) = \{ (\x{\mu}{},\F{}{\mu\nu},\F{}{\mu\nu,\sigma_1},
\dots,\F{}{\mu\nu,\sigma_1\cdots\sigma_q}) \},
\endEQ
where each $q$-jet
$(\x{\mu}{o},F_o\downindex{\mu\nu},F_o\downindex{\mu\nu,\sigma_1},
\dots,F_o\downindex{\mu\nu,\sigma_1\cdots\sigma_q})$ $\in J^q(F)$
is to be identified with 
a spacetime point $\x{\mu}{}=\x{\mu}{o}$
and values of the field tensor and its derivatives 
at $\x{\mu}{}=\x{\mu}{o}$,
\EQs
F_o\downindex{\mu\nu,\sigma_1\cdots\sigma_p} = 
\der{\sigma_1}\cdots\der{\sigma_p}\F{}{\mu\nu}(\x{}{o}), \quad
0\leq p\leq q , 
\label{jet}
\endEQs
where the notation \eqref{jet} with $p=0$
stands for $F_o\downindex{\mu\nu} = \F{}{\mu\nu}(\x{}{o})$.
Let $R(F)$ denote the solution space of \Meq/, 
which is the subspace of $J^1(F)$ defined by 
imposing the field equations \eqref{Meq} on $\F{}{\mu\nu}(x)$. 
The derivatives of the field equations \eqref{Meq} 
up to order $q$ 
define the $q$-fold prolonged solution space $R^q(F) \subset J^{q+1}(F)$
associated with \Meq/. 
Let $D_\mu$ be the total derivative operator
\EQ
D_\mu = 
\der{\mu}
+ \sum_{q\geq 0} 
\F{}{\alpha\beta,\mu\nu_1\dots\nu_q}
\Fjetder{\alpha\beta}{}{\nu_1\dots\nu_q}{}, 
\label{totalderop}
\endEQ
where 
$\Fjetder{\alpha\beta}{}{\tau_1\cdots\tau_r}{}$
are the partial differential operators satisfying
\EQs
\Fjetder{\alpha\beta}{}{\tau_1\cdots\tau_r}{} 
\F{}{\mu\nu,\sigma_1\cdots\sigma_q}
= \cases{ \id{[\mu}{\alpha}\id{\nu]}{\beta}
\id{(\sigma_1}{\tau_1}\cdots\id{\sigma_q)}{\tau_r} , &if $r=q$,\cr 
0 , &if $r\neq q$.}  
\endEQs

A local conserved current of \Meq/ is a vector function $\dens{\mu}$
defined on some $J^q(F)$ satisfying
\EQ
D_\mu\Psi^\mu=0
\qquad\text{ on $R^{q}(F)$}.
\label{conslaw}
\endEQ
We refer to the integer $q$ 
as the order of $\dens{\mu}$.
The conserved current \eqref{conslaw} is trivial if 
\EQ
\dens{\mu} =D_{\nu} \curl{\mu\nu}\quad
\text{on some $R^{p}(F)$},
\label{trivial}
\endEQ
where 
$\curl{\mu\nu}=-\curl{\nu\mu}$
are some functions on $J^p(F)$.
Two conserved currents are considered equivalent 
if their difference is a trivial conserved current. 
We refer to the class of conserved currents equivalent to
a current $\dens{\mu}$ as the conservation law associated with $\dens{\mu}$.
The smallest among the orders of these equivalent currents is called 
the order of the \conslaw/. 

We now write down the stress-energy, zilch, and chiral \conslaw/s
of \Meq/.
On $M^4$
let $\KV{\mu}{}$ be a conformal \Kvec/ \cite{Wald}
and $\KY{}{\mu\nu}=-\KY{}{\nu\mu}$ be a conformal \KYten/ 
\cite{DietzRudiger}. 
These are characterized, respectively, 
by the equations 
\EQ
\coder{(\mu} \KV{\nu)}{} = 
\frac{1}{4} \invflat{\mu\nu} \der{\sigma} \KV{\sigma}{} , \qquad
\coder{(\mu}\KY{}{\nu)\alpha} = 
\frac{1}{3}\invflat{\mu\nu} \der{\sigma}\KY{}{\sigma\alpha} 
+\frac{1}{3}\invflat{\alpha(\mu} \der{\sigma}\KY{}{\nu)\sigma} . 
\label{KVeqKYeq}
\endEQ
The solutions are polynomials in the spacetime coordinates $\x{\mu}{}$, 
\EQs
&& \KV{\mu}{} = 
\alpha_1\upindex{\mu} 
+\alpha_2\updownindices{\mu\nu}{} \x{}{\nu}
+\alpha_3 \x{\mu}{}
+\alpha_4\upindex{\sigma} \x{}{\sigma}\x{\mu}{}
- \frac{1}{2}\alpha_4\upindex{\mu} \x{\sigma}{}\x{}{\sigma},
\label{Kvecs}
\\
&& \KY{}{\mu\nu} =
\alpha_5\upindex{\mu\nu} 
+\alpha_6\upindex{[\mu} \x{\nu]}{}
+\vol{\sigma\tau}{\mu\nu}
\alpha_7\upindex{\sigma} \x{\tau}{}
+\alpha_8\upindex{\sigma[\mu} \x{\nu]}{}\x{}{\sigma} 
+ \frac{1}{4}\alpha_8\upindex{\mu\nu} \x{\sigma}{}\x{}{\sigma},
\label{KYtens}
\endEQs
with constant coefficients
\EQ\label{consts}
\alpha_1\upindex{\mu}, 
\alpha_2\upindex{\mu\nu}= -\alpha_2\upindex{\nu\mu}, 
\alpha_3, 
\alpha_4\upindex{\sigma}, 
\alpha_5\upindex{\mu\nu}= -\alpha_5\upindex{\nu\mu}, 
\alpha_6\upindex{\mu}, 
\alpha_7\upindex{\sigma},
\alpha_8\upindex{\mu\nu}= -\alpha_8\upindex{\nu\mu}.
\endEQ
There are 15 linearly independent conformal \Kvec/s \eqref{Kvecs}
and 20 linearly independent conformal \KYten/s \eqref{KYtens} on $M^4$.

The stress-energy and zilch \conslaw/s are, respectively, given by
\EQs
\curr{\mu}{\rm T}(F;\xi) =&&
\F{\mu\sigma}{} \F{}{\nu\sigma} \KV{\nu}{} 
-\frac{1}{4} \F{\nu\sigma}{} \F{}{\nu\sigma} \KV{\mu}{} , 
\label{Tconslaw}\\
\curr{\mu}{\rm Z}(F;\xi) =&&
\duF{\mu\sigma}{} (\Lie{\KV{}{}} \F{}{\nu\sigma}) \KV{\nu}{} 
-\F{\mu\sigma}{} (\Lie{\KV{}{}} \duF{}{\nu\sigma}) \KV{\nu}{} , 
\label{Zconslaw}
\endEQs
where 
\EQ\label{LieF}
\Lie{\xi} \F{}{\alpha\beta} = 
\KV{\sigma}{} \F{}{\alpha\beta,\sigma} 
- 2 (\der{[\alpha} \KV{\sigma}{}) \F{}{\beta]\sigma} ,\quad
\Lie{\xi} \duF{}{\alpha\beta} = 
\KV{\sigma}{} \duF{}{\alpha\beta,\sigma} 
- 2 (\der{[\alpha} \KV{\sigma}{}) \duF{}{\beta]\sigma} 
\endEQ
are the standard Lie derivatives of
the electromagnetic field tensor and its dual, 
with respect to the vector field $\KV{\sigma}{}$ on $M^4$. 

The chiral \conslaw/s are given by 
\EQs
\curr{\mu}{\rm V}(F;\xi,Y) = &&
\F{}{\nu\sigma} ( \coD{\mu}\Lie{\KV{}{}}\F{}{\alpha\beta} ) 
\KY{}{\nu\sigma\alpha\beta}
+4 \F{[\mu}{\sigma} ( \D{\nu}\Lie{\KV{}{}}\F{}{\alpha\beta} )
\KY{}{\nu]\sigma\alpha\beta}
\nonumber\\&&
+\frac{3}{5} \F{}{\nu\sigma} (\Lie{\KV{}{}}\F{}{\alpha\beta} ) 
\coder{\mu} \KY{}{\nu\sigma\alpha\beta}
+\frac{12}{5} \F{[\mu}{\sigma} ( \Lie{\KV{}{}}\F{}{\alpha\beta} )
\der{\nu}\KY{}{\nu]\sigma\alpha\beta} , 
\label{Vconslaw}
\endEQs
where
\EQ\label{KYTdef}
\KY{}{\nu\sigma\alpha\beta} = 
\KY{}{\nu\sigma} \KY{}{\alpha\beta}
- \KY{}{\nu[\alpha} \KY{}{\beta]\sigma}
-3 \invflat{[\nu|[\alpha} \KY{\tau}{\hp{\tau}\beta]} \KY{}{\tau|\sigma]}
+\frac{1}{2} \invflat{\nu[\alpha} \invflat{\beta]\sigma} 
\KY{\tau\lambda}{} \KY{}{\tau\lambda} . 
\endEQ
The current \eqref{Vconslaw} is equivalent to 
the first order current
\EQs
\curr{\mu}{\rm\hat V}(F;\xi,Y) = &&
-\Fder{}{\nu\sigma}{\mu} (\Lie{\KV{}{}}\F{}{\alpha\beta})
\KY{}{\nu\sigma\alpha\beta}
-4 \F{}{\nu\sigma,\tau} (\Lie{\KV{}{}}\F{}{\alpha\beta})
\KY{}{\nu\sigma\alpha[\mu} \invflat{\tau]\beta} 
\nonumber\\&&
-4\F{}{\nu\sigma} (\Lie{\KV{}{}}\F{}{\alpha\beta})
\der{\tau} \KY{}{\nu\sigma\alpha[\mu} \invflat{\tau]\beta} 
+\frac{3}{5} 
\F{}{\nu\sigma} (\Lie{\KV{}{}}\F{}{\alpha\beta} ) 
\coder{\mu} \KY{}{\nu\sigma\alpha\beta}
\nonumber\\&&
-\frac{8}{5} 
\F{[\mu}{\sigma} (\Lie{\KV{}{}}\F{}{\alpha\beta})
\der{\nu} \KY{}{\nu]\sigma\alpha\beta} . 
\label{minimalVconslaw}
\endEQs

In addition to the quadratic conserved currents
\eqref{Tconslaw}, \eqref{Zconslaw}, \eqrefs{Vconslaw}{minimalVconslaw},
\Meq/ also possess linear conserved currents 
given by 
\EQ\label{Wconslaw}
\curr{\mu}{\rm W}(F;W,\tilde W) =
\W{\nu} \F{\mu\nu}{} + \tW{\nu} \duF{\mu\nu}{} , 
\endEQ
where the vector functions $\W{\nu}(x),\tW{\nu}(x)$
satisfy the adjoint \Meq/
\EQ\label{adjointME}
\der{[\mu} \W{\nu]} + *\der{[\mu} \tW{\nu]} =0 . 
\endEQ

The previous linear and quadratic conserved currents 
each have higher order extensions 
given in terms of repeated Lie derivatives
on the electromagnetic field tensor. 
Let 
\EQ\label{Fextension}
\nF{n}{\zeta}{\mu\nu}(x) = 
(\Lie{\zeta})^n \F{}{\mu\nu}(x) , 
\endEQ
where $\F{}{\mu\nu}(x)$ satisfies \eqref{Meq}. 
It follows from the linearity and conformal invariance of \Meq/ \eqref{Meq}
that if $\zeta^\sigma$ is a conformal \Kvec/ 
then $\nF{n}{\zeta}{\mu\nu}(x)$ satisfies \Meq/.
Consequently, the replacement of 
$\F{}{\mu\nu}(x)$ by $\nF{n}{\zeta}{\mu\nu}(x)$
in any conserved current of order $q$ produces 
a conserved current of order $q+n$, for all $n\ge 1$. 

\Proclaim{ Theorem~2.1. }{
Let $\KV{}{}$ be a conformal \Kvec/ \eqref{Kvecs}
and $\KY{}{}$ be a conformal \KYten/ \eqref{KYtens}, 
let $\W{\nu}(x),\tW{\nu}(x)$ be solutions of \eqref{adjointME},
and define
\EQs
&& \curr{(n)\mu}{\rm T}(F;\xi) = 
\curr{\mu}{\rm T}(\nF{n}{\xi}{};\xi) , 
\label{extTconslaw}\\
&& \curr{(n)\mu}{\rm Z}(F;\xi) = 
\curr{\mu}{\rm Z}(\nF{n}{\xi}{};\xi) , 
\label{extZconslaw}\\
&& \curr{(n)\mu}{\rm\hat V}(F;\xi,Y) = 
\curr{\mu}{\rm\hat V}(\nF{n}{\xi}{};\xi,Y) , 
\label{extVconslaw}\\
&& \curr{(n)\mu}{\rm W}(F;\xi,W,\tilde W) = 
\curr{\mu}{\rm W}(\nF{n}{\xi}{};W,\tilde W) . 
\label{extWconslaw}
\endEQs
These are non-trivial conserved currents of \Meq/ \eqref{Meq}
of order $n$, $n+1$, $n+1$, $n$, respectively. }

\Proclaim{ Remark :}{
The current 
$\curr{(n)\mu}{\rm W}(F;\xi,W,\tilde W)$ is equivalent to 
$\curr{\mu}{\rm W}(F;(-\Lie{\xi})^n W,(-\Lie{\xi})^n \tilde W)$,
which is of order $0$. }

The extended currents in Theorem~2.1 
can be obviously generalized using Lie derivatives 
with respect to distinct conformal \Kvec/s 
$\othKV{}{1},\ldots,\othKV{}{n}$
in place of a single repeated conformal \Kvec/ in \eqref{Fextension}. 
However, 
this generalization does not lead to any additional independent currents. 

\Proclaim{ Theorem~2.2. }{
Every local \conslaw/ \eqref{conslaw} of order $q\ge 0$
of \Meq/ \eqref{Meq} 
is equivalent to a 
linear combination of the currents
\EQ
\curr{(n_1)\mu}{\rm T}(F;\xi), \quad
\curr{(n_2)\mu}{\rm Z}(F;\xi), \quad
\curr{(n_3)\mu}{\rm\hat V}(F;\xi,Y), \quad
\curr{\mu}{\rm W}(F;W,\tilde W) , 
\nonumber
\endEQ
involving a sum over 
conformal \Kvec/s $\xi$ and conformal \KYten/s $Y$
for each $n_1,n_2,n_3$, 
with $0\le n_1\le q$, $0\le n_2\le q-1$, $0\le n_3\le q-1$,
and solutions $W,\tilde W$ of 
the adjoint \Meq/ \eqref{adjointME}. }

Through substitution of the expressions
\eqref{Kvecs} for conformal \Kvec/s $\KV{}{}$ 
and \eqref{KYtens} for conformal \KYten/s $\KY{}{}$,
the currents \eqref{extTconslaw}, \eqref{extZconslaw}, \eqref{extVconslaw}
become homogeneous polynomials of degree $2n+1,2n+2,2n+3$
in the arbitrary constants \eqref{consts}. 
The coefficient of each monomial of these constants 
yields a conserved current,
some of which, however, are not independent. 
A complete, explicit basis of independent currents of order $q\ge 0$
is given later in Theorem~6.5 by using the null tetrad formalism 
for conformal \Kvec/s and conformal \KYten/s 
as described in \secref{conversion}. 

\Proclaim{ Proposition~2.3. }{
The set of stress-energy \conslaw/s \eqref{Tconslaw}
is a 15 dimensional vector space that admits a basis
in which 4 have no explicit $\x{\mu}{}$ dependence, 
7 are linear, and 4 are quadratic,
in the highest degree terms in $\x{\mu}{}$;
\vskip0pt
The set of zilch \conslaw/s \eqref{Zconslaw}
is a 84 dimensional vector space that admits a basis 
in which 9 have no explicit $\x{\mu}{}$ dependence, 
20 are linear, 26 are quadratic, 20 are cubic, and 9 are quartic,
in the highest degree terms in $\x{\mu}{}$;
\vskip0pt
The set of chiral \conslaw/s \eqref{Vconslaw}
is a 378 dimensional vector space that admits a basis 
in which 24 have no explicit $\x{\mu}{}$ dependence, 
54 are linear, 72 are quadratic, 78 are cubic, 
72 are quartic, 54 are quintic, and 24 are sextic, 
in the highest degree terms in $\x{\mu}{}$. 
\vskip0pt
In general, for $n\ge 0$, there are 
$\frac{1}{3}(n+1)^2(2n+3)^2(4n+5)$
linearly independent \conslaw/s 
arising from \eqref{extTconslaw},
$\frac{1}{3}(n+2)^2(2n+3)^2(4n+7)$
linearly independent \conslaw/s 
arising from \eqref{extZconslaw},
and $\frac{2}{3}(n+1)(n+3)(2n+3)(2n+7)(4n+9)$
linearly independent \conslaw/s 
arising from \eqref{extVconslaw}. }

Theorems~2.1 and~2.2 combined with Proposition~2.3 and Theorem~6.5
give a complete and explicit classification of
{\it all} non-trivial local \conslaw/s of \Meq/. 

We obtain conserved tensors 
from the stress-energy, zilch, and chiral conservation laws 
by first setting 
$\KV{\nu}{}$ to be a constant vector, 
$\KY{}{\sigma\tau}$ to be a constant skew-tensor,
and then factoring out 
$\KV{\nu}{},\KY{}{\sigma\tau\alpha\beta}$.
In \eqrefs{Tconslaw}{Zconslaw}, 
this directly leads to 
\EQs
\T{T}{\mu}{\nu}(F)= &&
\F{\mu\sigma}{} \F{}{\nu\sigma}
- \frac{1}{4} \id{\nu}{\mu} \F{\tau\sigma}{} \F{}{\tau\sigma} , 
\label{Ttens}\\
\T{Z}{\mu}{\nu\rho}(F)= &&
\F{\mu\sigma}{} \duF{}{\sigma(\nu,\rho)}
- \duF{\mu\sigma}{} \F{}{\sigma(\nu,\rho)} , 
\label{Ztens}
\endEQs
which are, respectively, 
the well-known stress-energy tensor and Lipkin's zilch tensor. 
In \eqref{minimalVconslaw}, 
after some lengthy manipulations, we extract the expression 
\EQs\label{Vtens}
\T{V}{\mu}{\nu\alpha\beta\sigma\tau}(F) =
&&
\Fder{}{\alpha\beta}{\mu} \F{}{\sigma\tau,\nu} 
+\Fder{}{\sigma\tau}{\mu} \F{}{\alpha\beta,\nu} 
- 2\Fder{}{[\alpha|[\sigma}{\mu} \F{}{\tau]|\beta],\nu}
+3 \flat{[\alpha| [\sigma} \Fder{}{\tau]\gamma}{\mu} \F{\gamma}{|\beta],\nu}
\nonumber\\&&
+3 \flat{[\sigma|[\alpha} \Fder{}{\beta]\gamma}{\mu} \F{\gamma}{|\tau],\nu}
+\flat{\alpha[\sigma} \flat{\tau]\beta}
\Fder{}{\gamma\lambda}{\mu} \F{\gamma\lambda}{,\nu} 
-\id{\nu}{\mu} 
( \Fder{}{\alpha\beta}{\gamma} \F{}{\sigma\tau,\gamma} 
\nonumber\\&&
- \Fder{}{[\alpha|[\sigma}{\gamma} \F{}{\tau]|\beta],\gamma}
+3 \flat{[\alpha| [\sigma} 
\Fder{}{\tau]\gamma}{\lambda} \F{\gamma}{|\beta],\lambda}
+\frac{1}{2} \flat{\alpha[\sigma} \flat{\tau]\beta}
\Fder{}{\gamma\lambda}{\rho} \F{\gamma\lambda}{,\rho} ) , 
\endEQs
which we call the chiral tensor. 

\Proclaim{ Theorem~2.4. }{
On the solutions of \Meq/ \eqref{Meq},
the tensors 
$\T{T}{\mu\nu}{}(\nF{n}{\xi}{})$, 
$\T{Z}{\mu\nu\rho}{}(\nF{n}{\xi}{})$, 
$\T{V}{\mu\nu\alpha\beta\sigma\tau}{}(\nF{n}{\xi}{})$,
for $n\ge 0$,
have the properties
\EQs
&& \D{\mu} \T{T}{\mu\nu}{}=0, \quad
\T{T}{\mu\nu}{} = \T{T}{(\mu\nu)}{}, \quad
\T{T}{\mu}{\mu} = 0 , 
\\
&& \D{\mu} \T{Z}{\mu\nu\rho}{}=0, \quad
\T{Z}{\mu\nu\rho}{} = \T{Z}{(\mu\nu\rho)}{}, \quad
\T{Z}{\mu\rho}{\rho} = 0 , 
\\
&& \D{\mu} \T{V}{\mu\nu\alpha\beta\sigma\tau}{}= 0, \ 
\T{V}{\mu\nu\alpha\beta\sigma\tau}{}=
\T{V}{(\mu\nu)[\sigma\tau][\alpha\beta]}{}, \ 
\T{V}{\mu\nu[\alpha\beta\sigma\tau]}{}=0, \ 
\T{V}{\mu\nu\alpha\beta\sigma}{\beta} =0, 
\\&&  
\T{V}{\tau[\nu\alpha\beta]\sigma}{\tau} =0, \quad
\T{V}{\tau(\nu\sigma)}{\tau\alpha\beta} =
-\frac{1}{2} \T{V}{\tau\rho(\nu}{\tau\rho[\alpha} \id{\beta]}{\sigma)} . 
\endEQs
Moreover, under the duality transformation \eqref{duality},
$\T{T}{\mu\nu}{}$ and $\T{Z}{\mu\nu\rho}{}$
have even parity, \ie/, are invariant, 
while 
$\T{V}{\mu\nu\alpha\beta\sigma\tau}{}$
has odd parity, \ie/, is chiral. 
The same invariance properties extend to the currents 
$\curr{(n)\mu}{\rm T}(F;\xi)$, 
$\curr{(n)\mu}{\rm Z}(F;\xi)$, 
$\curr{(n)\mu}{\rm\hat V}(F;\xi,Y)$ for all $n\ge 0$. }

\section{Preliminaries}
\label{method}

Given a conserved current \eqref{conslaw} of order $q$,
one can show that there are functions 
$R{}_{\mu}{}^{\sigma_1\cdots\sigma_p}$,
$\tilde R{}_{\mu}{}^{\sigma_1\cdots\sigma_p}$
on $J^{q+1}(F)$ 
so that $\dens{\mu}$ identically satisfies
\EQ
\D{\mu} \dens{\mu} = \sum_{0\leq p\leq q} \Big( 
R{}_{\mu}{}^{\sigma_1\cdots\sigma_p} 
\F{\mu\nu}{,\nu\sigma_1\cdots\sigma_p}
+ \tilde R{}_{\mu}{}^{\sigma_1\cdots\sigma_p} 
\duF{\mu\nu}{,\nu\sigma_1\cdots\sigma_p}
\Big) .
\endEQ
An application of the standard integration by parts procedure \cite{Olver}
then leads to an equivalent conserved current, 
which for convenience we again denote by $\dens{\mu}$
and which has order at most $2q-1$, 
identically satisfying
\EQ
\D{\mu}\dens{\mu} = 
\Q{\nu} \F{\mu\nu}{,\mu} + \tQ{\nu} \duF{\mu\nu}{,\mu}
\label{charconslaw}
\endEQ
for some functions $\Q{\nu},\tQ{\nu}$ on $J^{r}(F)$
for some $r\leq 2q$. 
We refer to the pair $\bigQ{}=(\Q{\nu},\tQ{\nu})$ 
as the characteristic of the conserved current $\dens{\mu}$
and call the integer $r$ the order of $\bigQ$.
If $\H{\mu}{}$ is a conserved current equivalent to $\dens{\mu}$
then $\bigQ$ is called a characteristic {\it admitted} by $\H{\mu}{}$.

A function $H$ defined on some $J^p(F)$
is a total divergence if and only if 
it is annihilated by the Euler operators
$\E{\mu\nu}(H) = 
\sum_{0\leq k\leq p}
(-1)^k \D{\sigma_1}\cdots\D{\sigma_k}
\Fjetder{\mu\nu}{}{\sigma_1\cdots\sigma_k}{} (H)$. 
Hence, 
all characteristics $\bigQ$ of order $r$ 
are determined by the equation
\EQ\label{eulereq}
\E{\mu\nu}( \Q{\beta} \F{\alpha\beta}{,\alpha} 
+ \tQ{\beta} \duF{\alpha\beta}{,\alpha} )
=0 
\qquad\eqtext{ on some $J^{p}(F)$. }
\endEQ
After some manipulations, this equation yields
\EQs
\D{[\mu}\Q{\nu]} +*\D{[\mu}\tQ{\nu]} = && 
\sum_{0\leq k\leq r}
(-1)^k \D{\sigma_1}\cdots\D{\sigma_k} 
( \Qtens{\beta\mu\nu}{\sigma_1\dots\sigma_k} \F{\alpha\beta}{,\alpha} 
+ \tQtens{\beta\mu\nu}{\sigma_1\dots\sigma_k} \duF{\alpha\beta}{,\alpha} )
\label{Qdeteq}
\endEQs
on $J^{p}(F)$, where
\EQ
\Qtens{\beta}{\mu\nu\sigma_1\cdots\sigma_k} =
\Fjetder{\mu\nu}{}{\sigma_1\cdots\sigma_k}{} \Q{\beta}, 
\quad
\tQtens{\beta}{\mu\nu\sigma_1\cdots\sigma_k} =
\Fjetder{\mu\nu}{}{\sigma_1\cdots\sigma_k}{} \tQ{\beta},
\quad 0\leq k\leq r, 
\endEQ
are the coefficients of the Fr\'echet derivative of $\Q{\beta},\tQ{\beta}$.
The solutions of the determining equations \eqref{Qdeteq}
for $r\ge 0$
are the characteristics for all conserved currents of \Meq/. 
Furthermore, given a solution $\bigQ{}=(\Q{\nu},\tQ{\nu})$, 
one can invert the Euler operator equations \eqref{eulereq}
by applying a standard homotopy operator (see \cite{Olver,AncoBluman1})
to obtain an explicit integral formula 
for a current $\dens{\mu}$ 
in the characteristic form \eqref{charconslaw}.

From the determining equations \eqref{Qdeteq},
we see that, on solutions of \Meq/, 
all characteristics $\bigQ$ of order $r$
satisfy 
\EQ
\D{[\mu} \Q{\nu]} + *\D{[\mu} \tQ{\nu]} =0
\qquad\eqtext{on $R^{r}(F)$} . 
\label{Qadsymmeq}
\endEQ
These equations are 
the adjoint of the determining equations for symmetries of \Meq/
\cite{survey,symmpaper}. 
We refer to them as the {\it adjoint symmetry equations}
and we call functions $\bigP=(\P{\nu},\tP{\nu})$ defined on $J^r(F)$
satisfying 
\EQ
\D{[\mu} \P{\nu]} + *\D{[\mu} \tP{\nu]} =0
\qquad\eqtext{on $R^{r}(F)$}
\label{Pdeteq}
\endEQ
{\it adjoint symmetries} of order $r$ of \Meq/.
Note that the gradients
\EQ
\P{\nu} = D_{\nu}\X,
\quad 
\tP{\nu} = D_{\nu}\tX,
\label{gaugeP}
\endEQ
for any functions $\X,\tX$ on some $J^p(F)$
are trivially a solution of \eqref{Pdeteq}.
We call $\bigP$ an {\it adjoint gauge symmetry}
if it agrees with \eqref{gaugeP} on $R^p(F)$,
and we consider two adjoint symmetries to be equivalent 
if their difference is an adjoint gauge symmetry. 
The order $r$ of an adjoint symmetry $\bigP$ is called {\it minimal}
if it is the smallest among the orders of all adjoint symmetries
equivalent to $\bigP$. 
If $\bigP$ is not equivalent to an adjoint gauge symmetry 
then we call $\bigP$ non-trivial. 

One can easily show that if a characteristic $\bigQ$
agrees with gradient expressions \eqref{gaugeP}
when restricted to solutions of \Meq/,
then it determines a trivial conserved current \eqref{trivial}
with $\curl{\mu\nu}= \X \F{\mu\nu}{} + \tX\duF{\mu\nu}{}$. 
The resulting \conslaw/ reflects the well-known divergence identities
\EQ
\D{\mu}( \F{\mu\nu}{,\nu} )=0 ,\quad \D{\mu}( \duF{\mu\nu}{,\nu} ) =0 , 
\qquad\eqtext{ on $J^2(F)$ }, 
\endEQ
which express the conservation of 
electric and magnetic charges in the free \Meq/.
Consequently, 
we call a characteristic $\bigQ$ trivial if 
\EQ
\Q{\nu} = D_{\nu}\X,
\quad 
\tQ{\nu} = D_{\nu}\tX,
\qquad\eqtext{ on $R^p(F)$ }
\label{gaugeQ}
\endEQ
for some functions $\X$, $\tX$ on $J^p(F)$, 
and we consider two characteristics to be equivalent 
if their difference is a trivial characteristic. 

Typically, as advocated \eg/ in \Refs{Olver,AncoBluman1}, 
the classification of conserved currents is based on 
first solving the adjoint symmetry equations
and then verifying which of the solutions satisfy 
the determining equations for characteristics. 
However, a serious complication arises for \Meq/. 
As we will see in \secref{currents}, 
in contrast to the evolutionary PDEs studied in \Refs{Olver,AncoBluman1}, 
\Meq/ possess equivalence classes of non-trivial adjoint symmetries 
all of which fail to satisfy the determining equations \eqref{Qdeteq} 
and, hence, are not characteristics of conserved currents. 
More importantly, 
adjoint symmetries that are equivalent to 
the characteristic of a conserved current 
typically also fail to satisfy \eqref{Qdeteq}. 
Thus, for a complete classification of characteristics,
one needs not only to find the equivalence classes of adjoint symmetries,
but also to determine whether each class admits a representative
that satisfies the determining equations \eqref{Qdeteq}. 

Here we circumvent these difficulties 
by employing a variant of the standard integral formula
for constructing a conserved current from its characteristic
\cite{AncoBluman3,AncoBluman4}.
Let 
\EQ
\H{\mu}{}(\bigP) =
\int_0^1 d\lambda \Big(
\P{\nu}
(x,\lambda F, \lambda \partial F, \ldots, \lambda \partial^q F) 
\F{\mu\nu}{} + 
\tP{\nu}
(x,\lambda F, \lambda \partial F, \ldots, \lambda \partial^q F) 
\duF{\mu\nu}{} \Big),
\label{Intformula}
\endEQ
where $\bigP=( \P{\nu}(x,F,\partial F,\ldots,\partial^q F)$,
$\tP{\nu}(x,F,\partial F,\ldots,\partial^q F) )$
is a pair of functions defined on some $J^q(F)$,
and $\partial^p F$ stands collectively for the variables 
$\F{}{\mu\nu,\sigma_1\cdots\sigma_p}$, $p\ge 0$. 

\Proclaim{ Proposition~3.1. }{
Let $\bigP= (\P{\nu},\tP{\nu})$ 
be an adjoint symmetry of order $q$ of \Meq/. 
Then $\H{\mu}{}(\bigP)$ is a conserved current of order $q$ of \Meq/. 
If $\bigP$ is equivalent to the characteristic $\bigQ$ 
of a conserved current $\dens{\mu}$, 
then the current $\H{\mu}{}(\bigP)$ is equivalent to $\dens{\mu}$. 
In particular, if $\bigP$ is equivalent to a trivial characteristic, 
then $\H{\mu}{}(\bigP)$ is a trivial current. }

\proclaim{Proof} The proof of the first and third claims 
amounts to a straightforward computation and will be omitted. 
As to the second claim, 
suppose that $\bigQ= (\Q{\nu},\tQ{\nu})$ 
is the characteristic of a conserved current $\dens{\mu}$.
Using \eqref{Qdeteq}, we see that
\EQs
\D{\mu}\H{\mu}{}(\bigQ) = && 
\int_0^1  d\lambda  \Big(
\Q{\nu}
(x,\lambda F, \lambda \partial F, \ldots, \lambda \partial^q F) 
\F{\mu\nu}{,\mu}
+\tQ{\nu}
(x,\lambda F, \lambda \partial F, \ldots, \lambda \partial^q F) 
\duF{\mu\nu}{,\mu}
\nonumber\\&&\qquad
+ \lambda \sum_{0\leq k\leq q} (-1)^k
\D{\sigma_1} \cdots \D{\sigma_k} \Big( 
\Qtens{\nu\alpha\beta}{\sigma_1\dots\sigma_k}
(x,\lambda F, \lambda \partial F, \ldots, \lambda \partial^q F) 
\F{\mu\nu}{,\mu}
\nonumber\\&&\qquad
+ \tQtens{\nu\alpha\beta}{\sigma_1\dots\sigma_k}
(x,\lambda F, \lambda \partial F, \ldots, \lambda \partial^q F) 
\duF{\mu\nu}{,\mu} 
\Big) \F{\alpha\beta}{} \Big) .
\nonumber
\endEQs
A repeated integration by parts yields the expression
\EQs
\D{\mu} \H{\mu}{}(\bigQ) = &&
\int_0^1 d\lambda \Big( 
\Big( \Q{\nu}
(x,\lambda F, \lambda \partial F, \ldots, \lambda \partial^q F) 
\nonumber\\&&
+\lambda \sum_{0\leq k\leq q} 
\Qtens{\nu\alpha\beta}{\sigma_1\dots\sigma_k}
(x,\lambda F, \lambda \partial F, \ldots, \lambda \partial^q F) 
\F{\alpha\beta}{,\sigma_1\dots\sigma_k} \Big)
\F{\mu\nu}{,\nu} 
\nonumber\\&&
+\Big( \tQ{\nu}
(x,\lambda F, \lambda \partial F, \ldots, \lambda \partial^q F) 
\nonumber\\&&
+\lambda \sum_{0\leq k\leq q} 
\tQtens{\nu\alpha\beta}{\sigma_1\dots\sigma_k}
(x,\lambda F, \lambda \partial F, \ldots, \lambda \partial^q F) 
\duF{\alpha\beta}{,\sigma_1\dots\sigma_k} \Big)
\duF{\mu\nu}{,\nu} 
\Big) +\D{\nu}\triv{\nu}{},
\nonumber
\endEQs
where $\triv{\nu}{}$ vanishes on solutions of \Meq/.
Now, an application of the fundamental theorem of calculus to 
the above integral gives 
\EQ
\D{\mu} \H{\mu}{}(\bigQ)=
\Q{\nu} \F{\mu\nu}{,\mu} + \tQ{\nu}\duF{\mu\nu}{,\mu}
+\D{\nu}\triv{\nu}{}
\qquad\eqtext{ on some $J^{p+1}(F)$. }
\nonumber
\endEQ
Thus by \eqref{charconslaw},
the equation 
$\D{\mu} \H{\mu}{}(\bigQ) = \D{\mu} (\dens{\mu} + \triv{\mu}{})$
holds identically on $J^{p+1}(F)$.
Consequently, we have (see, \eg/ \cite{exactbicomplex1,exactbicomplex2})
\EQ
\H{\mu}{}(\bigQ) = \dens{\mu} + \triv{\mu}{} + D_\nu \curl{\mu\nu}
\qquad\eqtext{ on $J^{p}(F)$, }
\endEQ
for some functions $\curl{\mu\nu} = -\curl{\nu\mu}$. 
Thus, since $\triv{\mu}{}=0$ on $R^{p-1}(F)$,
we see that $\H{\mu}{}(\bigQ)$ and $\dens{\mu}$ 
are equivalent conserved currents. 
\endproof

We emphasize that, as a consequence of Proposition~3.1,  
one can {\it completely} classify 
all non-trivial local \conslaw/s of \Meq/ by the following steps: 
\vskip0pt
(i) classify up to equivalence all adjoint symmetries of \Meq/;
\hfill\vskip0pt
(ii) use formula \eqref{Intformula} to construct 
the conserved currents arising from 
the equivalence classes of adjoint symmetries found in step (i);
\hfill\vskip0pt
(iii) determine all equivalence classes of the conserved currents 
found in step (ii). 

We carry out step (i) in \secref{adjointsymms}
and steps (ii), (iii) in \secref{currents}.
In step (iii) we first 
calculate a characteristic admitted by each conserved current in step (ii)
and then we determine the equivalence classes of these characteristics; 
finally, we find all equivalence classes of conserved currents 
by employing the following result. 

\Proclaim{Theorem~3.2. }{
There is a one-to-one correspondence between
equivalence classes of conserved currents 
and equivalence classes of characteristics
for \Meq/ \eqref{Meq}. }

\Proclaim{Corollary~3.3. }{
Let $\bigP= (\P{\nu},\tP{\nu})$ 
be an adjoint symmetry of \Meq/. 
If $\bigP$ is not equivalent to the characteristic $\bigQ$ of
the conserved current $\H{\mu}{}(\bigP)$,
then $\bigP$ is not equivalent to the characteristic of
any non-trivial conserved current of \Meq/. }

To prove Theorem~3.2
we start with a preliminary result. 

\Proclaim{Lemma~3.4. }{
Suppose $H$ is a function defined on $J^q(F)$ with the form
\EQ
H = \sum_{0\leq p\leq q-1}(
M_\mu{}^{\sigma_1\cdots\sigma_p}
\F{\mu\nu}{,\nu\sigma_1\cdots\sigma_p}
+{\tilde M}_\mu{}^{\sigma_1\cdots\sigma_p}
\duF{\mu\nu}{,\nu\sigma_1\cdots\sigma_p}),
\label{Geqn}
\endEQ
where 
$M_\mu{}^{\sigma_1\cdots\sigma_p} = 
M_\mu{}^{(\sigma_1\cdots\sigma_p)}$,
${\tilde M}_\mu{}^{\sigma_1\cdots\sigma_p} = 
{\tilde M}_\mu{}^{(\sigma_1\cdots\sigma_p)}$, 
$p\ge 0$, 
are some differential functions on $J^q(F)$.
If $H$ vanishes identically on $J^q(F)$, 
then, for $p=0$, 
\EQ
M_\mu=0,\qquad {\tilde M}_\mu=0\quad \text{ on $R^{q-1}(F)$} , 
\label{MERelations1}
\endEQ
and there are functions 
$N^{\sigma_1\cdots\sigma_p} = 
N^{(\sigma_1\cdots\sigma_p)}$, 
${\tilde N}^{\sigma_1\cdots\sigma_p} = 
{\tilde N}^{(\sigma_1\cdots\sigma_p)}$
on $J^q(F)$ so that for all $1\leq p\leq q-1$,
\EQ
M_\mu{}^{\sigma_1\cdots\sigma_p} = 
\delta_{\mu}^{(\sigma_1}N^{\sigma_2\cdots\sigma_p)},
\qquad
{\tilde M}_\mu{}^{\sigma_1\cdots\sigma_p} = 
\delta_{\mu}^{(\sigma_1}{\tilde N}^{\sigma_2\cdots\sigma_p)}
\quad\text{ on $R^{q-1}(F)$} . 
\label{MERelations2}
\endEQ}

\proclaim{Proof of Lemma~3.4. } 
First note that we only need to prove the first equation in 
both \eqrefs{MERelations1}{MERelations2}
since the second equation follows from the first one by duality.

Apply the partial derivative operator
$\Fjetder{\alpha\beta}{}{\gamma_1\dots\gamma_{p}}{}$ 
to \eqref{Geqn} 
and restrict the resulting expression to $R^{q-1}(F)$
to obtain 
\EQ
M^{\alpha(\gamma_1\cdots\gamma_{p-1}}\eta^{\gamma_{p})\beta}-
M^{\beta(\gamma_1\cdots\gamma_{p-1}}\eta^{\gamma_{p})\alpha}+
{\tilde M}^{\nu(\gamma_1\cdots\gamma_{p-1}}
\vol{\nu}{\gamma_{p})\alpha\beta} =0
\quad\text{on $R^{q-1}(F)$}.
\label{yht1}
\endEQ
When $p=1$, equation \eqref{yht1} immediately shows that 
$M_\mu=0$ 
on $R^{q-1}(F)$.
Now suppose that $p\geq 2$. 
In equation \eqref{yht1}, 
on one hand, symmetrize over the
indices $\alpha$, $\gamma_1$, \dots, $\gamma_{p}$ and,
on the other hand, 
contract over the indices $\beta$, $\gamma_{p}$ and then
symmetrize over $\alpha$, $\gamma_1$, \dots, $\gamma_{p-1}$ to see that
\EQs
&& M^{(\alpha\gamma_1\cdots\gamma_{p-1}}\eta^{\gamma_{p})\beta}=
M^{\beta(\gamma_1\cdots\gamma_{p-1}}\eta^{\gamma_{p}\alpha)} ,\ 
M^{(\alpha\gamma_1\cdots\gamma_{p-1})} =
\frac{p-1}{p+2}M_{\tau}{}^{\tau(\gamma_1\cdots\gamma_{p-2}}
\eta^{\gamma_{p-1}\alpha)} , 
\endEQs
on $R^{q-1}(F)$. 
By combining these equations, we have that
\EQ
M^{\beta(\gamma_1\cdots\gamma_{p-1}}\eta^{\gamma_{p}\alpha)}-
\frac{p-1}{p+2}M_\tau{}^{\tau(\gamma_1\cdots\gamma_{p-2}}
\eta^{\gamma_{p-1}\alpha}\eta^{\gamma_{p})\beta}=0 
\quad\text{on $R^{q-1}(F)$}.
\label{yht4}
\endEQ
Next choose a covector $X_\gamma\in T^*(M^4)$. 
Then equation \eqref{yht4} 
yields
\EQ
(M^{\beta\gamma_1\cdots\gamma_{p-1}}-
\frac{p-1}{p+2}M_\tau{}^{\tau\gamma_1\cdots\gamma_{p-2}}
\eta^{\gamma_{p-1}\beta})
X_{\gamma_1}\cdots X_{\gamma_{p-1}}=0
\quad\text{on $R^{q-1}(F)$},
\label{yht5}
\endEQ
whenever $X_\gamma X^\gamma\neq0$. 
By continuity, equation \eqref{yht5}
holds for all $X_\gamma\in T^*(M^4)$
and thus
\EQ
M^{\beta\gamma_1\cdots\gamma_{p-1}}
=\frac{p-1}{p+2}M_\tau{}^{\tau(\gamma_1\cdots\gamma_{p-2}}
\eta^{\gamma_{p-1})\beta} 
\eqtext{ on $R^{q-1}(F)$}. 
\nonumber
\endEQ
This proves the first equation in \eqref{MERelations2} 
with 
$N^{\sigma_1\cdots\sigma_{p-2}}=
\frac{p-1}{p+2}
M_\tau{}^{\tau\sigma_1\cdots\sigma_{p-2}}$.
\endproof

\proclaim{Proof of Theorem~3.2. } 
By Proposition~3.1, 
a conserved current with a trivial characteristic is itself trivial. 
Conversely, suppose that
$\dens{\mu}$ is a trivial conserved current 
in the characteristic form \eqref{charconslaw}. 
Since 
$\dens{\mu} = \D{\nu}\curl{\mu\nu}$ on some $R^{q}(F)$,
there are  functions 
$R_\nu{}^{\tau_1\cdots\tau_p}$,
${\tilde R}_\nu{}^{\tau_1\cdots\tau_p}$ 
on $J^{q+1}(F)$ so that
\EQ
\dens{\mu} = \D{\nu}\curl{\mu\nu} 
+ \sum_{0\leq p\leq q}(
R_\nu{}^{\mu\sigma_1\cdots\sigma_p}
\F{\tau\nu}{,\tau\sigma_1\cdots\sigma_p}+
{\tilde R}_\nu{}^{\mu\sigma_1\cdots\sigma_p}
\duF{\tau\nu}{,\tau\sigma_1\cdots\sigma_p})
\quad\text{on $J^{q+1}(F)$}.
\label{yht6}
\endEQ
Note that we can manipulate the term 
$R_\nu{}^{\mu\sigma_1\cdots\sigma_p}
\F{\tau\nu}{,\tau\sigma_1\cdots\sigma_p}$
by using the identities
\EQs
&&(R_\nu{}^{\mu\sigma_1\sigma_2\cdots\sigma_{p}}
-R_\nu{}^{(\mu\sigma_1\sigma_2\cdots\sigma_{p})})
\F{\tau\nu}{,\tau\sigma_1\sigma_2\cdots\sigma_{p}}=
\frac{2p}{p+1} 
R_\nu{}^{[\mu\sigma_1]\sigma_2\cdots\sigma_{p}}
\F{\tau\nu}{,\tau\sigma_1\sigma_2\cdots\sigma_{p}}
\nonumber\\
&&
=\D{\sigma_1} \Big(
\frac{2p}{p+1} R_\nu{}^{[\mu\sigma_1]\sigma_2\cdots\sigma_{p}}
\F{\tau\nu}{,\tau\sigma_2\cdots\sigma_{p}} \Big)
-\Big( \frac{2p}{p+1}
\D{\sigma_1} R_\nu{}^{[\mu\sigma_1]\sigma_2\cdots\sigma_{p}} \Big)
\F{\tau\nu}{,\tau\sigma_2\cdots\sigma_{p}} . 
\label{yht7}
\endEQs
The two terms in the final equality in \eqref{yht7} 
can be incorporated, respectively, into the terms 
$\D{\mu}\curl{\mu\nu}$ 
and 
$R_\nu{}^{\mu\sigma_1\cdots\sigma_{p-1}}
\F{\tau\nu}{,\tau\sigma_1\cdots\sigma_{p-1}}$
in \eqref{yht6}. 
Hence it is clear that by proceeding inductively we can arrange the 
coefficient functions 
$R_\nu{}^{\mu\sigma_1\cdots\sigma_p}$,
${\tilde R}_\nu{}^{\mu\sigma_1\cdots\sigma_p}$
in \eqref{yht6} to be symmetric in their upper indices:
\EQ
R_\nu{}^{\mu\sigma_1\cdots\sigma_p}=
R_\nu{}^{(\mu\sigma_1\cdots\sigma_p)},
\qquad
{\tilde R}_\nu{}^{\mu\sigma_1\cdots\sigma_p}=
{\tilde R}_\nu{}^{(\mu\sigma_1\cdots\sigma_p)},
\quad 1\leq p\leq q. 
\label{yht8}
\endEQ
Now, since $\dens{\mu}$ satisfies \eqref{charconslaw}, 
we have by \eqref{yht6} that 
\EQs
&& \Q{\nu}  \F{\mu\nu}{,\mu} + \tQ{\nu} \duF{\mu\nu}{,\mu}=
\nonumber\\
&&\quad
\sum_{0\leq p\leq q} \bigg(
( D_\mu R_\nu{}^{\mu\sigma_1\cdots\sigma_p}
+R_\nu{}^{\sigma_1\cdots\sigma_p} )
\F{\tau\nu}{,\tau\sigma_1\cdots\sigma_p}
+(D_\mu {\tilde R}_\nu{}^{\mu\sigma_1\cdots\sigma_p}
+{\tilde R}_\nu{}^{\sigma_1\cdots\sigma_p})
\duF{\tau\nu}{,\tau\sigma_1\cdots\sigma_p} \bigg)
\label{yht9}
\endEQs
on $J^{q+2}(F)$, 
where
\EQs
R_\nu{}^{\tau_1\cdots\tau_p}=0, \qquad
{\tilde R}_\nu{}^{\tau_1\cdots\tau_p}=0,
\qquad \text{if $p\geq q+1$ or $p=0$.}
\nonumber
\endEQs
Then Lemma~3.4 together with equations \eqrefs{yht8}{yht9} 
implies that
there are functions 
$N^{\sigma_1\cdots\sigma_p}$, 
${\tilde N}^{\sigma_1\cdots\sigma_p}$ 
so that on $R^{q+1}(F)$, 
\EQ
\D{\mu} R_\nu{}^{\mu} = \Q{\nu},
\qquad
\D{\mu} R_\nu{}^{\mu\sigma_1\cdots\sigma_p}
+R_\nu{}^{\sigma_1\cdots\sigma_p} 
=\delta_\nu^{(\sigma_1} N^{\sigma_2\cdots\sigma_p)},
\quad
1\le p\le q,
\endEQ
and similarly, 
\EQ
\D{\mu} {\tilde R}_\nu{}^{\mu} = \tQ{\nu},
\qquad
\D{\mu} {\tilde R}_\nu{}^{\mu\sigma_1\cdots\sigma_p}
+{\tilde R}_\nu{}^{\sigma_1\cdots\sigma_p} 
=\delta_\nu^{(\sigma_1} {\tilde N}^{\sigma_2\cdots\sigma_p)},
\quad 
1\le p\le q. 
\endEQ
It is easy to see that these equations lead to 
\EQs
&& \Q{\mu} = D_\mu \Big( \sum_{0\leq p\leq q-1}(-1)^p
D_{\sigma_1}\cdots D_{\sigma_p}
N^{\sigma_1\cdots\sigma_p} \Big) , 
\nonumber\\
&& 
\tQ{\mu} = D_\mu \Big( \sum_{0\leq p\leq q-1}(-1)^p
D_{\sigma_1}\cdots D_{\sigma_p}
{\tilde N}^{\sigma_1\cdots\sigma_p} \Big) ,
\qquad\eqtext{ on $R^{q+1}(F)$ }. 
\nonumber
\endEQs
Thus $\bigQ = (\Q{\mu},\tQ{\mu})$ is a trivial characteristic.
\endproof

\section{Classification of adjoint symmetries}
\label{adjointsymms}

We solve the adjoint symmetry equations \eqref{Pdeteq} 
by spinorial methods. 
Fix a complex null tetrad \cite{Penrose} basis $\e{\mu}{AA'}$ 
for $\flat{\mu\nu}$, satisfying $\invflat{\mu\nu}\e{\mu}{AA'}\e{\nu}{BB'}
=\vol{}{AB}\vol{}{A'B'}$,
with the inverse $\inve{\mu}{AA'}$,
where $\vol{}{AB}$ is the spin metric. 
In spinor form \Meq/ \eqref{Meq} become
\EQ
\sder{B}{A'} \sF{}{AB}(x) =0, \quad \sder{B'}{A} \csF{}{A'B'}(x) =0 , 
\label{spinorMeq}
\endEQ
where $\sF{}{AB}$ is the electromagnetic spinor defined by
$\F{}{\mu\nu}\inve{\mu}{AA'}\inve{\nu}{BB'}=
\sF{}{AB}\vol{A'B'}{} + \csF{}{A'B'}\vol{AB}{}$ 
and $\der{AA'}=\inve{\mu}{AA'}\der{\mu}$ 
is the spinorial derivative operator.
The duality symmetry \eqref{duality} of \Meq/ 
corresponds to the transformation
\EQ
\sF{}{AB} \rightarrow -\i\sF{}{AB}, \quad
\csF{}{A'B'} \rightarrow \i\csF{}{A'B'}.
\label{sduality}
\endEQ

We let 
\EQs
&& \sFder{}{AB,}{C'_1\cdots C'_p}{C_1\cdots C_p} 
= \frac{1}{2} \F{}{\mu\nu,\sigma_1\cdots\sigma_p} 
\inve{\mu}{AA'}\inve{\nu A'}{B} 
\inve{\sigma_1 C'_1}{C_1}\cdots \inve{\sigma_p C'_p}{C_p} , 
\label{dervar}\\
&& \csFder{}{A'B',}{C_1\cdots C_p}{C'_1\cdots C'_p}  
= \frac{1}{2} \F{}{\mu\nu,\sigma_1\cdots\sigma_p} 
\inve{\mu}{AA'}\inve{\nu A}{B'}
\inve{\sigma_1 C_1}{C'_1}\cdots \inve{\sigma_p C_p}{C'_p} 
\label{cdervar}
\endEQs
denote the spinor components of the jet variables 
$\F{}{\mu\nu,\sigma_1\cdots\sigma_p}$, $p\ge 0$. 
We write 
\EQ	
\sFder{}{AB}{C'_1\cdots C'_p}{C_1\cdots C_p} = 
\sFder{}{(AB,}{(C'_1\cdots C'_p)}{\ C_1\cdots C_p)}, 
\qquad
\csFder{}{A'B'}{C_1\cdots C_p}{C'_1\cdots C'_p} = 
\csFder{}{(A'B',}{(C_1\cdots C_p)}{\ C'_1\cdots C'_p)}
\label{symmdervar}
\endEQ
for the symmetric derivative variables,
using the convention that \eqsref{dervar}{symmdervar} 
with $p=0$ stand for $\sF{}{AB},\csF{}{A'B'}$. 
Recall that the symmetric spinor variables 
$\sFder{}{AB}{C'_1\cdots C'_p}{C_1\cdots C_p}$
together with the independent variables $\x{CC'}{} = \e{\mu}{CC'}\x{\mu}{}$ 
form a coordinate system on the space of solutions of Maxwell's equations
(see \cite{Penrose}).
Thus, $J^q(F)$ and $R^{q-1}(F)\subset J^{q}(F)$
admit the spinor coordinates 
\EQ
J^q(F) =\{( 
\x{CC'}{},\sF{}{AB},\sFder{}{AB,}{C'_1}{C_1},
\ldots, \sFder{}{AB,}{C'_1\cdots C'_q}{C_1\cdots C_q} )\} , 
\endEQ
and
\EQ
R^{q-1}(F) =\{( 
\x{CC'}{},\sF{}{AB},\sFder{}{AB}{C'_1}{C_1},
\ldots, 
\sFder{}{AB}{C'_1\cdots C'_{q}}{C_1\cdots C_{q}} )\} . 
\endEQ
One can easily verify that \Meq/ are locally solvable, 
that is, 
for every $q$-jet
$( x_o^{CC'},\phi_o{}_{AB},\dots,
\phi_o{}_{ABC_1\cdots C_q}^{\hp{AB}C'_1\cdots C'_q} )
\in R^{q-1}(F)$, $q\geq 1$, 
there is a solution
$\sFsol{}{AB}=\sFsol{}{AB}(x)$ of \eqref{spinorMeq}
such that
$\sder{C'_1}{C_1}\cdots \sder{C'_p}{C_p} \sFsol{}{AB}(x_o)
= \phi_o{}_{ABC_1\cdots C_p}^{\hp{AB}C_1'\cdots C_p'}$, 
$0\leq p\leq q$.
This is a consequence of the the fact that 
$\sF{}{AB},\csF{}{A'B'}$ and their symmetrized derivatives 
form an ``exact set of fields'' for \Meq/, 
as discussed by Penrose \cite{Penrose}. 

Let
$\sjetder{AB}{}{C_1\cdots C_p}{C_1'\cdots C_p'}$
be the partial differential operators defined by 
\EQs
&& \sjetder{AB}{}{C_1\cdots C_p}{C'_1\cdots C'_p} 
(\sFder{}{DE,}{F_1'\cdots F_q'}{F_1\cdots F_q})= 
\cases{
\vol{(D}{(A}\vol{E)}{B)}\vol{(F_1}{(C_1}\cdots\vol{F_q)}{C_p)}
\vol{(C_1'}{(F_1'}\cdots\vol{C_p')}{F_q')}, 
&if $p=q$;\cr 
0, &if $p\neq q$;}
\label{sFderop}\\
&&
\sjetder{AB}{}{C_1\cdots C_p} {C_1'\cdots C_p'}
(\csFder{}{D'E',}{F_1\cdots F_q}{F'_1\cdots F'_q})= 0 , 
\label{sFderop'}
\endEQs
and let
$\csjetder{A'B'}{}{C'_1\cdots C'_p}{C_1\cdots C_p}$
be the complex conjugate operator
\EQ
\csjetder{A'B'}{}{C'_1\cdots C'_p}{C_1\cdots C_p}
= \overline{ \sjetder{AB}{}{C_1\cdots C_p}{C_1'\cdots C_p'} } .
\nonumber
\endEQ
Let $\sD{A}{A'}$ be the spinorial total derivative operator
on $J^\infty(F)$ given by 
\EQs
\sD{A}{A'} = \inve{\mu A'}{A} \D{\mu}
= \sder{A'}{A} +\sum_{q\geq 0}
( \sFder{}{DE,}{A'F'_1\dots F'_q}{AF_1\dots F_q} 
\sjetder{DE}{}{F_1\dots F_q}{F'_1\dots F'_q}
+ \csFder{D'E',}{}{A'F'_1\dots F'_q}{AF_1\dots F_q}
\csjetder{}{D'E'}{F_1\dots F_q}{F'_1\dots F'_q} ) . 
\endEQs
Note, due to the commutativity of partial derivatives, 
$\sD{A}{A'}$ satisfies the identities 
\EQ
\sD{(C}{A'} \sD{A)A'}{} =0, \qquad
\sD{(C'}{A} \sD{A')A}{}=0. 
\label{sDids}
\endEQ

In spinor form 
the adjoint symmetry equations \eqref{Pdeteq} simply reduce to
\EQ
\sD{(A'}{B} \sP{}{B')B} =0
\quad\eqtext{ on } R^r(F),
\label{deteq}
\endEQ
where we have written 
\EQ
\sP{}{AA'}= 
\P{\mu}\inve{\mu}{AA'} + \i\tP{\mu}\inve{\mu}{AA'} 
\quad\eqtext{ on $J^r(F)$ }. 
\label{ComplexChar}
\endEQ
We refer to the solutions $\sP{}{AA'}$ of equation \eqref{deteq} 
as spinorial adjoint symmetries of order $r$. 
Note that adjoint gauge symmetries \eqref{gaugeQ} 
correspond here to the gradient solutions
\EQ
\sP{}{AA'}= \D{A'A}{} \X , 
\label{gaugePspin}
\endEQ
which satisfy the curl equation $\sD{(C}{A'}\sP{}{A)A'}=0$. 

Our classification of adjoint symmetries $\sP{}{AA'}$ 
makes use of \Kspin/s. 
Recall that Killing spinors 
$\KS{A_1 \cdots A_k}{A'_1 \cdots A'_l} = 
\KS{(A_1 \cdots A_k)}{(A'_1 \cdots A'_l)}(x)$
of type $(k,l)$ 
are the solutions of the conformally invariant equations
\cite{Penrose}
\EQ
\sder{(B}{(B'} \KS{A_1 \cdots A_k)}{A'_1 \cdots A'_l)} =0 . 
\label{KSeq}
\endEQ
For $k=l=1$, 
this equation is the conformal \Kvec/ equation \cite{Penrose}, 
and for $k=0,l=2$, 
the self-dual conformal Killing-Yano equation \cite{DietzRudiger}. 
Hence a type $(1,1)$ \Kspin/ $\KS{A'}{A}$ corresponds 
to a complex conformal \Kvec/
$\KV{\mu}{} =\inve{\mu A'}{A} \KS{A}{A'}$,
while a type $(0,2)$ \Kspin/ $\KS{}{A'B'}$ corresponds 
to a complex conformal \KYten/
$\KY{}{\mu\nu} =\inve{\mu A'}{A} \inve{\nu AB'}{}\KS{}{A'B'}$
satisfying the self-duality condition 
$*\KY{}{\mu\nu}=i\KY{}{\mu\nu}$.
These are polynomial expressions in the spacetime coordinates $\x{CC'}{}$, 
specifically, 
\EQs
&& \KV{A}{A'} =
\spinor{\alpha_1}{A}{A'} + \alpha_2\x{A}{A'}
+ \spinor{\alpha_3}{}{A'B'}\x{AB'}{}
+ \spinor{\alpha_4}{AB}{}\x{}{A'B}
+\spinor{\alpha_5}{B'}{B}\x{AB}{}\x{}{A'B'},
\nonumber\\
&& \KY{A'B'}{} =
\spinor{\alpha_6}{}{A'B'} + \spinor{\alpha_7}{}{A(A'}\x{A}{B')}
+\spinor{\alpha_8}{}{AB}\x{A}{A'}\x{B}{B'},
\label{KS}
\endEQs
where 
$\spinor{\alpha_1}{}{AA'}$,
$\spinor{\alpha_2}{}{}$, 
$\spinor{\alpha_3}{}{A'B'}$,
$\spinor{\alpha_4}{}{AB}$,
$\spinor{\alpha_5}{}{BB'}$,
$\spinor{\alpha_6}{}{A'B'}$,
$\spinor{\alpha_7}{}{AA'}$,
$\spinor{\alpha_8}{}{AB}$
are constant symmetric spinors. 
The following Lemma, 
which will be pivotal in our classification analysis,
is a special case of 
the well-known factorization property \cite{Penrose}
of Killing spinors in Minkowski space $M^4$. 

\Proclaim{Lemma~4.1. }{
A symmetric spinor field 
$\spinor{\xi}{A_1 \cdots A_k}{A'_1 \cdots A'_k}$
is a Killing spinor of type $(k,k)$ 
if and only if it can be expressed as
a sum of symmetrized products of $k$ Killing spinors of type $(1,1)$. 
A symmetric spinor field
$\spinor{Y}{A_1 \cdots A_{k}}{A'_1 \cdots A'_{k+4}}$
is a Killing spinor of type $(k,k+4)$  
if and only if it can be expressed as				
a sum of symmetrized products of two Killing spinors of type $(0,2)$
and $k$ Killing spinors of type $(1,1)$.
There are 
$\frac{1}{12}(k+1)^2(k+2)^2(2k+3)$ 
linearly independent \Kspin/s of type $(k,k)$,
and 
$\frac{1}{12}(k+1)(k+2)(k+5)(k+6)(2k+7)$ 
linearly independent \Kspin/s of type $(k,k+4)$, 
over the complex numbers.}

Let $\othKV{CC'}{}$ be a \Kspin/ 
and define 
\EQ
\sLie{\othKV{}{}}{}{AB} 
= \frac{1}{2} \inve{\mu}{AA'} \inve{\nu A'}{B} 
( \othKV{\tau}{} \F{}{\mu\nu,\tau} 
- 2 (\der{[\mu} \othKV{\tau}{}) \F{}{\nu]\tau} )
= \othKV{CC'}{} \sFder{}{AB,}{}{CC'} 
+ \sder{}{C'(A}\othKV{CC'}{}\sF{}{B)C} , 
\endEQ
which stands for the spinor components of 
the Lie derivative \eqref{LieF} of $\F{}{\mu\nu}$
with respect to $\othKV{\tau}{}=\inve{\tau}{CC'}\othKV{CC'}{}$. 
As a consequence of 
the linearity and conformal invariance of \Meq/ \eqref{spinorMeq},
one easily sees that, for any solution $\sF{}{AB}(x)$ of \eqref{spinorMeq},
$\sLie{\othKV{}{}}{}{AB}(x)$ is also a solution. 
We remark that, geometrically,
$\sLie{\othKV{}{}}{}{AB}$ represents 
a conformally weighted Lie derivative of $\sF{}{AB}$ 
(see \cite{Penrose}). 

Given any adjoint symmetry of \Meq/, 
we can obtain a family of higher order adjoint symmetries
by the action of conformal symmetries on $\sF{}{AB}$. 
Let 
\EQ
\gen{\othKV{}{}} = 
( \sLie{\othKV{}{}}{}{AB} ) \szerojetder{AB}{}
+( \csLie{\othKV{}{}}{}{A'B'} ) \cszerojetder{A'B'}{} , 
\label{evolsymm}
\endEQ
which is the conformal symmetry in evolutionary form \cite{Olver}, 
defined on $J^1(F)$, 
and let $\pr\gen{\othKV{}{}}$ denote the prolongation of $\gen{\othKV{}{}}$
to $J^\infty(F)$. 
Then, if $\sP{}{AA'}{}$ is any adjoint symmetry of order $r$, 
the linearity of the adjoint symmetry equation \eqref{deteq}
implies that 
$\pr\gen{\othKV{}{}}\sP{}{AA'}{}$ 
is an adjoint symmetry of order $r+1$. 
This can be iterated any number of times
using conformal symmetry generators 
$\pr\gen{\othKV{}{1}}$, $\pr\gen{\othKV{}{2}}$, etc. 
Note that 
$[\pr\gen{\othKV{}{1}}, \pr\gen{\othKV{}{2}}]= 
\pr\gen{[\othKV{}{1}, \othKV{}{2}]}$, 
where $[\othKV{}{1}, \othKV{}{2}]$ 
denotes the commutator of conformal \Kvec/s 
$\othKV{}{1},\othKV{}{2}$, 
which is again a conformal \Kvec/. 

\Proclaim{Proposition~4.2 } {
Let $\KV{AA'}{}$, 
$\othKV{AA'}{1},\dots,\othKV{AA'}{p}$
and $\KS{}{A'B'C'D'}$ be Killing spinors. 
Then the spinor functions
\EQs
&& \zeroP{AA'}{}(\phi;\xi) = 
\KV{B}{A'}\sF{}{AB}, 
\label{loworderR}\\
&& \firstP{AA'}{}(\bar\phi;\kappa) = 
\KS{}{A'B'C'D'}\csFder{B'C'}{}{D'}{A}
+\frac{3}{5}(\sder{B'}{A}\KS{}{A'B'C'D'})\csF{C'D'}{},
\label{loworderS}
\endEQs
are adjoint symmetries of order $q=0$ and $q=1$, respectively.
Hence their extensions 
\EQs
\zeroP{AA'}{}[\xi,\zeta_1,\dots,\zeta_p] 
&& 
= \frac{1}{p!} \sum_{\s\in S_p}
\pr\gen{\zeta_{s(1)}}\cdots \pr\gen{\zeta_{s(p)}} 
\zeroP{AA'}{}(\phi;\xi) , 
\label{Radjsymm}\\
\firstP{AA'}{}[\kappa,\zeta_1,\dots,\zeta_p]
&& 
= \frac{1}{p!} \sum_{\s\in S_p}
\pr\gen{\zeta_{s(1)}}\cdots \pr\gen{\zeta_{s(p)}} 
\firstP{AA'}{}(\bar\phi;\kappa) , 
\label{Sadjsymm}
\endEQs
are adjoint symmetries of order $q=p$ and $q=p+1$, respectively,
where $S_p$ denotes the symmetric group on the index set $\{1,\ldots,p\}$. 
Furthermore, 
\eqrefs{Radjsymm}{Sadjsymm}
are equivalent to non-trivial adjoint symmetries
with highest order terms given by, respectively, 
\EQs
&& 
(-1)^p \KV{B}{(A'} \othKV{\hp{1}C_1}{1C_1'}\cdots\othKV{\hp{p}C_p}{pC_p')}
\sFder{}{AB}{C_1'\cdots C_p'}{C_1\cdots C_p} , \quad
(-1)^p \KS{}{(A'B'C'D'} \othKV{(E_1}{1E_1'}\cdots\othKV{\hp{p}E_p)}{pE_p')}
\csFder{B'C'}{}{D'E_1'\cdots E_p'}{AE_1\cdots E_p} , 
\label{highestorderRS}
\endEQs 
whenever $\xi,\zeta_1,\ldots,\zeta_p,\kappa$ are non-zero. }

The proof of Proposition~4.2 is based on straightforward computations
which will be omitted.

Hereafter we use the notation 
in \eqrefs{Radjsymm}{Sadjsymm} 
with $p=0$ to refer to \eqrefs{loworderR}{loworderS}. 
We now let $\omega\downindex{AA'}(x)$
be a spinor field satisfying 
\EQ
\sder{B}{(A'}\omega\downindex{B')B} = 0,
\label{Wadjsymm}
\endEQ
and we call $\W{AA'}[\omega] = \omega\downindex{AA'}$
an {\it elementary} adjoint symmetry of \Meq/ \eqref{spinorMeq}. 

\Proclaim{Theorem~4.3 }{
Every adjoint symmetry $\sP{}{AA'}$ of order $r$ of \Meq/ \eqref{spinorMeq}
is equivalent to a sum of 
an elementary adjoint symmetry $\W{AA'}[\omega]$
and a linear adjoint symmetry given by a sum of 
$\zeroP{AA'}{}[\xi,\zeta_{1},\dots,\zeta_{p}]$
and
$\firstP{AA'}{}[\kappa,\varrho_{1},\dots,\varrho_{q}]$, 
$0\leq p\leq r$, $0\leq q\leq r-1$,
involving 
type $(1,1)$ \Kspin/s $\xi,\zeta_{i},\varrho_{j}$
and type $(0,4)$ \Kspin/s $\kappa$
for each $p,q$. }

\proclaim{ Proof.}
By replacing $\sP{}{AA'}$ with an equivalent
adjoint symmetry we can assume that $\sP{}{AA'}$ depends only on 
$\x{CC'}{}$ and $\sFder{}{AB}{C'_1\cdots C'_p}{C_1\cdots C_p}$,
$0\leq p\leq r$, \ie/, 
$\sP{}{AA'}$ is a function in the coordinates of $R^{r-1}(F) \subset J^r(F)$. 

Let $\sFsol{}{AB}=\sFsol{}{AB}(\x{}{})$ be a solution of \Meq/.
Define the linearization operator
\EQ
\linop{\varphi} = \sum_{p\geq0} \Big( 
\sFdersol{}{AB}{K_1'\cdots K_p'}{K_1\cdots K_p}
\sjetder{AB}{}{K_1\cdots K_p}{K_1'\cdots K_p'}
+\csFdersol{}{A'B'}{K_1\cdots K_p}{K_1'\cdots K_p'}
\csjetder{A'B'}{}{K'_1\cdots K'_p}{K_1\cdots K_p}
\Big),
\label{linop}
\endEQ
where 
$\sFdersol{}{AB}{K_1'\cdots K_p'}{K_1\cdots K_p}
= \sder{K'_1}{K_1}\cdots\sder{K'_p}{K_p}\sFsol{}{AB}$, 
$\csFdersol{}{A'B'}{K_1\cdots K_p}{K_1'\cdots K_p'}
= \sder{K_1}{K'_1}\cdots\sder{K_p}{K'_p}\csFsol{}{A'B'}$,
which are each symmetric spinor fields. 
Note that, for any adjoint symmetry $\sP{}{AA'}$ of order $r$, 
the linearization 
$\sP{}{\varphi AA'} = \linop{\varphi}\sP{}{AA'}$
again satisfies the adjoint symmetry equations
\EQ
\sD{(A'}{B} \sP{}{\varphi B')B} =0 \quad\eqtext{ on $R^r(F)$. }
\label{lindeteq}
\endEQ

By the local solvability of \Meq/,
the coefficients of 
$\sFdersol{}{IJ}{K'_1\cdots K'_p}{K_1\cdots K_p}$, 
$\csFdersol{}{I'J'}{K_1\cdots K_p}{K'_1\cdots K_p'}$, 
$0\le p\le r+1$, 
in equation \eqref{lindeteq} must vanish. 
Thus, we find that for $p=r+1$ the coefficients
yield the equations
\EQ
\p{B'}{(AIJ}{K_1\cdots K_r)}{K_1'\cdots K_r'} = 0,
\quad
\tp{B'(A}{I'J'}{K_1'\cdots K_r'}{K_1\cdots K_r)} = 0,
\label{solvadjsymm1}
\endEQ
and for $1\leq p\leq r$, the coefficients yield the equations
\EQs
&& 
\sD{(A'}{B} \p{B')B}{IJ}{K_1\cdots K_p}{K_1'\cdots K_p'}
-\vol{(K_p'|(A'}{}\p{B')}{(K_pIJ}{K_1\cdots K_{p-1})}{K_1'\cdots K_{p-1}')} 
=0 \quad\eqtext{ on $R^r(F)$, }
\label{solvadjsymm2}\\
&&
\sD{(A'}{B} \tp{B')B}{I'J'}{K'_1\cdots K'_p}{K_1\cdots K_p}
+\vol{(A'}{(K'_p}\tp{B')(K_p}{I'J'}{K'_1\cdots K'_{p-1})}{K_1\cdots K_{p-1})} 
=0 \quad\eqtext{ on $R^r(F)$, }
\label{solvadjsymm2'}
\endEQs
while for $p=0$ the coefficients yield the equations
\EQ
\sD{(A'}{B} \p{B')B}{IJ}{}{}=0, \quad
\sD{(A'}{B} \tp{B')B}{I'J'}{}{} =0 \quad\eqtext{ on $R^r(F)$, }
\label{solvadjsymm2''}
\endEQ
where we have written
\EQ
\p{AA'}{IJ}{K_1\cdots K_p}{K_1'\cdots K_p'}
=\sjetder{IJ}{}{K_1\cdots K_p}{K_1'\cdots K_p'}\sP{}{AA'}, \quad
\tp{AA'}{I'J'}{K'_1\cdots K'_p}{K_1\cdots K_p}
=\csjetder{I'J'}{}{K'_1\cdots K'_p}{K_1\cdots K_p}\sP{}{AA'}, \quad
0\le p\le r . 
\label{lincoeffP}
\endEQ
Note that 
$\p{AA'}{IJ}{K_1\cdots K_p}{K_1'\cdots K_p'}$, 
$\tp{AA'}{I'J'}{K'_1\cdots K'_p}{K_1\cdots K_p}$
are spinor functions of order $r$ 
in the coordinates of $R^{r-1}(F)\subset J^r(F)$ 
and are symmetric separately in their primed and unprimed indices 
excluding $A$, $A'$.

By reduction of $\p{AA'}{IJ}{K_1\cdots K_p}{K_1'\cdots K_p'}$, 
$\tp{AA'}{I'J'}{K'_1\cdots K'_p}{K_1\cdots K_p}$
into symmetric components, 
we find that the solution to the equations in \eqref{solvadjsymm1} is
\EQs
&&
\p{AA'}{IJ}{K_1\cdots K_r}{K'_1\cdots K'_r}=
\vol{A}{(I}\p{1\,A'}{J}{K_1\cdots K_{r})}{K'_1\cdots K'_r}
+ \vol{A}{(I}\vol{A'(K_1'}{}\p{2\,}{J}{K_1\cdots K_r)}{K_2'\cdots K_{r}')} , 
\label{highestorder}\\
&&
\tp{AA'}{I'J'}{K'_1\cdots K'_r}{K_1\cdots K_r}=
\vol{A(K_1|}{}\tp{1\,A'}{I'J'}{K'_1\cdots K'_{r}}{|K_2\cdots K_r)}
+ \vol{A'}{(I'}\vol{A(K_1}{}\tp{2\,}{J'}{K'_1\cdots K'_r)}{K_2\cdots K_{r})} , 
\label{highestorder'}
\endEQs
where 
$\p{1\,A'}{J}{K_1\cdots K_{r}}{K'_1\cdots K'_r}$, 
$\p{2\,}{J}{K_1\cdots K_r}{K_2'\cdots K_{r}'}$,
$\tp{1\,A'}{I'J'}{K'_1\cdots K'_{r}}{K_2\cdots K_r}$,
$\tp{2\,}{J'}{K'_1\cdots K'_r}{K_2\cdots K_{r}}$
are symmetric spinor functions of order $r$ 
depending only on the coordinates of $R^{r-1}(F) \subset J^r(F)$.
Next we substitute the expressions \eqrefs{highestorder}{highestorder'}
into equations \eqrefs{solvadjsymm2}{solvadjsymm2'} with $p=r$ 
and symmetrize over primed indices in the resulting expressions. 
This yields the equations
\EQ
\sD{(A'}{(I} \p{1\,B'}{J}{K_1\cdots K_r)}{K_1'\cdots K_r')} =0,
\qquad
\sD{(K_1}{(A'} \tp{1\,}{B'I'J'}{K_1'\cdots K_r')}{K_2\cdots K_r)}=0 
\quad\text{on $R^r(F)$}.
\label{totalKilling}
\endEQ
A straightforward analysis of 
the highest order terms in \eqref{totalKilling} 
shows that the functions 
$\p{1\,B'}{J}{K_1\cdots K_r}{K_1'\cdots K_r'}$, 
$\tp{1\,}{B'I'J'}{K_1'\cdots K_r'}{K_2\cdots K_r}$
must only depend on $\x{CC'}{}$ 
and, consequently, are \Kspin/s of type 
$(r+1,r+1)$ and $(r-1,r+3)$, respectively.

From equation \eqref{highestorder} we have that
\EQs
\sP{}{\varphi AA'} &&= 
\p{1\,A'}{J}{K_1\cdots K_{r}}{K'_1\cdots K'_r}
\sFdersol{}{AJ}{K_1'\cdots K_{r}'}{K_1\cdots K_r}
+\vol{AK_1}{} \tp{1\,A'}{I'J'}{K_1'\cdots K_r'}{K_2\cdots K_r}
\csFdersol{}{I'J'}{K_1\cdots K_r}{K_1'\cdots K_r'}
\nonumber\\
&&\qquad 
+\vol{A'K'_1}{} \p{2\,}{J}{K_1\cdots K_r}{K_2'\cdots K_{r}'}
\sFdersol{}{AJ}{K_1'\cdots K_{r}'}{K_1\cdots K_r}
+\vol{AK_1}{} \tp{2\,}{J'}{K_1'\cdots K_r'}{K_2\cdots K_{r}}
\csFdersol{}{A'J'}{K_1\cdots K_{r}}{K_1'\cdots K_r'}
+H_{AA'},
\label{adjsymm1}
\endEQs
where $H_{AA'}$ involves the derivatives of 
$\sFsol{}{IJ},\csFsol{}{I'J'}$ of order at most $r-1$.
Now by Lemma~4.1, 
$\p{1\,A'}{J}{K_1\cdots K_{r}}{K'_1\cdots K'_r}$
and $\tp{1\,}{A'I'J'}{K_1'\cdots K_r'}{K_2\cdots K_r}$
are respectively sums of symmetrized products of 
$r+1$ type $(1,1)$ \Kspin/s, 
$\KV{(J}{(A'} \spinor{\zeta_1}{K_1}{K'_1}\cdots\spinor{\zeta_r}{K_r)}{K'_r)}$,
and of a type $(0,4)$ \Kspin/ and $r-1$ type $(1,1)$ \Kspin/s,
$\KS{(A'I'J'K'_1}{} 
\spinor{\varrho_1}{K'_2}{(K_2}\cdots\spinor{\varrho_{r-1}}{K'_r)}{K_r)}$. 
Consequently, by \eqref{highestorderRS} 
in Proposition~4.2, 
there is an adjoint symmetry $\sPhat{AA'}$ 
equivalent to a sum of adjoint symmetries
$\zeroP{AA'}{}[\xi;\zeta_1,\dots,\zeta_r]$, 
$\firstP{AA'}{}[\kappa;\varrho_1\,\dots,\varrho_{r-1}]$ 
such that the terms involving derivatives of order $r$ of 
the spinor functions $\sFsol{}{IJ},\csFsol{}{I'J'}$ 
in the linearization
$\linop{\varphi} \sPhat{AA'}$ 
agree with the first two terms on the right-hand side 
of equation \eqref{adjsymm1}. 
Thus, after subtracting $\sPhat{AA'}$ from $\sP{}{AA'}$ 
we get an adjoint symmetry, 
which we again denote by $\sP{}{AA'}$, 
with the property that 
\EQ
\sP{}{\varphi AA'} =
\vol{A'K_1'}{} \p{2\,}{J}{K_1\cdots K_r}{K_2'\cdots K_{r}'}
\sFdersol{}{AJ}{K_1'\cdots K_{r}'}{K_1\cdots K_r}
+\vol{AK_1}{} \tp{2\,}{J'}{K_1'\cdots K_r'}{K_2\cdots K_{r}}
\csFdersol{}{A'J'}{K_1\cdots K_{r}}{K_1'\cdots K_r'}
+{\hat H}_{AA'},
\label{adjsymm11}
\endEQ
where ${\hat H}_{AA'}$ 
involves the derivatives of $\sFsol{}{IJ},\csFsol{}{I'J'}$ 
of order at most $r-1$.

Now, in equation \eqref{adjsymm11}
observe that 
$\sFdersol{}{AJ}{A'K_2'\cdots K_{r}'}{K_1\cdots K_r}
= \sder{A'}{A}\sFdersol{}{JK_1}{K_2'\cdots K_{r}'}{K_2\cdots K_r}$, 
$\csFdersol{}{A'J'}{AK_2\cdots K_{r}}{K_1'\cdots K_r'}
= \sder{A}{A'}\csFdersol{}{J'K'_1}{K_2\cdots K_{r}}{K_2'\cdots K_r'}$. 
Consequently, 
our next goal is to show that there is a differential
function $\X$ of $\x{CC'}{}$ and 
$\sFder{}{AB}{C'_1\cdots C'_p}{C_1\cdots C_p}$, 
$0\le p\le r-1$,
so that the terms 
involving derivatives of order $r$ of 
$\sFsol{}{IJ},\csFsol{}{I'J'}$ 
in $\linop{\varphi} \D{AA'}{}\X$ and in \eqref{adjsymm11} 
agree. 
For this to hold, it suffices to show that the functions 
$\p{2\,}{IJ}{K_1\cdots K_{r-1}}{K_1'\cdots K_{r-1}}$, 
$\tp{2\,}{I'J'}{K_1'\cdots K_{r-1}'}{K_1\cdots K_{r-1}}$
are of order $r-1$ 
and satisfy the integrability conditions
\EQs
\sjetder{PQ}{}{R_1\cdots R_{r-1}}{R_1'\cdots R_{r-1}'}
\p{2\,}{IJ}{K_1\cdots K_{r-1}}{K_1'\cdots K_{r-1}'}
&=&
\sjetder{IJ}{}{K_1\cdots K_{r-1}}{K_1'\cdots K_{r-1}'}
\p{2\,}{PQ}{R_1\cdots R_{r-1}}{R_1'\cdots R_{r-1}'} , 
\label{intcond1}\\
\csjetder{P'Q'}{}{R'_1\cdots R_{r-1}'}{R_1\cdots R_{r-1}}
\p{2\,}{IJ}{K_1\cdots K_{r-1}}{K_1'\cdots K_{r-1}'}
&=&
\sjetder{IJ}{}{K_1\cdots K_{r-1}}{K_1'\cdots K_{r-1}'}
\tp{2\,}{P'Q'}{R_1'\cdots R_{r-1}'}{R_1\cdots R_{r-1}} , 
\label{intcond2}\\
\csjetder{P'Q'}{}{R'_1\cdots R_{r-1}'}{R_1\cdots R_{r-1}}
\tp{2\,}{I'J'}{K_1'\cdots K_{r-1}'}{K_1\cdots K_{r-1}}
&=&
\csjetder{I'J'}{}{K_1'\cdots K_{r-1}'}{K_1\cdots K_{r-1}}
\tp{2\,}{P'Q'}{R_1'\cdots R_{r-1}'}{R_1\cdots R_{r-1}} . 
\label{intcond3}
\endEQs
In fact, if equations \eqref{intcond1}, \eqref{intcond2}, \eqref{intcond3} 
hold, then one can verify that a function $\X$ with the desired properties
is given by 
\EQs
\X = &&
\int_0^1 \bigg(
\p{2\,}{IJ}{K_1\cdots K_{r-1}}{K_1'\cdots K_{r-1}'}
(x,\phi,\partial\phi,\ldots,\partial^{r-2}\phi,\lambda\partial^{r-1}\phi)
\sFder{}{IJ}{K_1'\cdots K_{r-1}'}{K_1\cdots K_{r-1}}
\nonumber\\&&\fewquad
+\tp{2\,}{I'J'}{K_1'\cdots K_{r-1}'}{K_1\cdots K_{r-1}}
(x,\phi,\partial\phi,\ldots,\partial^{r-2}\phi,\lambda\partial^{r-1}\phi)
\csFder{}{I'J'}{K_1\cdots K_{r-1}}{K_1'\cdots K_{r-1}'}
\bigg)\,d\lambda , 
\nonumber
\endEQs
where 
$\partial^p\phi$ stands collectively for the variables
$\sFder{}{AB}{C'_1\cdots C'_{p}}{C_1\cdots C_{p}}$, $p\geq 0$. 

The proofs of conditions 
\eqref{intcond1}, \eqref{intcond2}, \eqref{intcond3} 
are based on like computations. 
We will therefore prove the first one and omit the proofs of the others. 

From equation \eqref{adjsymm11} we have
\EQ
\p{AA'}{IJ}{K_1\cdots K_r}{K_1'\cdots K_{r}'} = 
\vol{A}{(I}\vol{A'(K_1'}{}
\p{2\,}{JK_1}{K_2\cdots K_r)}{K_2'\cdots K_{r}')} . 
\label{P2again}
\endEQ
Now substitute \eqref{P2again} into equation \eqref{solvadjsymm2}, 
contract on the indices $B',K'_1$, 
and symmetrize over the indices $A',K'_2,\ldots,K'_r$ to obtain
\EQ
\sD{(A'}{(I}\p{2\,}{JK_1}{K_2\cdots K_r)}{K_2'\cdots K_{r}')} 
=
- \p{(A'}{(IJK_1}{K_2\cdots K_{r})}{K_2'\cdots K_{r}')}
\qquad\text{on $R^r(F)$}.
\label{DP2}
\endEQ
The terms of highest order in \eqref{DP2} 
arise from 
$\p{2\,}{JK_1}{K_2\cdots K_r}{K_2'\cdots K_r'}$
which is of order at most $r$. 
Thus an application of 
$\sjetder{PQ}{}{R_1\cdots R_{r+1}}{R_1'\cdots R_{r+1}'}$
and its complex conjugate 
to \eqref{DP2} yields
\EQs
&& \sjetder{(PQ}{}{R_1\cdots R_{r}}{(R_1'\cdots R_{r}'}
\vol{}{R_{r+1})(I} \vol{R'_{r+1})(A'}{}
\p{2\,}{JK_1}{K_2\cdots K_r)}{K_2'\cdots K_r')}=0,
\label{solvadjsymm3}\\
&& \csjetder{(P'Q'}{}{R_1'\cdots R_{r}'|}{(R_1\cdots R_{r}}
\vol{R_{r+1})}{(I} 
\p{2\,}{JK_1}{K_2\cdots K_r)}{(K_2'\cdots K_r'}
\vol{A')}{|R'_{r+1})}
=0 . 
\label{solvadjsymm3'}
\endEQs
Now choose a pair of spinors $\so{}{A},\si{}{A}$
such that $\so{}{A} \si{A}{} \neq 0$. 
Then equations \eqrefs{solvadjsymm3}{solvadjsymm3'} yield
\EQs
&& \cso{R'_1}{}\cdots\cso{R'_r}{} \csi{K'_2}{} \cdots\csi{K'_r}{} 
\so{}{P}{} \so{}{Q} \so{}{R_1} \cdots \so{}{R_r}
\si{}{J} \si{}{K_1} \cdots \si{}{K_r} 
\sjetder{PQ}{}{R_1\cdots R_{r}}{R_1'\cdots R_{r}'}
\p{2\,}{JK_1}{K_2\cdots K_r}{K_2'\cdots K_r'}=0,
\\
&& \cso{}{P'} \cso{}{Q'} \cso{}{R'_1}\cdots\cso{}{R'_r} 
\csi{K'_2}{} \cdots \csi{K'_r}{}
\so{R_1}{} \cdots \so{R_r}{}
\si{}{J} \si{}{K_1} \cdots \si{}{K_r}
\csjetder{P'Q'}{}{R_1'\cdots R_{r}'}{R_1\cdots R_{r}}
\p{2\,}{JK_1}{K_2\cdots K_r}{K_2'\cdots K_r'}=0 . 
\endEQs
By continuity, these equations hold for all spinors $\so{}{A},\si{}{A}$,
and hence yield
\EQ
\sjetder{PQ}{}{R_1\cdots R_{r}}{R_1'\cdots R_{r}'}
\p{2\,}{IJ}{K_1\cdots K_{r-1}}{K_1'\cdots K_{r-1}'}=0,
\qquad 
\csjetder{P'Q'}{}{R_1'\cdots R_{r}'}{R_1\cdots R_{r}}
\p{2\,}{IJ}{K_1\cdots K_{r-1}}{K_1'\cdots K_{r-1}'}=0 . 
\endEQ
Thus $\p{2\,}{IJ}{K_1\cdots K_{r-1}}{K'_1\cdots K'_{r-1}}$ 
is of order $r-1$.

We now return to \eqref{DP2} and apply 
$\sjetder{PQ}{}{R_1\cdots R_r}{R_1'\cdots R_r'}$
to obtain 
\EQ
\vol{(A'|(R_1'}{} \vol{}{(I|(P} 
\sjetder{QR_1}{}{R_2\cdots R_{r})}{R_2'\cdots R_{r}')}
\p{2\,}{|JK_1}{K_2\cdots K_{r})}{|K_2'\cdots K_{r}')} =
\sjetder{PQ}{}{R_1\cdots R_r}{R_1'\cdots R_{r}'}
\p{(A'}{(IJK_1}{K_2\cdots K_{r})}{K_2'\cdots K_{r}')} . 
\label{solvadjsymm6}
\endEQ
On the right-hand side of this equation 
we first use \eqref{lincoeffP} 
and the commutativity of partial derivatives 
followed by substitution of \eqref{P2again} 
to conclude that
\EQ
\sjetder{PQ}{}{R_1\cdots R_r}{R_1'\cdots R_r'}
\p{A'}{IJK_1}{K_2\cdots K_{r}}{K_2'\cdots K_{r}'} =
\vol{(R_1'|A'}{} \vol{}{(P|I} 
\sjetder{JK_1}{}{K_2\cdots K_{r}}{K_2'\cdots K_{r}'}
\p{2\,}{|QR_1}{R_2\cdots R_{r})}{|R_2'\cdots R_{r}')}.
\label{solvadjsymm7}
\endEQ
Substitution of \eqref{solvadjsymm7} into \eqref{solvadjsymm6}
gives
\EQ
\vol{(A'|(R'_1}{} \vol{}{(I|(P} 
\sjetder{QR_1}{}{R_2\cdots R_{r})}{R_2'\cdots R_{r}')}
\p{2\,}{|JK_1}{K_2\cdots K_{r})}{|K_2'\cdots K_{r}')} =
\vol{(R'_1|(A'}{} \vol{}{(P|(I} 
\sjetder{JK_1}{}{K_2\cdots K_{r})}{K_2'\cdots K_{r}')}
\p{2\,}{|QR_1}{R_2\cdots R_{r})}{|R_2'\cdots R_{r}')}.
\label{solvadjsymm8}
\endEQ
Finally, an analysis similar to that for \eqref{solvadjsymm3}
using spinor pairs $\so{}{A},\si{}{A}$
directly leads from \eqref{solvadjsymm8}
to equation \eqref{intcond1}, 
which completes the proof of 
the integrability condition for existence of $\X$. 

In summary, 
we have shown that given an adjoint symmetry $\sP{}{AA'}$ of order $r$,
there is an adjoint symmetry $\sPhat{AA'}$ equivalent to a sum 
of $\zeroP{AA'}{}[\xi;\zeta_1,\dots,\zeta_r]$, 
$\firstP{AA'}{}[\kappa;\varrho_1,\dots,\varrho_{r-1}]$,
and a differential function $\X$ 
with the property that the linearization $\linop{\varphi} H_{AA'}$ 
of the difference
$H_{AA'} = \sP{}{AA'} - \sPhat{AA'} - \D{AA'}{}\X$
involves derivatives of $\sFsol{}{IJ},\csFsol{}{I'J'}$
only up to order $r-1$. 
Thus, due to the local solvability of \Meq/, 
it follows that $H_{AA'}$ when restricted to $R^{r-1}(F) \subset J^r(F)$ 
is of order at most $r-1$. 
By the linearity of the adjoint symmetry equation \eqref{deteq},
we then conclude that 
the adjoint symmetry $\sP{}{AA'}$ is equivalent to the sum of 
$\sPhat{AA'}$ and an adjoint symmetry ${\hat H}_{AA'}$ of order at most $r-1$,
where ${\hat H}_{AA'}$ is a function 
in the coordinates of $R^{r-1}(F) \subset J^r(F)$
given by replacing all the variables 
$\sFder{}{AB,}{K'_1\dots K'_p}{K_1\dots K_p}$
in $H_{AA'}$ by the symmetric variables
$\sFder{}{AB}{K'_1\dots K'_p}{K_1\dots K_p}$, $p\geq 0$. 

Now we proceed inductively by descent in the order of $\sP{}{AA'}$. 
In the last step we see that $\sP{}{AA'}$ is equivalent to a sum of 
the linear adjoint symmetries 
\eqref{Radjsymm}, $0\leq p\leq r$, 
\eqref{Sadjsymm}, $0\leq p\leq r-1$, 
and an elementary adjoint symmetry \eqref{Wadjsymm}. 
This completes the proof of the Theorem.\endproof

\section{Classification of Currents}
\label{currents}

In this section we give a complete classification of 
the local \conslaw/s of \Meq/
and explicitly exhibit a basis for them in terms of 
conformal \Kvec/s and conformal \KYten/s. 

We start with some preliminaries.
Let $\othKV{CC'}{}$, $\KV{CC'}{}$ be type $(1,1)$ \Kspin/s, 
let $\KS{}{A'B'C'D'}$ be a type $(4,0)$ \Kspin/. 
We define derivatives of $\KV{CC'}{}$ and $\KS{}{A'B'C'D'}$ 
with respect to $\othKV{CC'}{}$ by 
\EQs
&& 
\Lie{\othKV{}{}} \KV{CC'}{} =
\othKV{EE'}{}\sder{}{EE'}\KV{CC'}{} - \KV{EE'}{}\sder{}{EE'}\othKV{CC'}{} , 
\label{LieKV}\\
&& 
\Lie{\othKV{}{}} \KS{}{A'B'C'D'} = 
\othKV{EE'}{} \sder{}{EE'}\KS{}{A'B'C'D'} 
+ 2(\sder{}{F(A'} \othKV{E'F}{}) \KS{}{B'C'D')E'} 
- \frac{3}{2} (\sder{}{CC'} \othKV{CC'}{}) \KS{}{A'B'C'D'} . 
\label{LieKS}
\endEQs
It is straightforward to verify that, 
as a consequence of the conformal invariance of 
the \Kspin/ equation \eqref{KSeq}, 
$\Lie{\othKV{}{}} \KV{CC'}{}$ and $\Lie{\othKV{}{}} \KS{}{A'B'C'D'}$
are \Kspin/s. 

Write 
\EQ
\ME{AA'}{} = \sFder{}{AB,}{B}{A'},\qquad
\cME{AA'}{} = \csFder{}{A'B',}{B'}{A} , 
\label{MEvar}
\endEQ
so that the solution space $R(F)$ is given by $\ME{AA'}{}=0$. 
One can then easily verify 
using the conformal symmetry \eqref{evolsymm} of \Meq/ \eqref{spinorMeq}
that 
\EQs
\pr\gen{\othKV{}{}} \ME{AA'}{}
&& 
= \othKV{CC'}{} \sD{CC'}{} \ME{AA'}{} 
+ (\sder{}{AA'} \othKV{CC'}{}) \ME{CC'}{} 
+ (\frac{1}{2} \sder{}{CC'} \othKV{CC'}{}) \ME{AA'}{} 
\nonumber\\&&
=(\jetLie{\othKV{}{}} +\frac{1}{2}\div\othKV{}{}) \ME{AA'}{} , 
\label{LieME}
\endEQs
and similarly that 
\EQs
\pr\gen{\othKV{}{}} \ME{}{AA'}
&&
=  \othKV{CC'}{} \sD{CC'}{} \ME{}{AA'}
-(\sder{}{CC'} \othKV{AA'}{}) \ME{}{CC'} 
+ (\sder{}{CC'} \othKV{CC'}{}) \ME{}{AA'} , 
\nonumber\\&&
=(\jetLie{\othKV{}{}} +\div\othKV{}{}) \ME{}{AA'} , 
\label{LiecME}
\endEQs
where 
$\div \othKV{}{}= \sder{}{CC'} \othKV{CC'}{}$,
and where $\jetLie{\othKV{}{}}$ denotes the natural lift of
the standard spinorial Lie derivative operator \cite{Penrose}
to spinor functions on $J^\infty(F)$. 

\subsection{ Equivalence classes of conserved currents }

We now proceed to determine the equivalence classes of
conserved currents 
arising from the classification of adjoint symmetries 
in Theorem~4.3.

In spinor form a conserved current $\dens{AA'} =\dens{\mu} \e{\mu}{AA'}$
with the characteristic form \eqref{charconslaw} satisfies
\EQ
\sD{AA'}{} \dens{AA'} = \sQ{AA'}\cME{}{AA'} + \csQ{AA'}\ME{}{AA'} , 
\label{characteristic}
\endEQ
where $\sQ{AA'} = (\Q{\mu} + \i\tQ{\mu})\inve{\mu}{AA'}$
stands for the spinorial characteristic of $\dens{AA'}$.
If $\H{AA'}{}$ is a conserved current equivalent to $\dens{AA'}$,
then $\sQ{AA'}$ is a spinorial characteristic 
admitted by $\H{AA'}{}$. 
For any linear adjoint symmetry $\sP{}{AA'}$, 
the conserved current arising from formula \eqref{Intformula}
is given by 
\EQ\label{adjsymmcurrent}
\H{AA'}{} = \frac{1}{2}\sP{A}{B'} \csF{A'B'}{}
+\frac{1}{2} \csP{A'}{B} \sF{AB}{} . 
\endEQ

Let $\KV{CC'}{}$, $\othKV{CC'}{1}$, \dots, $\othKV{CC'}{p}$ 
be {\it real} \Kspin/s, 
let $\KS{}{A'B'C'D'}$ be a \Kspin/,
and let $\omega\downindex{CC'}$ be a spinor field 
satisfying \eqref{Wadjsymm}. 
The conserved currents obtained from the adjoint symmetries
$\zeroP{AA'}{}[\xi,\zeta_1,\dots,\zeta_p]$, 
$\i\zeroP{AA'}{}[\xi,\zeta_1,\dots,\zeta_p]$, 
$\firstP{AA'}{}[\kappa,\zeta_1,\dots,\zeta_p]$, 
$\W{AA'}[\omega]$
through formula \eqref{adjsymmcurrent}
are given by, respectively, 
\EQs
\TH{AA'}[\xi,\othKV{}1, \dots,\othKV{}p] 
&=& 
\frac{1}{2}\big( 
\zeroP{B'}{A}
[\xi,\othKV{}{1}, \dots,\othKV{}{p}]\csF{A'B'}{}
+ \zerocP{B}{A'}
[\xi,\othKV{}{1}, \dots,\othKV{}{p}]\sF{AB}{}\big),
\label{Tdensity}\\
\ZH{AA'}[\xi,\othKV{}1, \dots,\othKV{}p] 
&=& 
\frac{\i}{2}\big( 
\zeroP{B'}{A}
[\xi,\othKV{}{1}, \dots,\othKV{}{p}]\csF{A'B'}{}
- \zerocP{B}{A'}
[\xi,\othKV{}{1}, \dots,\othKV{}{p}]\sF{AB}{}\big),
\label{Zdensity}\\
\VH{AA'}[\kappa,\othKV{}1, \dots,\othKV{}p]
&=&
\frac{1}{2}\big(
\firstP{B'}{A}[\kappa,\othKV{}1, \dots,\othKV{}p]\csF{A'B'}{}
+ \firstcP{B}{A'}[\bar\kappa,\othKV{}1, \dots,\othKV{}p]\sF{AB}{} \big),
\label{Vdensity}\\
\WH{AA'}[\omega] 
&=& 
\omega\mixedindices{A}{B'}\csF{A'B'}{}
+\overline\omega\mixedindices{A'}{B}\sF{AB}{} .
\label{Wdensity}
\endEQs

\Proclaim{ Lemma~5.1. } {
The conserved currents 
\eqref{Tdensity}, \eqref{Zdensity}, \eqref{Vdensity}, \eqref{Wdensity}
admit, respectively, the spinorial characteristics
\EQs
\sTQ{AA'}[\xi,\othKV{}1, \dots,\othKV{}p] &=& 
\frac{1}{2}\bigg(
\zeroP{AA'}{}[\xi,\othKV{}1,\dots,\othKV{}p]
+(-1)^p \sum_{a=0}^p \sum_{\s\in S_p} 
\nonumber\\&&\qquad
\frac{1}{a!(p-a)!}
\zeroP{AA'}{}[\Lie{\othKV{}{s(1)}}\cdots \Lie{\othKV{}{s(a)}}\xi,
\othKV{}{s(a+1)},\dots,\othKV{}{s(p)}]
\bigg),
\label{Tchar}\\
\sZQ{AA'}[\xi, \othKV{}1, \dots,\othKV{}p] &=& 
\frac{\i}{2}\bigg(
\zeroP{AA'}{}[\xi,\othKV{}1,\dots,\othKV{}p]
+ (-1)^{p+1} \sum_{a=0}^p \sum_{\s\in S_p} 
\nonumber\\&&\qquad
\frac{1}{a!(p-a)!}
\zeroP{AA'}{}[\Lie{\othKV{}{s(1)}}\cdots \Lie{\othKV{}{s(a)}}\xi,
\othKV{}{s(a+1)},\dots,\othKV{}{s(p)}]
\bigg),
\label{Zchar}\\
\sVQ{AA'}[\kappa,\othKV{}1, \dots,\othKV{}p] &=&
\frac{1}{2}\bigg(
\firstP{AA'}{}[\kappa, \othKV{}1, \dots,\othKV{}p]
+(-1)^{p+1} \sum_{a=0}^p \sum_{\s\in S_p} 
\nonumber\\&&\qquad
\frac{1}{a!(p-a)!}
\firstP{AA'}{}[\Lie{\othKV{}{s(1)}}\cdots \Lie{\othKV{}{s(a)}}\kappa,
\othKV{}{s(a+1)},\dots,\othKV{}{s(p)}]
\bigg),
\label{Vchar}\\
\sWQ{AA'}[\omega] &=& 
\omega\downindex{AA'} , 
\label{Wchar}
\endEQs
where $S_p$ denotes the symmetric group on the index set $\{1,\ldots,p\}$. 
These characteristics are adjoint symmetries of order 
$q_T,q_Z,q_V,q_W$, where
\EQs
&& q_T=p,  \eqtext{ if $p$ is even, } \quad
q_T <p,  \eqtext{ if $p$ is odd, }
\label{TQorder}\\
&& q_Z=p,  \eqtext{ if $p$ is odd, }\quad 
q_Z <p,  \eqtext{ if $p$ is even, } 
\label{ZQorder}\\
&& q_V=p+1,  \eqtext{ if $p$ is odd, }\quad  
q_V <p+1,  \eqtext{ if $p$ is even, } 
\label{VQorder}
\endEQs
and $q_W=0$.
In particular, for $p=0$, 
$\sTQ{AA'}[\xi] = \zeroP{AA'}{}[\xi]$ is of order $q_T=0$,
and $\sZQ{AA'}[\xi] = \sVQ{AA'}[\kappa] =0$. }

\Proclaim{ Remark: }{
It follows from Corollary~3.3, Proposition~4.2, and Lemma~5.1 that 
\EQ
\zeroP{AA'}{}[\xi,\othKV{}1,\dots,\othKV{}{2r+1}], \quad
\i\zeroP{AA'}{}[\xi,\othKV{}1,\dots,\othKV{}{2r}], \quad
\firstP{AA'}{}[\kappa, \othKV{}1, \dots,\othKV{}{2r}], \quad
r\geq 0 
\label{TZVnotchar}
\endEQ
are non-trivial linear adjoint symmetries,
respectively of order $2r+2,2r+1,2r+1$, 
none of which is equivalent to the characteristic of 
any non-trivial \conslaw/ of \Meq/. }

To prove Lemma~5.1 
we begin with a preliminary Proposition whose proof 
is based on straightforward albeit somewhat lengthy 
computations and will be omitted.

\Proclaim{Proposition~5.2. } {
The adjoint symmetries \eqrefs{Radjsymm}{Sadjsymm}
satisfy the equations
\EQs
&& \sD{(A'}{B}\zeroP{B')B}{}[\xi] 
= \KV{B}{(A'}\ME{B')B}{},
\label{Radj}\\
&& \sD{(A'}{B}\firstP{B')B}{}[\kappa] 
= -\KS{}{A'B'C'D'} \sD{D}{C'} \cME{}{DD'}
-\frac{2}{5} \cME{}{DD'} \sder{C'}{D} \KS{}{A'B'C'D'}, 
\label{Sadj}\\
&& \frac{1}{p!} \sum_{\s\in S_p} 
\jetLie{\othKV{}{s(p)}}
\zeroP{AA'}{}[\xi,\othKV{}{s(1)},\dots,\othKV{}{s(p-1)}]
= \nonumber\\&& \fewquad 
\frac{1}{p!} \sum_{\s\in S_p} 
\zeroP{AA'}{}[\Lie{\othKV{}{s(p)}}\xi,
\othKV{}{s(1)},\dots,\othKV{}{s(p-1)}]
+ \zeroP{AA'}{}[\xi,\othKV{}1,\dots,\othKV{}p], 
\label{LieR}\\
&& \frac{1}{p!} \sum_{\s\in S_p} 
\jetLie{\othKV{}{s(p)}}
\firstP{AA'}{}[\kappa,\othKV{}{s(1)},\dots,\othKV{}{s(p-1)}]
= \nonumber\\&& \fewquad 
\frac{1}{p!} \sum_{\s\in S_p} 
\firstP{AA'}{}[\Lie{\othKV{}{s(p)}}\kappa,
\othKV{}{s(1)},\dots,\othKV{}{s(p-1)}]
+ \firstP{AA'}{}[\kappa,\othKV{}1,\dots,\othKV{}p]. 
\label{LieS}
\endEQs }

\proclaim{Proof of Lemma~5.1. } 
The proof is based on similar computations for each spinorial characteristic. 
We therefore will prove the claim only for the characteristic
$\sVQ{AA'}[\kappa,\othKV{}1, \dots,\othKV{}p]$ 
and omit the rest. 

First by \eqref{Vdensity} we have 
\EQs
\sD{AA'}{}\VH{AA'}[\kappa, \othKV{}1, \dots,\othKV{}p] 
= &&
\frac{1}{2}\bigl(
\ME{}{AB'} \firstP{AB'}{}[\kappa, \othKV{}1, \dots,\othKV{}p] 
+\cME{}{BA'} \firstcP{BA'}{}[\bar\kappa, \othKV{}1, \dots,\othKV{}p] 
\nonumber\\&&
-\csF{A'B'}{} \sD{A'}{A} 
\firstP{AB'}{}[\kappa, \othKV{}1, \dots,\othKV{}p]
-\sF{AB}{} \sD{A}{A'} 
\firstcP{BA'}{}[\bar\kappa, \othKV{}1, \dots,\othKV{}p] 
\bigr) . 
\label{charProp1}
\endEQs
We recall \eqref{Sadjsymm}
and use equations \eqrefs{Sadj}{LieME} 
together with the fact that 
$\sD{A'}{A}$ commutes with $\pr\gen{\othKV{}{}}$
to obtain 
\EQs
&& 
( \sD{(A'}{A} 
\firstP{B')A}{}[\kappa,\othKV{}1,\dots,\othKV{}p] ) \csF{A'B'}{} 
\nonumber\\&&
= \frac{1}{p!} \sum_{\s\in S_p} 
\Big( \pr\gen{\othKV{}{s(1)}}\cdots\pr\gen{\othKV{}{s(p)}}
\sD{(A'}{A} \firstP{B')A}{}[\kappa] 
\Big) \csF{A'B'}{} 
\nonumber\\&&
= \frac{1}{p!} \sum_{\s\in S_p} 
\Big( \pr\gen{\othKV{}{s(1)}}\cdots\pr\gen{\othKV{}{s(p)}}
(-\KS{}{A'B'C'D'} \sD{D}{C'} \cME{}{DD'} 
-\frac{2}{5}\sder{C'}{D} \KS{}{A'B'C'D'} \cME{}{DD'})
\Big) \csF{A'B'}{} 
\nonumber\\&&
=\frac{1}{p!} \sum_{\s\in S_p} 
\Big( 
-\sD{D}{C'} ( \KS{}{A'B'C'D'} 
( \pr\gen{\othKV{}{s(1)}}\cdots\pr\gen{\othKV{}{s(p)}} \cME{}{DD'} ) 
\csF{A'B'}{} )
\nonumber\\&&\fewquad
+ ( \pr\gen{\othKV{}{s(1)}}\cdots\pr\gen{\othKV{}{s(p)}} \cME{}{DD'} )
( \KS{}{A'B'C'D'} \csFder{A'B'}{}{C'}{D}  
+\frac{3}{5}\sder{C'}{D} \KS{}{A'B'C'D'} \csF{A'B'}{} ) \Big)
\nonumber\\&&
= \frac{1}{p!} \sum_{\s\in S_p} 
( \pr\gen{\othKV{}{s(1)}}\cdots\pr\gen{\othKV{}{s(p)}} \cME{}{DD'} )
\firstP{DD'}{}[\kappa]
+\sD{AA'}{} \triv{AA'}{1}
\nonumber\\&&
= \frac{1}{p!} \sum_{\s\in S_p} 
( (\jetLie{\othKV{}{s(1)}}+\div\othKV{}{}) \cdots
(\jetLie{\othKV{}{s(p)}}+\div\othKV{}{}) \cME{}{DD'} )
\firstP{DD'}{}[\kappa]
+\sD{AA'}{} \triv{AA'}{1} , 
\nonumber\\&&
\endEQs
where $\triv{AA'}{1}$ is a trivial conserved current. 
Next we integrate by parts
and use the identity 
\EQs
(\jetLie{\othKV{}{}} + \div\othKV{}{}) 
(\cME{}{DD'} \firstP{DD'}{}[\kappa])
= \sD{AA'}{}( \othKV{AA'}{} \cME{}{DD'} \firstP{DD'}{}[\kappa] ) 
\endEQs
to see that 
\EQs
&& ( \sD{(A'}{A} 
\firstP{B')A}{}[\kappa,\othKV{}1,\dots,\othKV{}p] ) \csF{A'B'}{} 
\nonumber\\&&
= \frac{1}{p!} \sum_{\s\in S_p} 
(-1)^{p} \cME{}{DD'} 
\jetLie{\othKV{}{s(1)}}\cdots\jetLie{\othKV{}{s(p)}} 
\firstP{DD'}{}[\kappa]
+\sD{AA'}{} \triv{AA'}{2} , 
\nonumber
\endEQs
where $\triv{AA'}{2}$ is a trivial conserved current.
Now a repeated application of \eqref{LieS} yields 
\EQs
&& 
( \sD{(A'}{A} 
\firstP{B')A}{}[\kappa,\othKV{}1,\dots,\othKV{}p] ) \csF{A'B'}{} 
= 
\nonumber\\&&
(-1)^{p} \sum_{a=0}^p \sum_{\s\in S_p} 
\frac{1}{a!(p-a)!}  
\cME{}{DD'} 
\firstP{DD'}{}[\Lie{\othKV{}{s(1)}}\cdots 
\Lie{\othKV{}{s(a)}}\kappa,\othKV{}{s(a+1)},\dots,\othKV{}{s(p)}]
+\sD{AA'}{} \triv{AA'}{2} . 
\label{charProp2}
\endEQs
We substitute equation \eqref{charProp2} and its complex conjugate 
into \eqref{charProp1} to conclude that
$\VH{AA'}[\kappa,\othKV{}1, \dots,\othKV{}p]$ 
admits the characteristic \eqref{Vchar}.

Finally, by equation \eqref{Vchar} and Proposition~4.2,
we see that $\sVQ{AA'}[\kappa,\othKV{}1, \dots,\othKV{}p]$ 
is equivalent to an adjoint symmetry with 
the highest order term
\EQ
\frac{1}{2}((-1)^p -1)
\othKV{C_1}{(C_1'}\cdots\othKV{C_{p}}{C_{p}'}\KS{}{A'D'E'F')}
\csFder{D'E'}{}{F'C_1'\cdots C_{p}'}{AC_1\cdots C_{p}} . 
\endEQ
Hence we immediately obtain \eqref{VQorder}. 
\endproof

A conserved current $\dens{AA'}$ of order $q$ is linear/quadratic 
if it can be expressed as a homogeneous linear/quadratic polynomial 
in the variables
$\sFder{}{IJ,}{K_1'\cdots K_p'}{K_1\cdots K_p}$, $0\le p\le q$, 
and complex conjugate variables. 
Let the weight of a monomial 
be the sum of the orders of these variables,
and let the weight of a linear/quadratic current $\dens{AA'}$ 
be the maximum of the weights of the monomials in $\dens{AA'}$. 
This weight is called {\it minimal} if 
it is the smallest among the weights of 
all quadratic currents equivalent to $\dens{AA'}$. 

\Proclaim{ Proposition~5.3. }{
A conserved current $\dens{AA'}$ is equivalent to 
a linear/quadratic current of minimal weight $w$
if and only if $\dens{AA'}$ admits a characteristic $\sQ{AA'}$ 
equivalent to an elementary/linear adjoint symmetry $\sP{}{AA'}$ 
of minimal order $w$. }

\proclaim{ Proof. }
First, by Proposition~3.1 and Theorem~3.2, 
the equivalence classes of linear/quadratic currents
correspond to the equivalence classes of elementary/linear characteristics.
Let $\dens{AA'}$ be a linear/quadratic current 
of minimal weight $w$,
and let $\sP{}{AA'}$ be an elementary/linear adjoint symmetry
of minimal order $p$ 
equivalent to a spinorial characteristic admitted by $\dens{AA'}$.  
We now show that $w=p$. 

By the standard integration by parts procedure \cite{Olver}, 
one can show that 
$\dens{AA'}$ admits the spinorial characteristic 
\EQs
\sQ{AA'} = && 
\sum_{r\geq 0} (-1)^r 
\vol{CA}{} \sD{E_1}{E'_1}\cdots\sD{E_r}{E'_r}  \Big( 
\frac{r+2}{r+3} 
\sjetder{(BC}{}{E_1\cdots E_r)}{(E'_1\cdots E'_r} \curr{}{A')B}
-\sjetder{(BC}{}{E_1\cdots E_r)}{E_1'\cdots [E_r'} \curr{}{A']B}
\Big) , 
\label{Qformula}
\endEQs
which is of order $q$
where $q+1$ equals the weight of $\sD{AA'}{} \dens{AA'}$. 
Since this weight is at most $w+1$, 
we have $q\leq w$. 
Now since $\sQ{AA'}$ is equivalent to $\sP{}{AA'}$,
which has minimal order $p$, 
it immediately follows that $p\leq q$ and hence $p\leq w$. 
On the other hand, 
by Proposition~3.1 
we have that $\dens{AA'}$ is equivalent to a current of weight $p$, 
and hence $w\leq p$ since $\dens{AA'}$ has minimal weight. 
Thus we conclude $w=p$. 
\endproof

\Proclaim{Theorem 5.4. }{
Every conserved current of \Meq/ \eqref{spinorMeq} is equivalent to 
the sum of a linear current and a quadratic current. 
The equivalence classes of linear currents 
are represented by the currents 
\EQ
\WH{AA'}[\omega] , 
\label{Wspan}
\endEQ
where $\omega$ satisfies equation \eqref{Wadjsymm}.
The equivalence classes of 
quadratic currents of weight $w$
are represented by sums of the currents
\EQs
&& \TH{AA'}[\KV{}{},\othKV{}{1}, \dots,\othKV{}{2r}],
\quad 0\leq r\leq [w/2],
\label{Tspan}\\
&& \ZH{AA'}[\KV{}{},\othKV{}{1}, \dots,\othKV{}{2r+1}],
\quad 0\le r\leq [(w-1)/2],
\label{Zspan}\\
&& \VH{AA'}[\KS{}{},\othKV{}{1}, \dots,\othKV{}{2r+1}],
\quad 0\le r\leq [w/2]-1 , 
\label{Vspan}
\endEQs
involving 
type $(1,1)$ real \Kspin/s $\KV{}{}$, $\othKV{}{i}$,
and type $(0,4)$ \Kspin/s $\KS{}{}$
for each $r$. 
In particular, 
up to quadratic currents of lower weight,
a quadratic current of minimal weight $w$ is equivalent to 
a sum of currents given by 
\eqref{Tspan} for $r=w/2$ and \eqref{Vspan} for $r=w/2-1$, 
if $w$ is even,
or \eqref{Zspan} for $r=(w-1)/2$, 
if $w$ is odd. }

\proclaim{Proof. } 
Let $\dens{AA'}$ be a conserved current
satisfying \eqref{characteristic}. 
Recall from \eqref{Qadsymmeq}
that the spinorial characteristic $\sQ{AA'}$ of $\dens{AA'}$ 
satisfies the adjoint symmetry equation \eqref{deteq} 
so that, by Theorem~4.3, 
$\sQ{AA'}$ is equivalent to 
a sum of the linear adjoint symmetries 
\eqref{Radjsymm}, \eqref{Sadjsymm}
and the elementary adjoint symmetry \eqref{Wadjsymm}. 
Thus by Proposition~3.1, $\dens{AA'}$ is equivalent to 
a sum of linear and quadratic conserved currents. 

Clearly, by Theorem~3.2 and Lemma~5.1, 
any linear current is equivalent to a current $\WH{AA'}[\omega]$. 

Next, by Lemma~5.1, 
the quadratic currents 
$\TH{AA'}[\xi,\othKV{}1, \dots,\othKV{}{p}]$,
$\ZH{AA'}[\xi,\othKV{}1, \dots,\othKV{}{p}]$, 
$\VH{AA'}[\kappa,\othKV{}1, \othKV{}2, \dots,\othKV{}{p-1}]$
of order $p \ge 0$
have the respective weights
\EQs
&& w_T=p,\quad w_Z <p,\quad w_V=p,\quad
\eqtext{ if $p$ is even, }
\label{evenweight}\\
&& w_T<p,\quad w_Z =p,\quad w_V<p,\quad
\eqtext{ if $p$ is odd. }
\label{oddweight}
\endEQs
Consequently, by Proposition~5.3 together with Theorem~3.2 and Lemma~5.1, 
we see by using induction on $p$ that 
the currents \eqref{Tspan}, \eqref{Zspan}, \eqref{Vspan}
span the equivalence classes of quadratic currents $\dens{AA'}$ 
of weight at most $w$. 
\endproof

\Proclaim{ Remark: }{
The currents $\TH{AA'}[\xi]$ 
yield the stress-energy \conslaw/s \eqref{Tconslaw}, 
the currents $\ZH{AA'}[\xi,\xi]$ 
yield the zilch \conslaw/s \eqref{Zconslaw}, 
and the currents $\VH{AA'}[\kappa,\xi]$ 
yield the chiral \conslaw/s \eqref{Vconslaw}, 
in spinorial form. }

Write $\vs{0}{W}$ for the real vector space of equivalence classes of 
linear currents \eqref{Wspan}, 
write $\vs{w}{}$ for the real vector space of equivalence classes of 
quadratic currents of weight at most $w$,
and write $\vs{2r}{T}$, $\vs{2r+1}{Z}$, $\vs{2r+2}{V}$
for the respective real vector spaces of equivalence classes of currents 
spanned by \eqref{Tspan}, \eqref{Zspan}, \eqref{Vspan}, 
for a fixed $r\ge 0$. 
We remark that these spans contain non-trivial lower weight currents
and consequently it is convenient to define the following
quotient spaces of quadratic currents. 
Let
\EQ
\qvs{0}{T}=\vs{0}{T} ,\quad 
\qvs{1}{Z}=\vs{1}{Z} ,\quad 
\qvs{2}{V}=\vs{2}{V} ,
\endEQ
and define, for $r\ge 1$, 
\EQ
\qvs{2r}{T}=\vs{2r}{T}/\vs{2r-2}{T} ,\quad  
\qvs{2r+1}{Z}=\vs{2r+1}{Z}/\vs{2r-1}{Z} ,\quad 
\qvs{2r+2}{V}=\vs{2r+2}{V}/\vs{2r}{V} . 
\endEQ
Finally, 
write $\ks{r}{R}$ for 
the real vector space of real Killing spinors of type $(r,r)$,
and write $\ks{r,s}{}$ for 
the complex vector space of Killing spinors of type $(r,s)$,
regarded as a real vector space. 

Let $\dens{AA'}$ be a quadratic conserved current 
in $\vs{2r}{T} \oplus \vs{2r}{V}$. 
By Theorem~3.2, Lemma~5.1, Proposition~4.2 
and the \Kspin/ factorization result of Lemma~4.1, 
we see that any current equivalent to $\dens{AA'}$ 
admits a spinorial characteristic $\sQ{AA'}$ which is 
equivalent to a linear adjoint symmetry of order $2r$ of the form 
\EQ
\sP{}{AA'}= 
\KV{BC_1\cdots C_{2r}}{A'C_1'\cdots C_{2r}'}
\sFder{}{AB}{C_1'\cdots C_{2r}'}{C_1\cdots C_{2r}}
+ \KS{C_1\cdots C_{2r-1}}{A'B'C'D'C_1'\cdots C_{2r-1}'}
\csFder{B'C'}{}{D'C_1'\cdots C_{2r-1}'}{AC_1\cdots C_{2r-1}}
+H_{AA'} , 
\label{leadingorder}
\endEQ
where 
$\KV{BC_1\cdots C_{2r}}{A'C_1'\cdots C_{2r}'}$
is a real \Kspin/ of type $(2r+1,2r+1)$,
$\KS{C_1\cdots C_{2r-1}}{A'B'C'D'C_1'\cdots C_{2r-1}'}$
is a \Kspin/ of type $(2r-1,2r+3)$, 
and where $H_{AA'}$ is of order less than $2r$. 
It follows from \eqref{leadingorder} 
and the identities \eqref{sDids} that 
the highest order terms 
in the spinorial curl $\sD{(C}{A'} \sQ{A)A'}$ of $\sQ{AA'}$ 
are given by 
\EQ
\KV{BC_1\cdots C_{2r}}{A'C_1'\cdots C_{2r}'}
\sFder{}{AB}{A'C_1'\cdots C_{2r}'}{CC_1\cdots C_{2r}}
+ \KS{C_1\cdots C_{2r-1}}{A'B'C'D'C_1'\cdots C_{2r-1}'}
\csFder{A'B'}{}{C'D'C_1'\cdots C_{2r-1}'}{ACC_1\cdots C_{2r-1}}
\label{curlleadingorder}
\endEQ
on $R^{2r+1}(F)$,
and hence the \Kspin/s 
in \eqref{leadingorder} are unique
for the class of conserved currents equivalent to $\dens{AA'}$. 
This yields a linear map 
\EQ\label{evenmap}
I_{2r} :
\vs{2r}{T}\oplus\vs{2r}{V}\to \ks{2r+1}{R}\oplus \ks{2r-1,2r+3}{} , \qquad
I_{2r}(\dens{}) = 
( \KV{BC_1\cdots C_{2r}}{A'C_1'\cdots C_{2r}'}, 
\KS{C_1\cdots C_{2r-1}}{A'B'C'D'C_1'\cdots C_{2r-1}'} ) , 
\endEQ
which is well defined on the equivalence classes of currents in 
$\vs{2r}{T}\oplus\vs{2r}{V}$. 

Next let $\dens{AA'}$ be a quadratic conserved current 
in $\vs{2r+1}{Z}$. 
Similarly as above, 
we can show that any current equivalent to $\dens{AA'}$ 
admits a characteristic $\sQ{AA'}$ which is 
equivalent to a linear adjoint symmetry of order $2r+1$ of the form
\EQ
\sP{}{AA'}= 
\i \KV{BC_1\cdots C_{2r+1}}{A'C_1'\cdots C_{2r+1}'}
\sFder{}{AB}{C_1'\cdots C_{2r+1}'}{C_1\cdots C_{2r+1}}
+H_{AA'} , 
\label{leadingorder'}
\endEQ
where 
$\KV{BC_1\cdots C_{2r+1}}{A'C_1'\cdots C_{2r+1}'}$
is a real Killing spinor of type $(2r+2,2r+2)$,
and where $H_{AA'}$ is of order less than $2r+1$. 
From \eqrefs{leadingorder'}{sDids}
the highest order terms in the spinorial curl of $\sQ{AA'}$ 
are given by 
\EQ
\i \KV{BC_1\cdots C_{2r+1}}{A'C_1'\cdots C_{2r+1}'}
\sFder{}{AB}{A'C_1'\cdots C_{2r+1}'}{CC_1\cdots C_{2r+1}}
\label{curlleadingorder'}
\endEQ
on $R^{2r+2}(F)$,
and hence the \Kspin/ in \eqref{leadingorder'} is unique
for the class of conserved currents equivalent to $\dens{AA'}$. 
This again yields a linear map 
\EQ\label{oddmap}
I_{2r+1} :
\vs{2r+1}{Z}\to \ks{2r+2}{R} , \qquad
I_{2r+1}(\dens{}) = \KV{BC_1\cdots C_{2r+1}}{A'C_1'\cdots C_{2r+1}'}, 
\endEQ
which is well defined on the equivalence classes of currents in 
$\vs{2r+1}{Z}$.

\Proclaim{ Theorem~5.5. }{
(i) The vector space $\vs{w}{}$ is isomorphic to the direct sum 
\EQ\label{iso}
\vs{w}{} \simeq
\oplus_{r=0}^{[w/2]} \qvs{2r}{T}
\oplus_{r=0}^{[(w-1)/2]} \qvs{2r+1}{Z}
\oplus_{r=0}^{[w/2-1]} \qvs{2r+2}{V}. 
\endEQ
Moreover, for $r\ge 0$, 
the vector spaces $\qvs{2r}{T}$, $\qvs{2r+1}{Z}$, $\qvs{2r+2}{V}$
are respectively isomorphic to 
$\ks{2r+1}{R}$, $\ks{2r+2}{R}$, $\ks{2r+1,2r+5}{}$
under the mappings \eqrefs{evenmap}{oddmap}. 
Consequently, these vector spaces of conserved currents 
have the following dimensions over the real numbers:
\begin{mathletters}
\EQs
\dim \qvs{2r}{T} &=& \frac{1}{3}(r+1)^2(2r+3)^2(4r+5), 
\\
\dim \qvs{2r+1}{Z} &=& \frac{1}{3}(r+2)^2(2r+3)^2(4r+7), 
\\
\dim \qvs{2r+2}{V} &=& \frac{2}{3}(r+1)(r+3)(2r+3)(2r+7)(4r+9). 
\endEQs
\end{mathletters}
(ii) The vector space $\vs{0}{W}$ is isomorphic to 
the space of solutions to \Meq/.}

\proclaim{ Proof. }
To prove part (ii) of the Theorem, 
note that by Lemma~5.1 and Theorem~5.4,
the equivalence classes of linear conserved currents 
are in one-to-one correspondence with spinor fields
$\omega\downindex{CC'}$ satisfying \eqref{Wadjsymm}
up to gradient terms $\der{CC'}{}\X$. 
But such spinors $\omega\downindex{CC'}$ correspond to potentials 
for the electromagnetic fields
via $\sF{}{AB} = \sder{B'}{(A} \omega\downindex{B)B'}$, 
while gradient terms $\omega\downindex{CC'}=\der{CC'}{}\X$
represent pure gauge potentials 
giving rise to the zero electromagnetic field, 
$\sF{}{AB} = \sder{B'}{(A} \sder{}{B)B'}\X =0$. 
This establishes an isomorphism between 
solutions of \Meq/
and equivalence classes of linear conserved currents. 

To prove part (i) of the Theorem, 
we note that the isomorphism \eqref{iso} follows from Theorem~5.4. 
Our goal now is to show that 
$\qvs{2r}{T} \simeq \ks{2r+1}{R}$, 
$\qvs{2r+1}{Z} \simeq \ks{2r+2}{R}$, 
$\qvs{2r+2}{V} \simeq \ks{2r+1,2r+5}{}$. 

We first note that by construction 
$I_{2r}$ and $I_{2r+1}$ are linear maps
which descend to 
$\qvs{2r}{T}\oplus\qvs{2r}{V}$
and $\qvs{2r+1}{Z}$, respectively. 
Hence, in order to prove that they are one-to-one, 
it is sufficient to show that their respective kernels 
are contained in $\vs{2r-2}{}$ and $\vs{2r-1}{}$. 
Let $\dens{AA'}$ be a quadratic current in $\vs{2r}{T}\oplus \vs{2r}{V}$
admitting a spinorial characteristic $\sQ{AA'}$ 
such that $I_{2r}(\dens{})=0$. 
Hence we have 
$\KV{BC_1\cdots C_{2r}}{A'C_1'\cdots C_{2r}'}=0$
and $\KS{C_1\cdots C_{2r-1}}{A'B'C'D'C_1'\cdots C_{2r-1}'}=0$
in \eqref{curlleadingorder}. 
It then follows from \eqref{leadingorder} 
that $\sQ{AA'}$ is equivalent to a linear adjoint symmetry $\sP{}{AA'}$
of order less than $2r$. 
By Proposition~5.3 we see that $\dens{AA'}$ is equivalent to 
a quadratic current of weight less than $2r$
and thus is contained in $\vs{2r-2}{T}\oplus \vs{2r-2}{V}$ 
by Theorem~5.4. 
Hence we conclude $\ker I_{2r} \subset \vs{2r-2}{}$.
By a similar argument we can show that 
$\ker I_{2r+1} \subset \vs{2r-1}{}$.

Therefore 
the mappings $I_{2r}$ and $I_{2r+1}$ are one-to-one. 
Next, by the factorization property of \Kspin/s in Lemma~4.1,
it immediately follows that the mappings are onto. 
This establishes the required isomorphisms. 

Finally, the dimension counts of the vector spaces follow from Lemma~4.1.
\endproof

\subsection{ A basis } 

We now present an explicit basis for
the vector space $\vs{w}{}$ of currents spanned by 
\eqref{Tspan}, \eqref{Zspan}, \eqref{Vspan}. 
We start by defining a basis of 
complex conformal \Kvec/s 
and self-dual conformal \KYten/s 
as follows. 

Fix a spinor basis $\{\so{A}{},\si{A}{}\}$, 
namely, 
$\so{A}{}$ and $\si{A}{}$ are linearly independent constant spinors
satisfying $\so{}{A}\si{A}{}=1$. 
Now let
\EQs
&& 
\KV{AA'}{0,1} = \so{A}{}\cso{A'}{}, \quad
\KV{AA'}{0,2} = \so{A}{}\csi{A'}{}, \quad
\cKV{AA'}{0,2} = \si{A}{}\cso{A'}{}, \quad
\KV{AA'}{0,3} = \si{A}{}\csi{A'}{}, 
\label{basisCKV0th}\\
&& 
\KV{AA'}{1,1} = \x{A}{B'}\cso{A'}{}\cso{B'}{}, \quad
\KV{AA'}{1,2} = \x{A}{B'}\cso{(A'}{}\csi{B')}{}, \quad
\KV{AA'}{1,3} = \x{A}{B'}\csi{A'}{}\csi{B'}{}, 
\label{basisCKV1st}\\
&& 
\cKV{AA'}{1,1} = \x{A'}{B}\so{A}{}\so{B}{}, \quad
\cKV{AA'}{1,2} = \x{A'}{B}\so{(A}{}\si{B)}{}, \quad
\cKV{AA'}{1,3} = \x{A'}{B}\si{A}{}\si{B}{}, 
\label{cbasisCKV1st}\\
&& 
\KV{AA'}{1,4} = \x{AA'}{}, 
\label{basisCKV1st'}\\
&& 
\KV{AA'}{2,1} = \x{A}{B'}\x{A'}{B}\so{B}{}\cso{B'}{}, \quad
\KV{AA'}{2,2} = \x{A}{B'}\x{A'}{B}\so{B}{}\csi{B'}{}, \quad
\cKV{AA'}{2,2} = \x{A}{B'}\x{A'}{B}\si{B}{}\cso{B'}{}, \quad
\KV{AA'}{2,3} = \x{A}{B'}\x{A'}{B}\si{B}{}\csi{B'}{}, 
\label{basisCKV2nd}
\endEQs
and 
\EQs
&& 
\KY{0,1}{AB} = \so{A}{}\so{B}{}, \quad
\KY{0,2}{AB} = \so{(A}{}\si{B)}{}, \quad
\KY{0,3}{AB} = \si{A}{}\si{B}{}, 
\label{basisKY0th}\\
&&
\KY{1,1}{AB} = \x{(A}{C'}\so{B)}{}\cso{C'}{}, \quad
\KY{1,2}{AB} = \x{(A}{C'}\so{B)}{}\csi{C'}{}, \quad
\KY{1,3}{AB} = \x{(A}{C'}\si{B)}{}\cso{C'}{}, \quad
\KY{1,4}{AB} = \x{(A}{C'}\si{B)}{}\csi{C'}{}, 
\label{basisKY1st}\\
&& 
\KY{2,1}{AB} = \x{(A}{C'}\x{B)}{D'}\cso{C'}{}\cso{D'}{}, \quad
\KY{2,2}{AB} = \x{(A}{C'}\x{B)}{D'}\cso{C'}{}\csi{D'}{}, \quad
\KY{2,3}{AB} = \x{(A}{C'}\x{B)}{D'}\csi{C'}{}\csi{D'}{} . 
\label{basisKY2nd}
\endEQs

From equations \eqrefs{Kvecs}{KYtens}
we have the following result. 

\Proclaim{ Proposition~5.6. }{
The set of 15 spinorial conformal \Kvec/s 
\eqsref{basisCKV0th}{basisCKV2nd}
is a basis for the complex vector space 
$\ks{1,1}{}$ of type $(1,1)$ \Kspin/s.
The set of 10 spinorial conformal \KYten/s
\eqsref{basisKY0th}{basisKY2nd}
is a basis for the complex vector space 
$\ks{2,0}{}$ of type $(2,0)$ \Kspin/s.
}

To proceed, 
we state a preliminary result which follows immediately from
Theorem~5.5, Proposition~5.6 and Lemma~4.1, 
and equation \eqref{sduality}. 

\Proclaim{ Proposition~5.7. }{
Let $\dens{AA'}$ be a current of weight $q$. 
Then $\dens{AA'}$ is equivalent to a quadratic current 
which is a polynomial of degree at most $2q+2$ in $\x{CC'}{}$
and admits a unique decomposition into a sum of monomials 
each of which possess either even parity or odd parity 
with respect to the duality transformation \eqref{sduality}.
}

We now define a basis for $\vs{w}{}$ 
inductively in terms of 
the weight $0\le q\le w$ and the degree $0\le p\le 2q+2$. 
Let $\vs{w}{}=\vs{w}{+} \oplus \vs{w}{-}$, 
where $\vs{w}{+}$ and $\vs{w}{-}$ denote the subspaces of 
currents with even parity $(+)$ and odd parity $(-)$ respectively. 
We consider $\vs{w}{+}$ and $\vs{w}{-}$ separately. 

To proceed, we first state the main results 
and then present the proofs afterwards. 


The basis for $\vs{w}{+}$ 
is indexed by $s=q+1$ and $p$ in addition to two pairs of integers
$(i,j)$ and $(n,n')$ satisfying 
\EQs
&& \max(0,p-s) \le i\le j\le p-i, \quad
0\le p \le 2s, 
\label{ijrange}
\quad\eqtext{ and } \\&&\nonumber\\
&& \cases{
0\le n\le s-p+2i, \quad 
0\le n'\le s-p+2j, & if $i<j$; \cr
0\le n\le n'\le s-p+2i, & if $i=j$. }
\label{nn'range}
\endEQs
Write
\EQ
l=s-p+i, \quad
l'=s-p+j . 
\label{ll'}
\endEQ
For each value of $i,j,n,n',p,q$, 
we define the stress-energy type currents when $q$ is even, 
\EQs
\curr{AA'}{T(i,j,n,n';p,q)}[\KV{}{1},\ldots,\KV{}{q+1}] 
= &&
\frac{1}{2}\big( 
\zeroP{B'}{A}[\KV{}{1}, \dots,\KV{}{q+1}]
+ \zeroP{B'}{A}[\cKV{}{1}, \dots,\cKV{}{q+1}] 
\big)\csF{A'B'}{}
\nonumber\\&&\qquad
+ \frac{1}{2}\big( 
\zerocP{B}{A'}[\KV{}{1}, \dots,\KV{}{q+1}]
+ \zerocP{B}{A'}[\cKV{}{1}, \dots,\cKV{}{q+1}] \big) \sF{AB}{}, 
\label{basisTdensity}
\endEQs
\EQs
\curr{AA'}{T'(i,j,n,n';p,q)}[\KV{}{1},\ldots,\KV{}{q+1}] 
= &&
\frac{\i}{2}\big( 
\zeroP{B'}{A}[\KV{}{1}, \dots,\KV{}{q+1}]
- \zeroP{B'}{A}[\cKV{}{1}, \dots,\cKV{}{q+1}] 
\big)\csF{A'B'}{}
\nonumber\\&&\qquad
+ \frac{\i}{2}\big( 
\zerocP{B}{A'}[\KV{}{1}, \dots,\KV{}{q+1}]
- \zerocP{B}{A'}[\cKV{}{1}, \dots,\cKV{}{q+1}] \big) \sF{AB}{}, 
\label{basisT'density}\\
&& \eqtext{ if $i\neq j$ or $n\neq n'$ }, 
\nonumber
\endEQs
and we also define the zilch type currents when $q$ is odd, 
\EQs
\curr{AA'}{Z(i,j,n,n';p,q)}[\KV{}{1},\ldots,\KV{}{q+1}] 
= &&
\frac{\i}{2}\big( 
\zeroP{B'}{A}[\KV{}{1}, \dots,\KV{}{q+1}]
+ \zeroP{B'}{A}[\cKV{}{1}, \dots,\cKV{}{q+1}] 
\big)\csF{A'B'}{}
\nonumber\\&&\qquad
- \frac{\i}{2}\big( 
\zerocP{B}{A'}[\KV{}{1}, \dots,\KV{}{q+1}]
+ \zerocP{B}{A'}[\cKV{}{1}, \dots,\cKV{}{q+1}] \big) \sF{AB}{}, 
\label{basisZdensity}
\endEQs
\EQs
\curr{AA'}{Z'(i,j,n,n';p,q)}[\KV{}{1},\ldots,\KV{}{q+1}] 
= &&
\frac{1}{2}\big( 
\zeroP{B'}{A}[\KV{}{1}, \dots,\KV{}{q+1}]
- \zeroP{B'}{A}[\cKV{}{1}, \dots,\cKV{}{q+1}] 
\big)\csF{A'B'}{}
\nonumber\\&&\qquad
- \frac{1}{2}\big( 
\zerocP{B}{A'}[\KV{}{1}, \dots,\KV{}{q+1}]
- \zerocP{B}{A'}[\cKV{}{1}, \dots,\cKV{}{q+1}] \big) \sF{AB}{}, 
\label{basisZ'density}\\
&& \eqtext{ if $i\neq j$ or $n\neq n'$ }, 
\nonumber
\endEQs
where 
$\zerocP{A}{A'}[\KV{}{1}, \dots,\KV{}{q+1}] = 
\overline{ \zeroP{A'}{A}[\cKV{}{1}, \dots,\cKV{}{q+1}] }$. 
In \eqsref{basisTdensity}{basisZ'density}
the set $\{ \KV{}{1},\ldots,\KV{}{q+1} \}$
consists of the conformal \Kvec/s 
\eqref{basisCKV0th}, \eqref{basisCKV1st}, \eqref{basisCKV1st'},
\eqref{basisCKV2nd}
with each $\KV{}{a,b}$ 
appearing $\#[\KV{}{a,b}]$ times 
according to the following count formulas:
\EQs
&&
\#[\KV{}{0,1}] = 
\min(n,l,n'), 
\label{count1}\\&&
\#[\KV{}{0,2}] = 
\max(0,\min(n-n',l-n')), 
\\&&
\#[\cKV{}{0,2}] = 
\max(0,\min(n'-n,l-n)), 
\\&&
\#[\KV{}{0,3}] = 
\max(0,\min(l-n',l-n)), 
\\
&&
\#[\KV{}{1,1}] = 
\max(0,\min(n'-l',l'-l)), 
\\&&
\#[\KV{}{1,2}] = 
\max(0,l'-l-|l'-n'|), 
\\&&
\#[\KV{}{1,3}] = 
\max(0,\min(l'-n',l'-l)), 
\\
&&
\#[\KV{}{1,4}] = s-(i+j+l+l')/2, 
\\
&&
\#[\cKV{}{1,1}] = \#[\cKV{}{1,2}] = \#[\cKV{}{1,3}] = 0, 
\\
&&
\#[\KV{}{2,1}] = 
\max(0,\min(n'-2l'+l,n-l)), 
\\&&
\#[\KV{}{2,2}] = 
\max(0,\min(n-n'+2(l'-l),n-l)), 
\\&&
\#[\cKV{}{2,2}] = 
\max(0,\min(n'-n+2(l-l'),n'-2l'+l)), 
\\&&
\#[\KV{}{2,3}] = 
i-\max(0,n'-2l'+l,n-l). 
\label{count15}
\endEQs
Note that 
$\#[\KV{}{0,1}]+ \#[\KV{}{0,2}]+ \#[\cKV{}{0,2}]+ \#[\KV{}{0,3}]=l$, 
$\#[\KV{}{1,1}]+ \#[\KV{}{1,2}]+ \#[\KV{}{1,3}]=j-i$, 
$\#[\KV{}{1,4}]=p-i-j$, 
$\#[\KV{}{2,1}]+ \#[\KV{}{2,2}]+ \#[\cKV{}{2,2}]+ \#[\KV{}{2,3}]=i$,
and so $\sum_{a,b} \#[\KV{}{a,b}]=s$.

\Proclaim{ Theorem~5.8 }{
The set of all currents \eqrefs{basisTdensity}{basisT'density}
for $0\le p\le 2q+2$, $q=2r$, $0\le r\le [w/2]$, 
together with \eqrefs{basisZdensity}{basisZ'density}
for $0\le p\le 2q+2$, $q=2r+1$, $0\le r\le [(w-1)/2]$, 
constitutes a basis for $\vs{w}{+}$. 
}


The basis for $\vs{w}{-}$ 
is indexed by $s=q-1$ and $p$ 
in addition to two pairs and a triplet of integers 
$(i,j,k)$, $(n,n')$ and $(m,m')$, satisfying
\EQs
&& p'=p-k, \quad
0\le k\le 4, \quad
0\le p'\le 2s, 
\label{kp'range}\\
&& 
\max(0,p'-s) \le i\le j\le p'-i, 
\label{ijrange'}\\&&\nonumber\\
&& 
\cases{
0\le n\le s-p'+2i, \quad
0\le n'\le s-p'+2j, &  if $i<j$; \cr
0\le n\le n'\le s-p'+2i, & if $i=j$; }
\label{nn'range'}\\&&\nonumber\\
&&
\cases{
0\le m\le 4-k, \quad
0\le m'\le k, & if $n=0$;  \cr
m= 4-k, \quad 
m'=k, & if $n>0$.  }
\label{mm'range}
\endEQs
Write
\EQ
l=s-p'+i, \quad
l'=s-p'+j, \quad
h=[k/2]. 
\label{ll's}
\endEQ
For each value of $i,j,k,n,n',m,m',p,q$, 
if $q$ is even, we define the chiral type currents 
\EQs
&& \curr{AA'}{+V(i,j,k,n,n',m,m';p,q)}
[\KY{1}{},\KY{2}{},\KV{}{1},\ldots,\KV{}{q-1}] = 
\frac{1}{2}\big( 
\firstP{B'}{A}[\KS{}{},\KV{}{1}, \dots,\KV{}{q-1}] 
+ \firstP{B'}{A}[\KS{}{},\cKV{}{1}, \dots,\cKV{}{q-1}] \big) 
\csF{A'B'}{}
\nonumber\\&&\fewquad
+ \frac{1}{2}\big( 
\firstcP{B}{A'}[\cKS{}{},\KV{}{1},\dots,\KV{}{q-1}] 
+ \firstcP{B}{A'}[\cKS{}{},\cKV{}{1},\dots,\cKV{}{q-1}] \big)
\sF{AB}{} , 
\label{basisVdensity}\\
&& \curr{AA'}{-V(i,j,k,n,n',m,m';p,q)}
[\KY{1}{},\KY{2}{},\KV{}{1},\ldots,\KV{}{q-1}] = 
\frac{\i}{2}\big( 
\firstP{B'}{A}[\KS{}{},\KV{}{1}, \dots,\KV{}{q-1}] 
+ \firstP{B'}{A}[\KS{}{},\cKV{}{1}, \dots,\cKV{}{q-1}] \big) 
\csF{A'B'}{}
\nonumber\\&&\fewquad
- \frac{\i}{2}\big( 
\firstcP{B}{A'}[\cKS{}{},\KV{}{1},\dots,\KV{}{q-1}] 
+ \firstcP{B}{A'}[\cKS{}{},\cKV{}{1},\dots,\cKV{}{q-1}] \big)
\sF{AB}{} , 
\label{basisVdensity'}
\endEQs
and 
\EQs
&& \curr{AA'}{+V'(i,j,k,n,n',m,m';p,q)}
[\KY{1}{},\KY{2}{},\KV{}{1},\ldots,\KV{}{q-1}] = 
\frac{\i}{2}\big( 
\firstP{B'}{A}[\KS{}{},\KV{}{1}, \dots,\KV{}{q-1}] 
- \firstP{B'}{A}[\KS{}{},\cKV{}{1}, \dots,\cKV{}{q-1}] \big) 
\csF{A'B'}{}
\nonumber\\&&\fewquad
+\frac{\i}{2}\big( 
\firstcP{B}{A'}[\cKS{}{},\KV{}{1},\dots,\KV{}{q-1}] 
- \firstcP{B}{A'}[\cKS{}{},\cKV{}{1},\dots,\cKV{}{q-1}] \big)
\sF{AB}{} , 
\label{basisV'density}\\
&& \curr{AA'}{-V'(i,j,k,n,n',m,m';p,q)}
[\KY{1}{},\KY{2}{},\KV{}{1},\ldots,\KV{}{q-1}] = 
\frac{1}{2}\big( 
\firstP{B'}{A}[\KS{}{},\KV{}{1}, \dots,\KV{}{q-1}] 
- \firstP{B'}{A}[\KS{}{},\cKV{}{1}, \dots,\cKV{}{q-1}] \big) 
\csF{A'B'}{}
\nonumber\\&&\fewquad
-\frac{1}{2}\big( 
\firstcP{B}{A'}[\cKS{}{},\KV{}{1},\dots,\KV{}{q-1}] 
- \firstcP{B}{A'}[\cKS{}{},\cKV{}{1},\dots,\cKV{}{q-1}] \big)
\sF{AB}{} , 
\label{basisV'density'}\\
&&\fewquad\nonumber
\eqtext{ if $i\neq j$ or $n\neq n'$ }, 
\endEQs
where $\KS{ABCD}{}= \KY{1}{(AB}\KY{2}{CD)}$,
and 
$\firstcP{A}{A'}[\cKS{}{},\KV{}{1},\dots,\KV{}{q-1}] 
= \overline{ \firstP{A'}{A}[\KS{}{},\cKV{}{1},\dots,\cKV{}{q-1}] }$. 
In \eqsref{basisVdensity}{basisV'density'}
the set $\{ \KV{}{1},\ldots,\KV{}{q-1} \}$
consists of the conformal \Kvec/s 
\eqref{basisCKV0th}, \eqref{basisCKV1st}, \eqref{basisCKV1st'},
\eqref{basisCKV2nd}
with each $\KV{}{a,b}$ 
appearing $\#[\KV{}{a,b}]$ times 
according to the previous count formulas \eqsref{count1}{count15}, 
and, in addition, the set $\{ \KY{1}{},\KY{2}{} \}$ 
consists of the conformal \KYten/s 
\eqref{basisKY0th}, \eqref{basisKY1st}, \eqref{basisKY2nd}
with each $\KY{a,b}{}$ appearing $\#[\KY{a,b}{}]$ times
according to the following count formulas:
\EQs
&&
\#[\KY{0,1}{}]= 
\min(2+h-k,\max(0,m+k-h-2)), 
\label{count1'}\\&&
\#[\KY{0,2}{}]= 
\max(0,\min(m,4+2h-2k-m)),
\\&&
\#[\KY{0,3}{}]= 
\max(0,2+h-k-m),
\\
&&
\#[\KY{1,1}{}]= 
\max(0,m+k-4+\min(k-2h,m')), 
\\&&
\#[\KY{1,2}{}]= 
\max(0,m+k-4+\max(0,k-2h-m')), 
\\&&
\#[\KY{1,3}{}]= 
\min(k-2h,m')-\max(0,m+k-4+\min(k-2h,m')), 
\\&&
\#[\KY{1,4}{}]= 
\max(0,k-2h-m')-\max(0,m+k-4+\max(0,k-2h-m')), 
\\
&&
\#[\KY{2,1}{}]= 
\max(0,m'-k+h),
\\&&
\#[\KY{2,2}{}]= 
\max(0,k-m'+\min(0,2m'-2k+2h)),
\\&&
\#[\KY{2,3}{}]= 
\max(0,h-\max(0,m'-k+2h)). 
\label{count10'}
\endEQs
Note that 
$\#[\KY{0,1}{}]+ \#[\KY{0,2}{}]+ \#[\KY{0,3}{}]=2-k+h$, 
$\#[\KY{1,1}{}]+ \#[\KY{1,2}{}]+ \#[\KY{1,3}{}]+ \#[\KY{1,4}{}]=k-2h$, 
$\#[\KY{2,1}{}]+ \#[\KY{2,2}{}]+ \#[\KY{2,3}{}]+ \#[\KY{2,4}{}] =h$,
and so $\sum_{a,b}\#[\KY{a,b}{}] =2$.

\Proclaim{ Theorem~5.9 }{
The set of all currents \eqsref{basisVdensity}{basisV'density'}
for $0\le p\le 2q+2$, $q=2r+2$, $0\le r\le [w/2]-1$, 
constitutes a basis for $\vs{w}{-}$. 
}

The proof of Theorems~5.8 and~5.9 relies on the construction of 
an explicit basis for \Kspin/s based on the factorization Lemma~4.1.

\Proclaim{ Lemma~5.10 }{
(i) For $s\geq 0$ a basis for $\ks{s,s}{R}$ is given by 
the set of symmetrized products of conformal \Kvec/s
\EQs
&& \KV{\ (A_1}{1(A'_1} \cdots \KV{\ A_s)}{s A'_s)}
+ \cKV{\ (A_1}{1(A'_1} \cdots \cKV{\ A_s)}{s A'_s)} , 
\label{symmprodKV}\\
&& \i\KV{\ (A_1}{1(A'_1} \cdots \KV{\ A_s)}{s A'_s)}
-\i\cKV{\ (A_1}{1(A'_1} \cdots \cKV{\ A_s)}{s A'_s)} , 
\quad\eqtext{ if $i\neq j$ or $n\neq n'$, }
\label{symmprodKV'}
\endEQs
where $\KV{}{1},\ldots,\KV{}{s}$ 
are chosen by the count formulas \eqsref{count1}{count15}
with $p,i,j,n,n'$ satisfying \eqrefs{ijrange}{nn'range}. 
\hfill\vskip0pt
(ii) For $s\geq 0$ a basis for $\ks{4+s,s}{}$ is given by 
the set of symmetrized products of conformal \Kvec/s and conformal \KYten/s
\EQs
&& \KY{1}{\ (AB} \KY{2}{\ CD} 
\KV{\ A_1}{1(A'_1} \cdots \KV{\ A_s)}{s A'_s)}
+ \KY{1}{\ (AB} \KY{2}{\ CD} 
\cKV{\ A_1}{1(A'_1} \cdots \cKV{\ A_s)}{s A'_s)} , 
\label{symmprodKYKV}\\
&& \i\KY{1}{\ (AB} \KY{2}{\ CD} 
\KV{\ A_1}{1(A'_1} \cdots \KV{\ A_s)}{s A'_s)}
-\i \KY{1}{\ (AB} \KY{2}{\ CD} 
\cKV{\ A_1}{1(A'_1} \cdots \cKV{\ A_s)}{s A'_s)}, 
\quad\eqtext{ if $i\neq j$ or $n\neq n'$, }
\label{symmprodKYKV'}
\endEQs
where $\KV{}{1},\ldots,\KV{}{s}$, $\KY{1}{},\KY{2}{}$
are chosen by the respective 
count formulas \eqsref{count1}{count15} and \eqsref{count1'}{count10'}
with $p,k,i,j,n,n',m,m'$ satisfying \eqsref{kp'range}{mm'range}. 
}

\proclaim{Proof. } 
Let $\KV{A_1\cdots A_s}{A'_1\cdots A'_s}$ be a 
\Kspin/ of type $(s,s)$. 
By the factorization Lemma~4.1, 
one can show that 
$\KV{A_1\cdots A_s}{A'_1\cdots A'_s}$ is 
a sum of linearly independent monomials in $\x{CC'}{}$
given by 
\EQ
\KV{A_1\cdots A_s}{A'_1\cdots A'_s} = 
\sum_{i,j,p} 
\spinor{\Gamma_{(i,j,p)}}{A_1\cdots A_s}{A'_1\cdots A'_s}(x,\gamma) , 
\label{monomials}
\endEQ
with $i,j,p$ satisfying 
$i\geq \max(0,p-s)$, $j\geq \max(0,p-s)$, $i+j\leq p$, 
and 
\EQ
\spinor{\Gamma_{(i,j,p)}}{A_1\cdots A_s}{A'_1\cdots A'_s}(x,\gamma)
= \x{(A_1\cdots A_{p-i}| E_1\cdots E_i}{E'_1\cdots E'_j (A'_1\cdots A'_{p-j}}
\spinor{\gamma}{|A_{p-i+1}\cdots A_s) B_1 \cdots B_i}
{A'_{p-j+1} \cdots A'_s) B'_1 \cdots B'_j}
\vol{E_1B_1}{}\cdots \vol{E_iB_i}{}
\vol{}{E'_1B'_1}\cdots \vol{}{E'_jB'_j} , 
\label{basismonomial1}
\endEQ
where 
$\x{B_1 \cdots B_k}{B'_1\cdots B'_k} =
\x{B_1}{B'_1} \cdots \x{B_k}{B'_k}$,
and where 
$\spinor{\gamma}{A_{1}\cdots A_{l} B_1 \cdots B_i}
{A'_{1} \cdots A'_{l'} B'_1 \cdots B'_j}$
is a constant symmetric spinor. 
Note that the complex conjugate of each monomial in \eqref{monomials} is 
$\spinor{{\bar\Gamma}_{(i,j,p)}}{A_1\cdots A_s}{A'_1\cdots A'_s}
(x,\bar\gamma)
= \spinor{\Gamma_{(j,i,p)}}{A_1\cdots A_s}{A'_1\cdots A'_s}
(x,\bar\gamma)$. 

Hence, by setting
\EQ
\spinor{\gamma{}_{(n,n')}}{C_1\cdots C_{l+i}}{C'_1 \cdots C'_{l'+j}} 
= \so{(C_1}{} \cdots\so{C_n}{} \si{C_{n+1}}{} \cdots \si{C_{l+i})}{}
\cso{}{(C'_1} \cdots\cso{}{C'_{n'}} \csi{}{C'_{n'+1}} \cdots\csi{}{C'_{l'+j})},
\label{factorKVproduct}
\endEQ
we obtain a basis for the real vector space $\ks{s,s}{R}$,
given by the \Kspin/s
\EQ
\KV{A_1\cdots A_s}{A'_1\cdots A'_s} = 
\sum_{i,j,p} \Big( 
\spinor{\Gamma_{(i,j,p)}}{A_1\cdots A_s}{A'_1\cdots A'_s}
(x,\gamma{}_{(n,n')})
+ \spinor{{\Gamma}_{(j,i,p)}}{A_1\cdots A_s}{A'_1\cdots A'_s}
(x,\bar\gamma{}_{(n,n')})
\Big) , 
\label{basisKS1}
\endEQ
and
\EQ
\KV{A_1\cdots A_s}{A'_1\cdots A'_s} = 
\sum_{i,j,p} \Big( 
\i\spinor{\Gamma_{(i,j,p)}}{A_1\cdots A_s}{A'_1\cdots A'_s}
(x,\gamma{}_{(n,n')})
-\i \spinor{{\Gamma}_{(j,i,p)}}{A_1\cdots A_s}{A'_1\cdots A'_s}
(x,\bar\gamma{}_{(n,n')})
\Big) , 
\eqtext{ when $i\neq j$ or $n\neq n'$, }
\label{basisKS'1}
\endEQ
where $i,j,p$ satisfy \eqref{ijrange}
and $n,n'$ satisfy \eqref{nn'range}. 

Now, by straightforward calculations one can verify that 
each conformal \Kvec/ product \eqref{symmprodKV}
for fixed $i,j,p,n,n'$ in the count formulas \eqsref{count1}{count15}
is equal to the monomial 
$\spinor{\Gamma_{(i,j,p)}}{A_1\cdots A_s}{A'_1\cdots A'_s}
(x,\gamma{}_{(n,n')})
+ \spinor{{\Gamma}_{(j,i,p)}}{A_1\cdots A_s}{A'_1\cdots A'_s}
(x,\bar\gamma{}_{(n,n')})$
plus a certain linear combination of monomials
$\spinor{\Gamma_{(i',j',p)}}{A_1\cdots A_s}{A'_1\cdots A'_s}
(x,\gamma{}_{(n,n')})
+ \spinor{{\Gamma}_{(j',i',p)}}{A_1\cdots A_s}{A'_1\cdots A'_s}
(x,\bar\gamma{}_{(n,n')})$
over smaller index values $0\le i'<i$, $i'\le j' <j$. 
Similar calculations hold for each conformal \Kvec/ product \eqref{symmprodKV'}
in terms of monomials 
$\i\spinor{\Gamma_{(i,j,p)}}{A_1\cdots A_s}{A'_1\cdots A'_s}
(x,\gamma{}_{(n,n')})
-\i \spinor{{\Gamma}_{(j,i,p)}}{A_1\cdots A_s}{A'_1\cdots A'_s}
(x,\bar\gamma{}_{(n,n')})
\neq 0$ 
for $i\neq j$ or $n\neq n'$. 
Therefore, by induction on $i,j,p,n,n'$, 
it follows that the set of conformal \Kvec/ products 
\eqrefs{symmprodKV}{symmprodKV'}
comprise a basis for $\ks{s,s}{R}$.

This completes the proof of part (i). 
The proof of part (ii) is similar and will be omitted. 
\endproof

\proclaim{ Proof of Theorems~5.8 and~5.9. }
Let $\dens{AA'}$ be any of the currents
\eqref{basisTdensity}, \eqref{basisT'density}, 
\eqref{basisZdensity}, \eqref{basisZ'density}. 
If $q$ is even, 
we see by Lemma~5.1 that the highest order terms 
in the spinorial characteristic \eqref{Tchar} admitted by $\dens{AA'}$
are given by \eqref{leadingorder} with $r=(s-1)/2$, 
where $\KV{BC_1\cdots C_{2r}}{A'C_1'\cdots C_{2r}'}$
is a \Kspin/ as given in Lemma~5.10(i), 
and where $\KS{C_1\cdots C_{2r-1}}{A'B'C'D'C_1'\cdots C_{2r-1}'}=0$. 
Similarly, 
if $q$ is odd, 
the highest order terms 
in the spinorial characteristic \eqref{Zchar} admitted by $\dens{AA'}$
are given by \eqref{leadingorder'} with $r=s/2-1$, 
where $\KV{BC_1\cdots C_{2r+1}}{A'C_1'\cdots C_{2r+1}'}$
is a \Kspin/ as above. 

Hence, the mappings \eqrefs{evenmap}{oddmap} provide an isomorphism
$\vs{w}{+} \simeq \oplus_{r=0}^{w} \ks{r+1}{R}$. 
Thus by Theorem~5.5 and Lemma~5.10 
the currents in Theorem~5.8 
comprise a basis for $\vs{w}{+}$. 

Next let $\dens{AA'}$ be any of the currents
\eqref{basisVdensity}, \eqref{basisVdensity'}, 
\eqref{basisV'density}, \eqref{basisV'density'}. 
We see by Lemma~5.1 that the highest order terms 
in the spinorial characteristic \eqref{Vchar} 
admitted by $\dens{AA'}$
are given by \eqref{leadingorder} with $r=(s-1)/2$, 
where 
$\KS{C_1\cdots C_{2r-1}}{A'B'C'D'C_1'\cdots C_{2r-1}'}$
is a \Kspin/ as given in Lemma~5.10(ii)
and $\KV{BC_1\cdots C_{2r}}{A'C_1'\cdots C_{2r}'}=0$.

Hence, the mapping \eqref{oddmap} provides an isomorphism
$\vs{w}{-} \simeq \oplus_{r=0}^{[w/2]} \ks{2r-1,3+2r}{}$. 
Thus by Theorem~5.5 and Lemma~5.10 the currents in Theorem~5.9
comprise a basis for $\vs{w}{-}$. 
\endproof

\subsection{ Classification algorithm }

To conclude, we remark that by Theorems~5.4 and~5.5, and Lemma~5.10, 
there is a simple algorithm 
for writing any given quadratic conserved current of \Meq/
explicitly as a sum of currents \eqref{Tspan}, \eqref{Zspan}, \eqref{Vspan}
up to a trivial current. 
Let $\dens{AA'}$ be a quadratic current of weight $w\ge 0$
and proceed by the following steps:
\vskip 0pt
(i) Calculate a spinorial characteristic $\sQ{AA'}$ for 
$\dens{AA'}$ by \eqref{Qformula}. 
\hfill\vskip 0pt
(ii) Calculate the spinorial curl $\sD{(C}{A'} \sQ{A)A'}$ 
and equate its highest order terms to expressions 
\eqref{curlleadingorder} or \eqref{curlleadingorder'}. 
There are two cases to consider:
Let $q$ denote the order of the curl expression. 
If $q$ is odd, the highest order terms must match \eqref{curlleadingorder}
for some \Kspin/s 
$\KV{BC_1\cdots C_{2r}}{A'C_1'\cdots C_{2r}'}$
and $\KS{C_1\cdots C_{2r-1}}{A'B'C'D'C_1'\cdots C_{2r-1}'}$ 
with $r=(q-1)/2$.
Then $\dens{AA'}$ is equivalent to a sum of quadratic currents
\eqrefs{Tspan}{Vspan} of weight $q-1$
plus lower order currents,
with the \Kspin/s $\KV{}{}$, $\othKV{}{1},\ldots,\othKV{}{2r}$
in \eqref{Tspan}
and the \Kspin/s $\KS{}{}$, $\othKV{}{1},\ldots,\othKV{}{2r-1}$
in \eqref{Vspan}
given by a respective \Kspin/ factorization of 
$\KV{BC_1\cdots C_{2r}}{A'C_1'\cdots C_{2r}'}$
and $\KS{C_1\cdots C_{2r-1}}{A'B'C'D'C_1'\cdots C_{2r-1}'}$ 
as in the proof of Lemma~5.10. 
Similarly, 
if $q$ is even, the highest order terms must match \eqref{curlleadingorder'}
for some \Kspin/
$\KV{BC_1\cdots C_{2r+1}}{A'C_1'\cdots C_{2r+1}'}$
with $r=q/2-1$.
Then $\dens{AA'}$ is equivalent to a sum of quadratic currents 
\eqref{Zspan} of weight $q-1$
plus lower order currents,
with the \Kspin/s $\KV{}{}$, $\othKV{}{1},\ldots,\othKV{}{2r+1}$
in \eqref{Zspan}
given by a \Kspin/ factorization of 
$\KV{BC_1\cdots C_{2r+1}}{A'C_1'\cdots C_{2r+1}'}$
as in the proof of Lemma~5.10. 
\hfill\vskip0pt
(iii) Subtract from $\dens{AA'}$ the quadratic current of minimal weight
$w_Q=q-1$ determined in step (ii)
and repeat these steps until $q=1$. 

By steps (i), (ii), (iii) 
we have expressed $\dens{AA'}$ as 
a sum of quadratic currents \eqref{Tspan}, \eqref{Zspan}, \eqref{Vspan}
of minimal weights at most $w_Q$ 
plus a trivial current.

\section{Correspondence between tensorial and spinorial currents}
\label{conversion}

\subsection{ Currents }

In this section we show that the tensorial currents in Theorem~2.1
span the vector space of equivalence classes of quadratic currents 
in Theorem~5.4.

Let 
$\KV{\mu}{} = \inve{\mu}{AA'} \KV{AA'}{}$ 
be a real conformal \Kvec/,
and let 
$\KY{}{\mu\nu} = \inve{\mu AA'}{} \inve{\nu BB'}{}$
$(\KY{A'B'}{} \vol{AB}{} + \cKY{AB}{} \vol{A'B'}{})$
be a real conformal \KYten/.
Let 
$\F{}{\mu\nu} = \e{\mu}{AA'} \e{\nu}{BB'}$
$(\sF{}{AB} \vol{A'B'}{} + \csF{}{A'B'} \vol{AB}{})$ 
be a solution of \Meq/. 
Write 
\EQ\label{nfoldF}
\snF{n}{\xi}{}{AB} = (\pr\gen{\xi})^n \sF{}{AB} ,\qquad
\csnF{n}{\xi}{}{A'B'} = (\pr\gen{\xi})^n \csF{}{A'B'} , 
\endEQ
and let 
$\xi^{(n)}=\{\KV{}{1},\ldots,\KV{}{n}\}$,
$\KV{}{1}=\cdots=\KV{}{n}=\xi$, 
denote the set consisting of $n$ copies of a conformal \Kvec/ $\KV{}{}$.
Then we have the identities
\EQ
\zeroP{AA'}{}[\xi^{(n+1)}]
= \zeroP{AA'}{}(\snF{n}{\xi}{}{};\xi),\quad 
\firstP{AA'}{}[\kappa,\xi^{(n)}]
= \firstP{AA'}{}(\csnF{n}{\xi}{}{};\kappa) . 
\label{pfoldzerofirst}
\endEQ

One can verify by a direct computation that 
the spinorial forms of the currents
\eqref{extTconslaw}, \eqref{extZconslaw}, \eqref{extVconslaw}
are given by, respectively, 
\EQs
&& \curr{(n)AA'}{\rm T}(F;\xi) = 
\zeroP{B'}{A}[\xi^{(n+1)}] \csnF{n}{\xi}{A'B'}{}
+ \zerocP{B}{A'}[\xi^{(n+1)}] \snF{n}{\xi}{AB}{} , 
\label{symmTdensity}\\
&& \curr{(n)AA'}{\rm Z}(F;\xi) = 
2 \i \zeroP{B'}{A}[\xi^{(n+2)}] \csnF{n}{\xi}{A'B'}{}
- 2 \i \zerocP{B}{A'}[\xi^{(n+2)}] \snF{n}{\xi}{AB}{} , 
\label{symmZdensity}\\
&& \curr{(n)AA'}{\rm V}(F;\xi,Y) = 
8\firstP{B'}{A}[\kappa,\xi^{(n+1)}] \csnF{n}{\xi}{A'B'}{}
+ 8\firstcP{B}{A'}[\kappa,\xi^{(n+1)}] \snF{n}{\xi}{AB}{} , 
\label{symmVdensity}
\endEQs
where $\KS{}{A'B'C'D'} = 3\KY{(A'B'}{}\KY{C'D')}{}$.

\Proclaim{ Proposition 6.1.  }{
The currents 
$\curr{(n)AA'}{\rm T}(F;\xi)$, 
$\curr{(n)AA'}{\rm Z}(F;\xi)$,
$\curr{(n)AA'}{\rm V}(F;\xi,Y)$
admit spinorial characteristics \eqref{characteristic}
given by, respectively,  
\EQs
\sTQ{(n)AA'}[\xi] &=&
(-1)^n 2 \zeroP{AA'}{}[\xi^{(2n+1)}] , \quad
\label{symmTchar}\\
\sZQ{(n)AA'}[\xi] &=&
(-1)^n 4 \i \zeroP{AA'}{}[\xi^{(2n+2)}] , \quad
\label{symmZchar}\\
\sVQ{(n)AA'}[\xi,Y] &=&
(-1)^n 16
\sum_{a=0}^n \frac{n!}{(n-a)!a!} 
\firstP{AA'}{}[(\Lie{\xi})^a \kappa, \xi^{(2n+1-a)}] . 
\label{symmVchar}
\endEQs
These characteristics are adjoint symmetries of order 
$q_T=2n,q_Z=2n+1,q_V=2n+2$, respectively. }

\proclaim{ Proof. }
The calculation of each characteristic is similar. 
We will prove \eqref{symmTchar} and omit the proofs of the other two. 

By Lemma~5.1 and equation \eqref{characteristic}, 
the spinorial characteristic of current \eqref{symmTdensity} for $n=0$ is 
$\sTQ{(0)AA'}[\xi] = 2\zeroP{AA'}{}[\xi]$
and thus we have 
\EQ
\sD{AA'}{} \curr{(0)AA'}{\rm T}(F;\xi) = 
2\zeroP{AA'}{}[\xi] \cME{}{AA'} 
+ 2\zerocP{AA'}{}[\xi] \ME{}{AA'} 
+\sD{AA'}{} \triv{AA'}{1} , 
\label{zeroTchar}
\endEQ
where $\triv{AA'}{1}$ is a trivial current. 
Now, 
by replacing $\sF{AB}{}$ by $\snF{n}{\xi}{AB}{}$ in \eqref{zeroTchar}
and then using equations \eqrefs{nfoldF}{pfoldzerofirst}
together with equation \eqref{LiecME}, 
we obtain
\EQs
&& \sD{AA'}{} \curr{(n)AA'}{\rm T}(F;\xi) 
\nonumber\\&&
= 2\zeroP{AA'}{}[\xi^{(n+1)}] 
(\pr\gen{\xi})^n \cME{}{AA'}
+ 2\zerocP{}{AA'}[\xi^{(n+1)}] 
(\pr\gen{\xi})^n \ME{}{AA'}
+\sD{AA'}{} \triv{AA'}{2}
\nonumber\\&& 
= 2\cME{}{AA'} 
(-\jetLie{\xi})^n \zeroP{AA'}{}[\xi^{(n+1)}] 
+ 2\ME{}{AA'}
(-\jetLie{\xi})^n \zerocP{AA'}{}[\xi^{(n+1)}] 
+ \sD{AA'}{} \triv{AA'}{3} ,
\endEQs
where $\triv{AA'}{2},\triv{AA'}{3}$ are trivial currents. 
By a repeated application of \eqref{LieR} 
and a comparison with \eqref{characteristic} 
we conclude that
$\curr{(n)AA'}{\rm T}(F;\xi)$ admits 
the spinorial characteristic \eqref{symmTchar}. 
\endproof

\Proclaim{ Lemma 6.2. }{
The currents
$\TH{AA'}[\xi,\zeta_{1}, \dots,\zeta_{2r}]$, 
$\ZH{AA'}[\xi,\zeta_{1}, \dots,\zeta_{2r+1}]$,
$\VH{AA'}[\kappa,\zeta_{1}, \dots,\zeta_{2r+1}]$
for $r\ge 0$ 
are, respectively, 
equivalent to a linear combination of the currents 
\EQs
\curr{(n)AA'}{\rm T}(F;\zeta) , \quad
\curr{(n)AA'}{\rm Z}(F;\zeta) , \quad
\curr{(n)AA'}{\rm V}(F;\zeta,Y) , \quad
0\le n\le r, 
\label{symmVspan}
\endEQs
involving a sum over 
conformal \Kvec/s $\zeta$ and conformal \KYten/s $Y$
for each $n$. }

\proclaim{ Proof. }
The computations to prove the equivalence 
are similar for each pair of currents. 
We will prove the claim for the pair 
$\TH{AA'}[\KV{}{},\othKV{}{1}, \dots,\othKV{}{2r}]$, 
$\curr{(n)AA'}{\rm T}(F;\zeta)$
and omit the other two. 

The image of the current 
$\TH{AA'}[\KV{}{},\othKV{}{1}, \dots,\othKV{}{2r}]$
under the map \eqref{evenmap}
is the \Kspin/ 
$\KV{(C'}{(C} \othKV{\hp{1}C'_1}{1C_1} \cdots 
\othKV{\hp{2r}C'_{2r})}{2rC_{2r})}$
which is a symmetric multilinear expression in 
$\xi,\zeta_1,\ldots,\zeta_{2r}$
and hence can be written as a linear combination of 
powers of \Kspin/s of the form 
$\spinor{\varrho}{(C'}{(C} \spinor{\varrho}{C'_1}{C_1} 
\cdots \spinor{\varrho}{C'_{2r})}{C_{2r})}$. 
Thus by Theorem~5.5, 
there is a linear combination of currents 
$\TH{AA'}[\varrho^{(2r+1)}]$
that is equivalent to the current 
$\TH{AA'}[\KV{}{},\othKV{}{1}, \dots,\othKV{}{2r}]$
up to a current of lower weight. 
Hence, by induction on $r$, 
we have that for all $r\geq 0$
the current 
$\TH{AA'}[\KV{}{},\othKV{}{1}, \dots,\othKV{}{2r}]$
is equivalent to a linear combination of currents
$\TH{AA'}[\varrho^{(2n+1)}]$, $0\leq n \leq r$,
each of which 
is equivalent to a multiple of a current \eqref{symmTdensity} 
by the relation 
$\sTQ{(n)AA'}[\varrho]= 2(-1)^n \sTQ{AA'}[\varrho^{(2n+1)}]$
and Proposition~3.1.
Thus, 
the current $\TH{AA'}[\KV{}{},\othKV{}{1}, \dots,\othKV{}{2r}]$
lies in the equivalence class of 
a linear combination of the currents 
$\curr{(n)AA'}{\rm T}(F;\zeta)$, $0\leq n\leq r$. 
\endproof

We conclude with the following Proposition whose proof is immediate. 

\Proclaim{ Proposition 6.3. }{
Under the duality transformation \eqref{sduality},
the currents 
$\curr{(n)AA'}{\rm T}(F;\xi)$, 
$\curr{(n)AA'}{\rm Z}(F;\xi)$,
$\curr{(n)AA'}{\rm V}(F;\xi,Y)$
transform as
\EQ
\curr{(n)AA'}{\rm T} \to \curr{(n)AA'}{\rm T}, \quad
\curr{(n)AA'}{\rm Z} \to \curr{(n)AA'}{\rm Z}, \quad
\curr{(n)AA'}{\rm V} \to -\curr{(n)AA'}{\rm V} . 
\endEQ }

\subsection{ A basis }
\label{basis}

Here we present a basis for 
tensorial currents given in Theorem~2.1
by using the basis for the spinorial currents given in Theorems~5.8 and~5.9.

First, by fixing a spinor basis $\{\so{A}{},\si{A}{}\}$
we obtain a null tetrad basis
for vectors in Minkowski space, 
\EQs
&& \l{\mu} = \inve{\mu}{AA'} \so{A}{} \cso{A'}{} , 
\n{\mu} = \inve{\mu}{AA'} \si{A}{} \csi{A'}{} , 
\m{\mu} = \inve{\mu}{AA'} \so{A}{} \csi{A'}{} , 
\cm{\mu} = \inve{\mu}{AA'} \cso{A'}{} \si{A}{} . 
\label{nulltetrad}
\endEQs
Products of the basis spinors $\so{A}{},\si{A}{}$
yield a basis for complex skew-tensors in Minkowski space, 
\EQs
&& \inve{\mu}{AA'}\inve{\nu}{BB'} \so{A}{}\so{B}{} \vol{}{A'B'}
= \l{[\mu} \m{\nu]} + \i *\l{[\mu} \m{\nu]} ,  
\label{spinprod}\\
&& \inve{\mu}{AA'}\inve{\nu}{BB'} \so{(A}{}\si{B)}{} \vol{}{A'B'}
= \l{[\mu} \n{\nu]} + \i *\l{[\mu} \n{\nu]} , 
\label{spinprod'}\\
&& \inve{\mu}{AA'}\inve{\nu}{BB'} \si{A}{}\si{B}{} \vol{}{A'B'}
= \n{[\mu} \cm{\nu]} + \i *\n{[\mu} \cm{\nu]} . 
\label{spinprod''}
\endEQs

In terms of expressions \eqsref{nulltetrad}{spinprod''}, 
the basis \eqsref{basisCKV0th}{basisCKV2nd} 
for complex conformal \Kvec/s 
and the basis \eqsref{basisKY0th}{basisKY2nd} 
for self-dual conformal \KYten/s 
takes the form 
\EQs
&& 
\KV{\mu}{0,1} = \l{\mu}, \quad
\KV{\mu}{0,2} = \m{\mu}, \quad
\cKV{\mu}{0,2} = \cm{\mu}, \quad
\KV{\mu}{0,3} = \n{\mu}, 
\label{tensbasisCKV0th}\\
&& 
\KV{\mu}{1,1} = \x{}{\nu}( 1 - \i * ) \l{[\mu}\cm{\nu]} , \quad
\KV{\mu}{1,2} = \x{}{\nu}( 1 - \i * )\l{[\mu}\n{\nu]} , \quad
\KV{\mu}{1,3} = \x{}{\nu}( 1 - \i * )\m{[\mu}\n{\nu]} , 
\label{tensbasisCKV1st}\\
&& 
\cKV{\mu}{1,1} = \x{}{\nu}( 1 + \i * ) \l{[\mu}\m{\nu]} , \quad
\cKV{\mu}{1,2} = \x{}{\nu}( 1 + \i * )\l{[\mu}\n{\nu]} , \quad
\cKV{\mu}{1,3} = \x{}{\nu}( 1 + \i * )\cm{[\mu}\n{\nu]} , 
\label{tenscbasisCKV1st}\\
&& 
\KV{\mu}{1,4} = \x{\mu}{\nu}, 
\label{tensbasisCKV1st'}\\
&& 
\KV{\mu}{2,1} = \x{\mu}{}\x{}{\nu}\l{\nu} 
-\frac{1}{2} \x{\nu}{}\x{}{\nu} \l{\mu},  \quad
\KV{\mu}{2,2} = \x{\mu}{}\x{}{\nu}\m{\nu} 
-\frac{1}{2} \x{\nu}{}\x{}{\nu}\m{\mu}, 
\\&&
\cKV{\mu}{2,2} = \x{\mu}{}\x{}{\nu}\cm{\nu} 
-\frac{1}{2} \x{\nu}{}\x{}{\nu}\cm{\mu}, \quad
\KV{\mu}{2,3} = \x{\mu}{}\x{}{\nu}\n{\nu} 
-\frac{1}{2} \x{\nu}{}\x{}{\nu}\n{\mu}, 
\label{tensbasisCKV2nd}
\endEQs
and 
\EQs
&& 
\KY{0,1}{\mu\nu} = 
( 1 + \i * ) \l{[\mu}\m{\nu]} , \quad
\KY{0,2}{\mu\nu} = 
(1 + \i * ) \l{[\mu}\n{\nu]} , \quad
\KY{0,3}{\mu\nu} = 
( 1+ \i * )\n{[\mu}\cm{\nu]} , 
\label{tensbasisKY0th}\\
&& 
\KY{1,1}{\mu\nu} = 
(1 + \i *) \x{[\mu}{}\l{\nu]} , \quad
\KY{1,2}{\mu\nu} = 
( 1+ \i * ) \x{[\mu}{}\m{\nu]} , 
\\&&
\KY{1,3}{\mu\nu} = 
( 1+ \i * ) \x{[\mu}{}\cm{\nu]} , \quad 
\KY{1,4}{\mu\nu} = 
( 1 + \i * )\x{[\mu}{}\n{\nu]} , 
\label{tensbasisKY1st}\\
&& 
\KY{2,1}{\mu\nu} = 
\x{\mu}{} \x{}{\sigma}( 1- \i * ) \l{[\nu}\m{\sigma]} 
- \x{\nu}{} \x{}{\sigma}( 1- \i * ) \l{[\mu}\m{\sigma]} 
+\frac{1}{2} \x{2}{} (1-\i *) \l{[\mu}\m{\nu]} , 
\\&&
\KY{2,2}{\mu\nu} = 
\x{\mu}{} \x{}{\sigma}( 1- \i * ) \l{[\nu}\n{\sigma]} 
- \x{\nu}{} \x{}{\sigma}( 1- \i * ) \l{[\mu}\n{\sigma]} 
+\frac{1}{2} \x{2}{} (1-\i *) \l{[\mu}\n{\nu]} , 
\\&&
\KY{2,3}{\mu\nu} = 
\x{\mu}{} \x{}{\sigma}( 1- \i * ) \cm{[\nu}\n{\sigma]} 
- \x{\nu}{} \x{}{\sigma}( 1- \i * ) \cm{[\mu}\n{\sigma]} 
+\frac{1}{2} \x{2}{} (1-\i *) \cm{[\mu}\n{\nu]} .
\label{tensbasisKY2nd}
\endEQs
Note that, geometrically, 
the \Kvec/s \eqsref{tensbasisCKV0th}{tensbasisCKV2nd}
describe 4 null translations, 6 null boosts, 1 dilation, 
4 null conformal transformations. 

\Proclaim{ Proposition~6.4. }{
Let $\KV{}{i}$ denote the conformal \Kvec/s in 
\eqsref{basisTdensity}{basisZ'density}
and \eqsref{basisVdensity}{basisV'density'}
written in the tensorial form 
\eqsref{tensbasisCKV0th}{tensbasisCKV2nd},
and let $\KY{i}{}$ denote the conformal \KYten/s 
in \eqsref{basisVdensity}{basisV'density'}
written in the tensorial form 
\eqsref{tensbasisKY0th}{tensbasisKY2nd}. 
Then the tensorial forms of the currents 
\eqsref{basisTdensity}{basisZ'density}, 
\eqsref{basisVdensity}{basisV'density'} 
are, respectively, given by 
\EQs
&& \TH{\mu} = 
\F{\mu\sigma}{} \nF{q}{\nu\sigma}{}
[\KV{}{1},\ldots,\KV{}{q}] \KV{\nu}{q+1} 
-\frac{1}{4} \F{\nu\sigma}{} \nF{q}{\nu\sigma}{} 
[\KV{}{1},\ldots,\KV{}{q}] \KV{\mu}{q+1} 
+c.c., 
\label{basisTconslaw}\\
&& \H{\mu}{T'} = 
\i\F{\mu\sigma}{} \nF{q}{\nu\sigma}{}
[\KV{}{1},\ldots,\KV{}{q}] \KV{\nu}{q+1} 
-\frac{\i}{4} \F{\nu\sigma}{} \nF{q}{\nu\sigma}{} 
[\KV{}{1},\ldots,\KV{}{q}] \KV{\mu}{q+1} 
+c.c., 
\label{basisT'conslaw}\\
&& \ZH{\mu} = 
\F{\mu\sigma}{} \dunF{q}{\nu\sigma}{}
[\KV{}{1},\ldots,\KV{}{q}] \KV{\nu}{q+1} 
- \duF{\mu\sigma}{} \nF{q}{\nu\sigma}{}
[\KV{}{1},\ldots,\KV{}{q}] \KV{\nu}{q+1} 
+c.c., 
\label{basisZconslaw}\\
&& \H{\mu}{Z'} = 
\i\F{\mu\sigma}{} \dunF{q}{\nu\sigma}{}
[\KV{}{1},\ldots,\KV{}{q}] \KV{\nu}{q+1}
-\i \duF{\mu\sigma}{} \nF{q}{\nu\sigma}{}
[\KV{}{1},\ldots,\KV{}{q}] \KV{\nu}{q+1}
+c.c., 
\label{basisZ'conslaw}\\
&& \H{\mu}{+V} = 
\F{}{\nu\sigma} ( \coD{\mu}\nF{q}{\alpha\beta}{}
[\KV{}{1},\ldots,\KV{}{q-1}] ) 
\KY{}{\nu\sigma\alpha\beta} 
+4 \F{[\mu}{\sigma} ( \D{\nu}\nF{q}{\alpha\beta}{} 
[\KV{}{1},\ldots,\KV{}{q-1}] ) 
\KY{}{\nu]\sigma\alpha\beta}
\nonumber\\&&\qquad
+\frac{3}{5} \F{}{\nu\sigma} \nF{q}{\alpha\beta}{}
[\KV{}{1},\ldots,\KV{}{q-1}] 
\coder{\mu} \KY{}{\nu\sigma\alpha\beta} 
+\frac{12}{5} \F{[\mu}{\sigma} \nF{q}{\alpha\beta}{}
[\KV{}{1},\ldots,\KV{}{q-1}]
\der{\nu}\KY{}{\nu]\sigma\alpha\beta} 
+c.c., 
\label{basisVconslaw}\\
&& \H{\mu}{+V'} = 
\i\F{}{\nu\sigma} ( \coD{\mu}\nF{q}{\alpha\beta}{}
[\KV{}{1},\ldots,\KV{}{q-1}] ) 
\KY{}{\nu\sigma\alpha\beta} 
+4\i \F{[\mu}{\sigma} ( \D{\nu}\nF{q}{\alpha\beta}{} 
[\KV{}{1},\ldots,\KV{}{q-1}] ) 
\KY{}{\nu]\sigma\alpha\beta}
\nonumber\\&&\qquad
+\frac{3\i}{5} \F{}{\nu\sigma} \nF{q}{\alpha\beta}{}
[\KV{}{1},\ldots,\KV{}{q-1}] 
\coder{\mu} \KY{}{\nu\sigma\alpha\beta} 
+\frac{12\i}{5} \F{[\mu}{\sigma} \nF{q}{\alpha\beta}{}
[\KV{}{1},\ldots,\KV{}{q-1}]
\der{\nu}\KY{}{\nu]\sigma\alpha\beta} 
+c.c., 
\label{basisV'conslaw}\\
&& \H{\mu}{-V} = 
\F{}{\nu\sigma} ( \coD{\mu}\dunF{q-1}{\alpha\beta}{}
[\KV{}{1},\ldots,\KV{}{q-1}] ) 
\KY{}{\nu\sigma\alpha\beta} 
+4 \F{[\mu}{\sigma} ( \D{\nu}\dunF{q-1}{\alpha\beta}{} 
[\KV{}{1},\ldots,\KV{}{q-1}] ) 
\KY{}{\nu]\sigma\alpha\beta}
\nonumber\\&&\qquad
+\frac{3}{5} \F{}{\nu\sigma} \dunF{q-1}{\alpha\beta}{}
[\KV{}{1},\ldots,\KV{}{q-1}] 
\coder{\mu} \KY{}{\nu\sigma\alpha\beta} 
+\frac{12}{5} \F{[\mu}{\sigma} \dunF{q-1}{\alpha\beta}{}
[\KV{}{1},\ldots,\KV{}{q-1}]
\der{\nu}\KY{}{\nu]\sigma\alpha\beta} 
+c.c., 
\label{basisVconslaw'}\\
&& \H{\mu}{-V'} = 
\i\F{}{\nu\sigma} ( \coD{\mu}\dunF{q-1}{\alpha\beta}{}
[\KV{}{1},\ldots,\KV{}{q-1}] ) 
\KY{}{\nu\sigma\alpha\beta} 
+4\i \F{[\mu}{\sigma} ( \D{\nu}\dunF{q-1}{\alpha\beta}{} 
[\KV{}{1},\ldots,\KV{}{q-1}] ) 
\KY{}{\nu]\sigma\alpha\beta}
\nonumber\\&&\qquad
+\frac{3\i}{5} \F{}{\nu\sigma} \dunF{q-1}{\alpha\beta}{}
[\KV{}{1},\ldots,\KV{}{q-1}] 
\coder{\mu} \KY{}{\nu\sigma\alpha\beta} 
+\frac{12\i}{5} \F{[\mu}{\sigma} \dunF{q-1}{\alpha\beta}{}
[\KV{}{1},\ldots,\KV{}{q-1}]
\der{\nu}\KY{}{\nu]\sigma\alpha\beta} 
+ c.c. , 
\label{basisV'conslaw'}
\endEQs
where
\EQs
\nF{n}{\mu\nu}{}[\KV{}{1},\ldots,\KV{}{n}] && 
= \frac{1}{n!} \sum_{\s\in S_n} 
\Lie{\KV{}{s(1)}} \cdots \Lie{\KV{}{s(n)}} \F{}{\mu\nu} , 
\label{Fextension'}\\
\KY{}{\nu\sigma\alpha\beta} &&
= \frac{1}{2} \sum_{\s\in S_2} \Big( 
\KY{s(1)}{\nu\sigma} \KY{s(2)}{\alpha\beta}
- \KY{s(1)}{\nu[\alpha} \KY{s(2)}{\beta]\sigma}
-3 \invflat{[\nu|[\alpha} 
\KY{s(1)\tau}{\beta]} \KY{s(2)}{\tau|\sigma]}
\Big)
\nonumber\\&&\fewquad
+\frac{1}{2} \invflat{\nu[\alpha} \invflat{\beta]\sigma} 
\KY{1\tau\lambda}{} \KY{2}{\tau\lambda} . 
\endEQs
Here $c.c.$ stands for the complex conjugate of
all preceding terms in an expression. 
}

The proof of Proposition~6.4 is a straightforward computation
and will be omitted. 

Now, by converting the basis of currents in Theorems~5.8 and~5.9 
into tensorial form using Proposition~6.4, 
we obtain the following tensorial basis 
for the vector space of quadratic currents 
$\vs{w}{} =\vs{w}{+} \oplus \vs{w}{-}$ 
of weight at most $w$. 

\Proclaim{ Theorem~6.5 }{
A tensorial basis for $\vs{w}{+}$ is given by 
the set of all currents 
\eqrefs{basisTconslaw}{basisT'conslaw}
for $0\le p\le 2q+2$, $q=2r$, $0\le r\le [w/2]$, 
and 
\eqrefs{basisZconslaw}{basisZ'conslaw} 
for $0\le p\le 2q+2$, $q=2r+1$, $0\le r\le [(w-1)/2]$, 
in which the conformal \Kvec/s are given by 
the count formulas \eqsref{count1}{count15}
indexed by $i,j,n,n'$ satisfying \eqrefs{ijrange}{nn'range}, 
with currents \eqrefs{basisT'conslaw}{basisZ'conslaw} 
restricted to $i\ne j$ or $n\neq n'$. 
A tensorial basis for $\vs{w}{-}$ is given by 
the set of all currents 
\eqsref{basisVconslaw}{basisV'conslaw'}
for $0\le p\le 2q+2$, $q=2r+2$, $0\le r\le [w/2]-1$, 
in which the conformal \Kvec/s and conformal \KYten/s are given by 
the respective count formulas 
\eqsref{count1}{count15} and \eqsref{count1'}{count10'}
indexed by $i,j,k,n,n',m,m'$ satisfying \eqsref{kp'range}{mm'range}, 
with currents \eqrefs{basisV'conslaw}{basisV'conslaw'} 
restricted to $i\ne j$ or $n\neq n'$. 
}

We remark that 
the tensorial basis currents in Theorem~6.5 can be expressed
as linear combinations of the currents in Theorem~2.2
by the procedure used in the proof of Lemma~6.2.

\section{Concluding Remarks}
\label{conclude}

In this paper 
we classify all local \conslaw/s of \Meq/ in Minkowski space
in a systematic fashion by classifying their characteristics.
Even though \Meq/ are a degenerate system of PDEs, 
we are able to establish a one-to-one correspondence between 
classes of equivalent conserved currents 
and classes of equivalent characteristics. 
We find the characteristics by solving 
the adjoint symmetry equations of \Meq/ 
by means of spinorial methods, 
leading, essentially, to a one-to-one correspondence between 
classes of adjoint symmetries 
and Killing spinors of certain type. 
Interestingly, 
we find classes of adjoint symmetries 
that are not equivalent to characteristics 
and hence do not correspond to conserved currents. 
We also identify a recursion structure within the spaces of 
adjoint symmetries and conserved currents
which is induced by 
Lie derivatives with respect to conformal \Kvec/s. 
The use of spinorial methods allows us to obtain 
all conserved currents explicitly,
in a unified manner in coordinate invariant form in terms of \Kspin/s,
and this leads to the identification of 
new chiral conserved currents along with an associated conserved tensor. 
In addition, 
by means of a factorization of \Kspin/s in Minkowski space,
we exhibit a basis for conserved currents of any order or weight. 

Moreover, 
our classification extends to \conslaw/s of \Meq/ 
$\covcoder{\mu} \F{}{\mu\nu}(x)=0$
and $\covcoder{\mu} \duF{}{\mu\nu}(x)=0$
in a curved background metric $\g{\mu\nu}$. 
Here $\covder{\mu}$ and $*$ stand for 
the torsion-free covariant derivative 
and Hodge star operator 
associated to $\g{\mu\nu}$. 
All local \conslaw/s 
continue to arise from adjoint symmetries of \Meq/ 
through the integral formula \eqref{Intformula}. 
The adjoint symmetries can be obtained in a straightforward manner
in spinor form and involve \Kspin/s of the curved metric. 
Interestingly, 
the \Kspin/ equations now possess integrability conditions \cite{Penrose}
which lead to restrictions on the curvature tensor of $\g{\mu\nu}$.
Furthermore, additional curvature conditions arise
from the determining equations for the adjoint symmetries. 

Consequently, non-trivial \conslaw/s of \Meq/
exist only for certain classes of metrics $\g{\mu\nu}$.
A complete analysis of the curvature conditions will be explored elsewhere.
Of particular interest is the family of black-hole spacetime metrics, 
since \Meq/ admit non-trivial symmetries \cite{Kalninsetal}
in addition to the symmetries due to spacetime isometries 
in the Kerr spacetime metric. 
The methods of \secsref{method}{currents}
can be expected to resolve the issue of whether \Meq/ 
possess corresponding \conslaw/s. 

The relation between the local \conslaw/s and local symmetries of \Meq/,
in flat and curved spacetime, will be explored in a subsequent paper
\cite{symmpaper}. 

Finally, our methods can be extended to the analysis of local \conslaw/s 
of other physical field equations,
in particular, 
the linearized gravity wave equation on flat and curved spacetimes.

\end{document}